\documentclass[aps,prd,twocolumn,superscriptaddress,nofootinbib,longbibliography]{revtex4-2}

\usepackage[english]{babel}

\usepackage{graphicx}
\usepackage{dcolumn}
\usepackage{bm}
\usepackage{lipsum}
\usepackage{amsmath,amssymb}
\usepackage{dsfont}
\usepackage{float}
\usepackage{array}
\makeatletter
\let\newfloat\newfloat@ltx
\makeatother
\usepackage{algpseudocode}
\usepackage{multirow}
\usepackage{colortbl}
\usepackage{booktabs}
\usepackage{stmaryrd}
\newcolumntype{C}{>{$}c<{$}}

\AtBeginDocument{%
  \heavyrulewidth=.08em
  \lightrulewidth=.05em
  \cmidrulewidth=.03em
  \belowrulesep=.65ex
  \belowbottomsep=0pt
  \aboverulesep=.4ex
  \abovetopsep=0pt
  \cmidrulesep=\doublerulesep
  \cmidrulekern=.5em
  \defaultaddspace=.5em
}

\usepackage{verbatim}

\usepackage[normalem]{ulem}
\usepackage[utf8]{inputenc}
\usepackage{epigraph}
\usepackage{siunitx}
\usepackage[english]{babel}
\usepackage{listings}
\usepackage{graphicx}
\usepackage[utf8]{inputenc}
\usepackage{listings}
\usepackage{matlab-prettifier}
\usepackage{xcolor}
\usepackage{amsfonts}
\usepackage{amsmath}
\usepackage{amssymb}
\usepackage{amsthm}
\usepackage{empheq} 
\newcolumntype{P}[1]{>{\centering\arraybackslash}p{#1}}
\newcolumntype{M}[1]{>{\centering\arraybackslash}m{#1}}
\usepackage{braket}
\usepackage{xcolor}
\usepackage[colorlinks = true,
            linkcolor = blue,
            urlcolor  = blue,
            citecolor = blue,
            anchorcolor = blue]{hyperref}
\usepackage{comment}
\usepackage{hyperref}
\usepackage{orcidlink}
\usepackage{tcolorbox} 
\usepackage{color}

\newcommand{\be}{\begin{equation}}
\newcommand{\ee}{\end{equation}}

\newcommand{\bW}{\boldsymbol{W}}
\newcommand{\bSigma}{\boldsymbol{\Sigma}}
\newcommand{\bn}{\boldsymbol{n}}
\newcommand{\bv}{\boldsymbol{v}}
\newcommand{\bM}{\boldsymbol{M}}
\newcommand{\bN}{\boldsymbol{N}}
\newcommand{\blambda}{\boldsymbol{\lambda}}
\newcommand{\bR}{\boldsymbol{R}}
\newcommand{\bP}{\boldsymbol{P}}
\newcommand{\bF}{\boldsymbol{F}}
\newcommand{\bQ}{\boldsymbol{Q}}
\newcommand{\bh}{\boldsymbol{h}}
\newcommand{\bd}{\boldsymbol{d}}
\newcommand{\bx}{\boldsymbol{x}}
\newcommand{\bTheta}{\boldsymbol{\Theta}}
\newcommand{\btheta}{\boldsymbol{\theta}}
\newcommand{\bUpsilon}{\boldsymbol{\Upsilon}}
\newcommand{\bXi}{\boldsymbol{\Xi}}
\newcommand{\bGamma}{\boldsymbol{\Gamma}}
\providecommand{\bC}{\boldsymbol{C}}
\providecommand{\br}{\boldsymbol{r}}
\newcommand{\fpnew}[1]{\textcolor{orange}{#1}}

\usepackage{fontawesome5}
\newcommand{\gaplikerepo}{https://github.com/gaplike/gaplike}
\newcommand{\gaplikedocs}{https://gaplike.github.io/gaplike}
\newcommand{\gapliketag}{0.2.0}
\newcommand{\gaplikedoi}{10.5281/zenodo.22141966}
\DeclareRobustCommand{\srclink}[1]{%
  \href{\gaplikerepo/blob/\gapliketag/paper/#1}{\faGithub\,\texttt{\detokenize{#1}}}}

\begin{document}

\title{Mind the second gap: The impact of gaps on noise and signal parameter inference}
\author{Ollie Burke\,\orcidlink{0000-0003-2393-209X}}
\email{ollie.burke@glasgow.ac.uk}
\affiliation{School of Physics and Astronomy, University of Glasgow, Glasgow G12 8QQ, United Kingdom}

\author{Federico Pozzoli\,\orcidlink{0009-0009-6265-584X}}
\email{federico.pozzoli@aei.mpg.de}
\affiliation{Max Planck Institute for Gravitational Physics (Albert Einstein Institute), Am Mühlenberg 1, Potsdam 14476, Germany}

\author{Martina Muratore\,\orcidlink{0000-0002-9630-5698}}
\affiliation{Max Planck Institute for Gravitational Physics (Albert Einstein Institute), Am Mühlenberg 1, Potsdam 14476, Germany}

\author{Jonathan R. Gair\,\orcidlink{0000-0002-1671-3668}}
\affiliation{Max Planck Institute for Gravitational Physics (Albert Einstein Institute), Am Mühlenberg 1, Potsdam 14476, Germany}
\begin{abstract}
Gravitational-wave data is never analysed without some conditioning: filtering
and tapering act multiplicatively on the noise, destroying the stationarity on
which the Whittle likelihood --- the backbone of nearly all gravitational-wave
inference --- rests. The problem is particularly evident for LISA, where gaps are
expected to overlap almost all signals and the noise must be estimated jointly with those astrophysical
signals.
We provide an analytical and numerical framework to assess the impact of noise mis-modelling when
signal and noise parameters are inferred jointly in the presence of gaps. From
a Fisher-matrix formalism verified against Bayesian inference, we obtain closed
forms for the
model uncertainties, their maximum-likelihood estimates and the true
scatter in the latter, for various approximations to the likelihood. The Fisher-based linear formalism is extended through the Godambe--White curvature to the non-linear regime
that severe mis-modelling produces.
We survey gap families from long, well-tapered interruptions to short,
high-rate ones, injecting massive black-hole binaries. Except in the case of frequent gaps, we show that the windowed covariance is well approximated by its leading diagonal ---
the window convolved with the power spectral density --- for the estimation of both signal and
noise parameters. By omitting the convolution, the signal sector remains unbiased but the noise parameters, in particular 
the optical-metrology noise which is dominant at high frequency,  
is displaced by over an order of magnitude in
power. At the highest glitch rates the diagonal approximation stays accurate yet loses an order of
magnitude in precision. To remedy this, we propose an exact time-domain likelihood of the surviving samples which can be efficiently (and accurately solved) via preconditioned conjugate gradient methods. 
\end{abstract}

\maketitle

\tableofcontents

\section{Introduction}

Noise inference within gravitational wave (GW) astronomy is fundamental for ensuring robust parameter estimation of astrophysical signals. Ground-based detectors within the umbrella of the LIGO--Virgo--KAGRA (LVK) collaboration observe signal-free stretches of data allowing for the noise properties to be estimated prior to signal recovery~\cite{Allen:2005fk, Veitch:2014wba}. Future ground-based detectors such as the Einstein Telescope (ET)~\cite{Abac_2026} or Cosmic Explorer~\cite{reitze2019cosmicexploreruscontribution}, as well as the space-based Laser Interferometer Space Antenna (LISA)~\cite{LISA:2024hlh} will not have this luxury. For future generation detectors, the noise properties must be inferred jointly with the characteristics of the astrophysical sources 
lurking within the data set. Noise estimation 
will therefore be performed jointly with inference on
the many overlapping signals present in the GW data stream~\cite{Littenberg:2023xpl, Strub:2024kbe, Katz:2024oqg, deng2025modular}. 
There is therefore a need to understand (1) source and noise parameter correlations and (2) the impact of signal and noise mis-modelling on parameter estimation. There is extensive literature on source mis-modelling -- it is well known that using unfaithful waveform templates 
to characterise GW signals in a data stream may result in biases on the parameter estimates~\cite{Cutler:2007mi, Miller:2005qu, Flanagan:1997kp}. Noise mis-modelling, on the other hand, is significantly less researched with potentially very dangerous consequences if misunderstood. The primary focus of this work is to understand the impact of noise-mismodelling on parameter estimation in the context of \emph{both} signal and noise parameter estimation.

Future generation detectors will be subject to 
a plethora of features that disturb the usual simplifying assumptions that are placed on the noise process. Literature within the GW field typically 
assumes that the noise process is both Gaussian and stationary with the added restraint of being circulant. 
The primary reason for this is that, under these assumptions, the (frequency-domain) likelihood takes an exceptionally simple form 
-- the Whittle likelihood ~\cite{whittle:1957,Finn:1992wt}. 
Realistic detectors generate data 
plagued with non-Gaussian transients (glitches)~\cite{Boumerdassi:2025gvf,  Muratore:2025knh,Houba:2024tyn, sauter2025maximumlikelihooddetectioninstrumental}, periods of data interruptions (gaps) ~\cite{Dey:2021dem,Pearson_2026,Burke:2025bun,Wang:2024ovi,Wang:2024ovi,Carre:2010ra,Blelly:2021oim, Baghi:2019eqo}, and inherent time-evolving structure in the noise properties (non-stationary features)~\cite{Cornish:2026tjt,Edwards:2020tlp,Zackay:2019kkv,2025arXiv251103604B}. Each of these features render the Whittle-based likelihood only approximate for parameter inference, potentially resulting in biased parameter estimates and/or altered parameter uncertainties that could lead to false conclusions on the characteristics  of the sources. 

The work of ~\cite{Burke:2025bun} focused on the impact of noise mis-modelling on signal parameter estimation with the primary focus on missing data. When data gaps are handled via apodization (windowing) techniques, as demonstrated in ~\cite{Talbot:2021igi,Edy:2021par,Burke:2025bun}, mis-specifying the model noise-covariance matrix may still result in incorrect parameter uncertainties. The Whittle-based covariance matrix requires a correction by convolving the window with the noise power spectral density (PSD) in order to account for the leakage effects introduced by finite time-series -- a feature the Whittle-likelihood does not account for. The work of~\cite{Burke:2025bun} concluded that a Whittle-based likelihood could be used for parameter inference provided the taper was smooth enough to concentrate power in the main diagonal of the covariance matrix. However, one limitation of this work is that the underlying PSD 
was assumed to be perfectly known and the focus was on the impact on signal parameter estimation. In this work, we will generalise~\cite{Burke:2025bun} to the case where both the signal and noise parameters are unknown, reflecting a more realistic scenario for future generation GW detectors. 

The methodology described in this article is applicable to \emph{any} time-domain function that tampers with the noise process. Be it data gaps as a result of instrumental malfunctions (or maintenance), glitches, modulations and time-domain filters. Techniques described in this work are also applicable to the key algorithm that will underpin search and recovery of GWs in LISA, the Global Fit. The Global-Fit is the data analysis technique that simultaneously infers the noise properties and resolvable astrophysical signals present in the data stream~\cite{Littenberg:2023xpl, Strub:2024kbe, Katz:2024oqg, deng2025modular}. Techniques to accelerate the wall-time of the likelihood function will require approximations and safe model mis-specification that do not bias the parameter estimation of either the signal or noise parameters. The work presented here will provide a toolkit of suitable (fast) likelihood functions that can be used to infer signal and noise parameters in the presence of missing data.

The remainder of this manuscript is organised as follows. We first fix our
notational conventions in Sec.~\ref{sec:conventions} and introduce the
fundamentals of GW data analysis in Sec.~\ref{sec:gwda}, with stationary and
non-stationary treatments given in Secs.~\ref{subsec:stationary_noise} and
\ref{subsec:tampered_noise_process}. The core methodology is developed in
Sec.~\ref{sec:methodology}: the joint signal/noise Fisher matrix in
Sec.~\ref{sec:joint_fisher_matrix}, the Hessian-based likelihood in
Sec.~\ref{sec:hessian_based_likelihood}, noise mis-modelling in
Sec.~\ref{sec:mismodelling_fisher_matrix}, the metrics that quantify its
impact in Sec.~\ref{subsec:metrics_scatter_to_width}, and the nonlinear
completion of that formalism, needed when the mis-modelling is severe, in
Sec.~\ref{sec:Godambe_White_formalism}.
Section~\ref{sec:approximations_windowed_fd_covariance} presents a hierarchy
of approximations to the windowed covariance matrix. Our massive black hole
binary (MBHB) configuration, noise model and gap scenarios are described in
Sec.~\ref{sec:setup_mbh_gaps_simulation}. The two routes to an affordable
likelihood follow: the frequency-domain techniques of
Sec.~\ref{sec:numerics}, and the time-domain conjugate-gradient solver of
Sec.~\ref{subsec:td_exact}. We verify our analytical calculations against
Bayesian inference in Sec.~\ref{sec:verification}, with results detailed in
Sec.~\ref{sec:results}. We summarise our findings and discuss their
implications for LISA noise inference in Sec.~\ref{sec:conclusions}, ending
with a discussion on future work. Supporting derivations and identities are
collected in the appendices.

As a final note -- we are aware that this paper is rather long and rather involved. For those with little time to read the details, the main conclusion is that to deal with gaps in the data, we suggest to taper the edges of the gaps and model the frequency domain covariance matrix as a diagonal matrix with elements given by 
\begin{equation}\label{eq:intro_best_cheap_approximation_covariance}
\tilde{\bSigma}(f;\blambda) = \frac{1}{2}\int |\tilde{w}(f - u)|^2 S_{n}(u;\blambda)\,\text{d}u
\end{equation}
for $w$ the time domain and $\tilde{w}$ the frequency domain window functions, with zeros representing missing data, and $S_n$ the noise power spectral density. Equation~\eqref{eq:intro_best_cheap_approximation_covariance} gives the exact variance of each frequency bin for any PSD. What it discards are the leakage-induced correlations \emph{between} bins, and these matter when the spectrum has sharp structure. In our case that structure comes from the zeros of the TDI transfer function.

\subsection{Conventions}\label{sec:conventions}

We analyse a continuous signal $x(t)$ through a finite record of $N$ samples taken at
uniformly spaced times $t_i = i\,\Delta t$, for $i = 0,1,\dots,N-1$, so that $x(t)$ is
represented by the collection of discrete samples $\{x(t_i)\}_{i=0}^{N-1}$. The cadence $\Delta t = t_{i+1} - t_i$ fixes the
spacing between neighbouring samples, with the total time of observation 
$T_{\rm obs} = N\Delta t$. 

Our convention for the continuous Fourier transform (CFT) and its inverse (ICFT) is
\begin{equation}\label{eq:CFT}
    \tilde{x}(f) = \mathcal{F}[x(t)] = \int_{-\infty}^{\infty} x(t)\,e^{-2\pi i f t}\,\mathrm{d}t\,,
\end{equation}
\begin{equation}\label{eq:ICFT}
    x(t) = \mathcal{F}^{-1}[\tilde{x}(f)] = \int_{-\infty}^{\infty} \tilde{x}(f)\,e^{2\pi i f t}\,\mathrm{d}f\,.
\end{equation}
Every signal considered here is real, so its transform obeys the Hermitian symmetry
$\tilde{x}(-f) = \tilde{x}(f)^{*}$; the negative-frequency content is thus fully determined
by the positive frequencies.

In the discrete setting we use boldface for the sampled arrays. The vector $\boldsymbol{x}$
holds the time-domain (TD) data, $\boldsymbol{x}_i = x(t_i)$, while $\tilde{\boldsymbol{x}}$
holds the frequency-domain (FD) data at resolution $\Delta f = 1/T = 1/(N\Delta t)$. We
follow the standard FFT bin ordering, labelling the FD vector from $0$ to $N-1$ with the
upper half carrying the negative frequencies: $\tilde{\boldsymbol{x}}_i = \tilde{x}(i\Delta f)$
for $i = 0,\dots,N/2-1$, and $\tilde{\boldsymbol{x}}_i = \tilde{x}\big((i-N)\Delta f\big)$
for $i = N/2,\dots,N-1$. For an even sample count, $N\in 2\mathbb{Z}^{+}$, the discrete
Fourier transform (DFT) pair relating the two representations reads
\begin{align}\label{eq:DFT}
    \tilde{\boldsymbol{x}}_j &= \Delta t \sum_{i=0}^{N-1} \boldsymbol{x}_i\, e^{-2\pi i f_j t_i}\,, \\
    \boldsymbol{x}_i &= \Delta f \sum_{j=0}^{N-1} \tilde{\boldsymbol{x}}_j\, e^{2\pi i f_j t_i}\,, \label{eq:inverseDFT}
\end{align}
with $t_i = i\Delta t$ and $f_j = j\Delta f$; the relation $\Delta t\,\Delta f = 1/N$ lets us
write the phase as $t_i f_j = ij/N$.

It is convenient to view the DFT as a linear map between the two domains,
\begin{subequations}
\begin{align}
    \tilde{\boldsymbol{x}} &= \Delta t\,\sqrt{N}\,\boldsymbol{P}\,\boldsymbol{x}\,, \label{eq:defDFT}\\
    \boldsymbol{x} &= \frac{1}{\Delta t\,\sqrt{N}}\,\boldsymbol{P}^{\dagger}\,\tilde{\boldsymbol{x}}\,, \label{eq:defIDFT}
\end{align}
\end{subequations}
where the DFT matrix $\boldsymbol{P}$ is assembled from powers of the primitive $N$-th root
of unity $\omega \equiv e^{2\pi i/N}$,
\begin{equation}\label{eq:def_P_jk_matrix}
    \boldsymbol{P}_{jk} = \frac{1}{\sqrt{N}}\,\omega^{-jk}\,.
\end{equation}
By construction $\boldsymbol{P}$ is unitary, $\boldsymbol{P}^{\dagger} = \boldsymbol{P}^{-1}$
(with ${}^{\dagger}$ denoting the Hermitian conjugate), and symmetric,
$\boldsymbol{P}^{T} = \boldsymbol{P}$. We normalise $\boldsymbol{P}$ and
$\boldsymbol{P}^{\dagger}$ symmetrically so that both are dimensionless, the dimensionful
factors being carried instead by $\boldsymbol{x}$ and $\tilde{\boldsymbol{x}}$ in the TD and
FD respectively. This matrix picture is a bookkeeping device for analytic manipulations
only: in practice $\boldsymbol{P}$ is never formed explicitly, since the FFT evaluates the
transform in $\mathcal{O}(N\log_2 N)$ operations rather than the $\mathcal{O}(N^2)$ cost of
a dense matrix--vector product.

Angle brackets denote an ensemble average over realisations of the relevant random process.
For a wide-sense stationary, ergodic continuous process $X(t)$ the two-point correlation
function is $\Sigma(t,t') = \langle X(t)X^{*}(t')\rangle$, whose discrete counterpart is the
data covariance matrix
\begin{equation}\label{eq:cov_def}
    \boldsymbol{\Sigma} = \mathbb{E}_{\boldsymbol{x}}\!\left[\boldsymbol{x}\boldsymbol{x}^{\dagger}\right]
    \approx \langle \boldsymbol{x}\boldsymbol{x}^{\dagger}\rangle\,,
\end{equation}
where $\mathbb{E}_{\boldsymbol{x}}[\,\cdot\,]$ is the expectation with respect to the process
that generates the data $\boldsymbol{x}$.

We finally distinguish two families of parameters. The astrophysical signal is described by
$\boldsymbol{\theta} = \{\theta^{a}\}$, with Latin indices $a = 1,\dots,d$ ranging over the
$d$ signal parameters and derivatives abbreviated $\partial_a := \partial/\partial\theta^{a}$.
The noise parameters are described by  $\boldsymbol{\blambda} = \{\lambda^{\alpha}\}$, 
which enter through
the covariance $\boldsymbol{\Sigma} = \boldsymbol{\Sigma}(\boldsymbol{\blambda})$; Greek indices
$\alpha,\beta,\rho,\sigma,\dots$ label the noise parameters, with
$\partial_\alpha := \partial/\partial\lambda^{\alpha}$. We will use $\bTheta = \boldsymbol{\theta} \cup \boldsymbol{\lambda}$ to indicate the full noise and signal parameter set.  Capitalized 
latin indices $A, B, C,\ldots$ represent the elements  $ \Theta^{A} \in \bTheta$ that span the full parameter space. A hat marks a maximum-likelihood
estimate (for instance $\Delta\hat{\boldsymbol{\blambda}}$), while a tilde marks a
frequency-domain quantity, as introduced above.

\section{Gravitational Wave Data Analysis}\label{sec:gwda}
\subsection{Stationary noise}\label{subsec:stationary_noise}
The data stream from a GW detector can be modelled as a linear superposition of a GW signal $h$ and a noise process $n$ governed by parameters $(\boldsymbol{\theta},\boldsymbol{\blambda})$ 
\begin{equation}
\boldsymbol{d} = \boldsymbol{h}(t;\boldsymbol{\theta}) + \boldsymbol{n}(t;\boldsymbol{\blambda})\,.\label{eq:data_model}
\end{equation}

In the case of the LISA instrument or multi-detector configuration, there are multiple (potentially correlated) data sets involving multiple data channels. For the sake of simplicity, we will consider a single data channel but all results presented in this manuscript are readily generalised to account for multiple data channels. 

The astronomy is encoded in the signal parameters $\boldsymbol{\theta}$ whereas the probabilistic noise process $\bn$ encodes parameters $\blambda$ which govern the statistical nature of the noise. It is common within GW data analysis to assume that the noise process is both \emph{stationary} and \emph{circulant}. For a stationary noise process $\bn$, the auto-covariance matrix in the time-domain is simply a function of the lag between two time-points -- resulting in a Toeplitz structure. Similarly, for a circulant noise process, the auto-covariance matrix is a function of the lag between two time-points modulo the total number of samples. In other words, the covariance matrix is periodic in time and all statistics governing the noise process are encapsulated within the observation window. For large data sets whose noise correlation times are far shorter than the observation time, Toeplitz matrices are asymptotically equivalent to circulant matrices~\cite{whittle:1957, gray2006toeplitz}, which is why the Whittle-approximation is used for GW data analysis with long data stretches.

Under the circulant condition, the noise covariance matrix in the frequency domain $\tilde{\bSigma}$ takes the familiar form~\cite{Burke:2025bun,Finn:1992wt}
\begin{equation}\label{eq:diagonal_cov_circulant}
\tilde{\bSigma}_{ij} = (\mathbb{E}[\tilde{\boldsymbol{n}}\tilde{\boldsymbol{n}}^{\dagger}])_{ij}= \delta_{ij}\frac{S_n[i]}{2\Delta f}\,.
\end{equation}
where $S_{n}[i]$ is the noise PSD, a function representing the power of the noise as a function of frequency. The noise PSD is a function of the noise parameters $\boldsymbol{\blambda}$, which are to be inferred from the data. This result is a direct consequence of the Wiener-Khinchin theorem~\cite{wiener1930generalized,khintchine1934korrelationstheorie}, stating that the Fourier transform of the auto-covariance matrix in the time domain is equivalent to the noise power spectral density. Asserting the circulant condition yields a diagonal covariance matrix in the frequency domain, the key result for GW data analysis. This is all discussed at length in~\cite{Burke:2025bun}, so we will not repeat the discussion here.  



In our work, we will assume that our time-domain noise covariance matrix follows a zero-mean multivariate gaussian distribution $\bn \sim \mathcal{N}(\boldsymbol{0},\bSigma)$ with covariance matrix $\bSigma$. Defining residuals $\bn = \bd - \bh$, the log-likelihood takes the form 
\begin{equation}
\log\mathcal{L}(\boldsymbol{\blambda}) \propto -\frac{1}{2}\boldsymbol{n}^T \bSigma^{-1}\boldsymbol{n} - \frac{1}{2}\log\det\bSigma
\end{equation}
Conversion from the time-domain to frequency-domain is straightforward using the DFT matrix. We have the two identities
\begin{align}
\text{Time Domain:} \ \bSigma &=  \frac{1}{N\Delta t^2}\bm{P}^{\dagger}\tilde{\bm{\Sigma}}\bm{P}\,.\label{eq:defIDFTcov} \\
\text{Frequency Domain:} \ \tilde{\bSigma} & = N \Delta t^2 \bm{P}\bm{\Sigma}\bm{P}^\dagger\,.
    \label{eq:defFDcov}
\end{align}
As described in ~\cite{Burke:2025bun}, if the noise process is stationary and circulant, then the Whittle-likelihood follows~\cite{whittle:1957,Finn:1992wt,Burke:2025bun}
\begin{equation}
    \ln \mathcal{L} (\bm{n} | \bm{\theta}) \propto -\frac{1}{2} \tilde{\bm{n}}^\dagger \tilde{\bm{\Sigma}}^{-1} \tilde{\bm{n}} - \frac{1}{2} \ln \det \left(\frac{\Delta f}{\Delta t} \tilde{\bm{\Sigma}} \right) \,,
    \label{eq:defFDlike_matrix}
\end{equation}
or explicitly in terms of the noise PSD $S_n(f;\boldsymbol{\lambda})$, using \eqref{eq:diagonal_cov_circulant}
\begin{align}
    \ln \mathcal{L} (\bm{n}) &\propto -2 \Delta f \sum_{j=1}^{N/2-1} \frac{\left| \tilde{\bm{n}}_j \right|^2}{S_n^j} - \sum_{j=1}^{N/2-1} \ln \frac{S_n^j}{\Delta t} \nonumber\\
    & - \frac{\Delta f}{S_n^0} \left| \tilde{\bm{n}}_0 \right|^2 - \frac{\Delta f}{S_n^{\frac{N}{2}}} \left| \tilde{\bm{n}}_{\frac{N}{2}} \right|^2 \nonumber\\
    & - \frac{1}{2} \ln \frac{S_n^0}{\Delta t} - \frac{1}{2} \ln \frac{S_n^{\frac{N}{2}}}{\Delta t} - \frac{N}{2}\ln 2\pi \,.
    \label{eq:whittle_likelihood}
\end{align}
Where we have outlined the DC and Nyquist components ($j = 0, j = N/2$) and defined both $h:= h(\btheta)$ and $S_n := S_n(\blambda)$ for notational simplicity. For signal parameter estimation, one would replace $\tilde{\bm{n}}$ with the residuals $\tilde{\bm{d}} - \tilde{\bm{h}}(\btheta)$ in the above expression.

Equation \eqref{eq:whittle_likelihood} is the Whittle likelihood function~\cite{whittle:1957}, which is the standard likelihood function used for GW parameter estimation.

\subsection{Tampering with the noise process}\label{subsec:tampered_noise_process}


The purpose of a window function is to mitigate the effects of spectral leakage in the frequency domain. Spectral leakage occurs when a signal is not periodic within the observation window, leading to power from one frequency bin leaking into adjacent bins. This can distort the estimated PSD and affect parameter estimation. In our analysis, we will model window functions as a multiplicative operation on the time-domain data, represented by a diagonal matrix $\bW$ with entries corresponding to the window function values at each time sample. For simplicity, we will operate in the time-domain with all time-domain quantities convertible to the frequency domain via the DFT matrix. 

Although window functions have favourable properties with regards to reducing bin-leakage, they also have undesirable qualities as well -- they directly impact the statistical features of the noise process. For a $\bn$ a circulant (or Toeplitz) noise process, the result of applying a window function $\bW$ to the data stream renders the noise process non-stationary and, as a by-product, non-circulant. This effect renders the usual Whittle likelihood function invalid and a new likelihood function must be derived. 

Windowed data is written $\bW\boldsymbol{d}$, and the associated windowed noise covariance matrix is 
\begin{equation}
\mathbb{E}\left[(\bW \bn)(\bW \bn)^T\right] = \bW \bSigma \bW\,. 
\end{equation}
The frequency-domain representation follows by conjugating with the DFT matrix, 
\begin{align}
\widetilde{\bW\bSigma\bW} = \bP^{\dagger}(\bW\bSigma\bW)\bP &= \bP^{\dagger}\bW\bP \bP^{\dagger}\bSigma\bP\bP^{\dagger}\bW\bP \nonumber \\
&= \tilde{\bW}\tilde{\bSigma}\tilde{\bW}\,,
\end{align}
and has components
\begin{equation}\label{eq:windowed_frequency_domain_covariance}
    \big[\widetilde{\bW\bSigma\bW}\big]_{ij} = \frac{\Delta f}{2}\sum_{p=0}^{N-1} \tilde{w}[\overline{i-p}]\,\tilde{w}^{*}[\overline{j-p}]\,S_{n}[p]\,,
\end{equation}
where an overline denotes reduction modulo $N$, $\overline{n} \equiv n \bmod N$, mapping every frequency index back into $\{0,1,\dots,N-1\}$. Equation~\eqref{eq:windowed_frequency_domain_covariance} can be evaluated efficiently using Algorithm~1 of Ref.~\cite{Burke:2025bun}.

The presence of off diagonal components in the windowed covariance matrix renders the Whittle likelihood function invalid. The correlations present must be accounted for and the cheap-to-evaluate Whittle-likelihood needs to be modified. To ease notation, we will set $\bQ = (\bW\bSigma\bW)$.  

For the windowed noise process, the log-likelihood is given by
\begin{equation}\label{eq:log_likelihood_Q_TD}
    \log \mathcal{L}(\boldsymbol{\lambda}) = -\frac{1}{2}(\bW\bn)^T \bQ^+(\bW\bn) - \frac{1}{2}\log\det(\bQ) . 
\end{equation}
Here $\log\det$ is understood as the pseudo-determinant of the singular matrix $\bQ$ (the product of its nonzero eigenvalues, i.e.\ the determinant restricted to the observed subspace). Here $\bQ^+ = (\bW\bSigma\bW)^{+}$ denotes the pseudo-inverse of $\bQ = \bW\bSigma\bW$, which satisfies the identities \eqref{eq:pseudo_inv_def_1} in appendix \ref{app:identities}. Finally, from Ref.~\cite{Burke:2025bun}, Eq. (37), we have over the observed subspace of samples (i.e., non-gated\footnote{We will refer to ``gated'' data with sharp cutoffs separating observed and unobserved data (modelled through a rectangular window). Windowed data is the same but with a smooth taper applied to each gap segment.} data) that 
\begin{equation}\label{eq:pseudo_inv_gate}
   (\bW\bSigma\bW)^+(\bW\bSigma\bW) = \boldsymbol{\Pi}_{\rm obs} \neq \mathbb{I}\,,
\end{equation}
where $\boldsymbol{\Pi}_{\rm obs}$ is the orthogonal projector onto the observed (non-gated) samples. This simply means that the pseudo-inverse of the gated noise process explicitly maps vector quantities onto observed data only, essentially annihilating vector quantities during the gated segment. We note that $\boldsymbol{A}^{+}\boldsymbol{A}$ is always the orthogonal projector onto the row space of $\boldsymbol{A}$ [Eq.~\eqref{eq:pseudo_inv_def_1}]; what is special to the gating window is that this projector coincides with coordinate deletion.

For completeness, the log-likelihood in the frequency domain takes the form
\begin{equation}\label{eq:windowed_likelihood_fd_alt}
	\log \mathcal{L} (\bm{n} | \bm{\theta}) = -\frac{1}{2}(\widetilde{\bm{W} \bm{n}})^{\dagger}( \tilde{\bQ})^{+} (\widetilde{\bm{W}\bm{n}}) - \frac{1}{2}\log\det(\tilde{\bQ}) \,,
\end{equation}
for $\tilde{\bQ} = \tilde{\bm{W}} \tilde{\bm{\Sigma}} \tilde{\bm{W}}$, $\widetilde{\bm{W} \bm{n}} = \tilde{\bm{W}} \tilde{\bm{n}}$ and we have used the fact that $\bP$ is unitary, so that $\det{\bP}\,\det{\bP^{\dagger}} = 1$.

For our analysis, we will assume that the PSD $S_{n}(f)$ exhibits no time-dependent drifts, such that the underlying noise process (without gaps) is purely circulant. Time-dependence in the PSD as a consequence of orbital variations (time-changing light travel times in the LISA armlengths) or onboard time-dependent disturbances are outside the scope of this work. Outside gap segments with different PSDs, assuming each segment is locally stationary, then a segmented Whittle-like approach could be conducted (see ~\cite{Burke:2025bun}). For marginal slow drifting PSDs, time-frequency methods such as short-time fourier transforms~\cite{bandopadhyay2026globaltimefrequencysearchstellarmass, Speri_2026} or Wilson-Daubechies Meyer (WDM) based analyses ~\cite{Vajpeyi:2026msr, cornish2026nonstationarynoisegravitationalwave,Cornish_2020, Digman:2022jmp} could be used. The work presented here is directly applicable to time-frequency based analysis since all time-domain and frequency-domain algebraic quantities can be translated to the WDM basis via suitable linear transformations.

\section{Noise and Signal Fisher Matrix Methodology}\label{sec:methodology}

Throughout this section we carry out all derivations in the time domain, where the pseudo-inverse and trace manipulations are most transparent. Every result maps to the frequency domain by conjugation with the unitary DFT matrix $\bP$ of Eq.~\eqref{eq:def_P_jk_matrix}. Because the quantities of interest are traces of products of covariances and their derivatives, the balanced $\bP$ and $\bP^{\dagger}$ factors cancel pairwise --- using the cyclic property~\eqref{eq:trace_identity} and $\bP^{\dagger}\bP = \mathbb{I}$ --- so the frequency-domain expression is obtained simply by replacing each time-domain quantity with its tilded counterpart, e.g. $\text{Tr}[\boldsymbol{A}\boldsymbol{B}] = \text{Tr}[\tilde{\boldsymbol{A}}\tilde{\boldsymbol{B}}]$. We demonstrate this conversion explicitly once for the noise Fisher matrix below [Eq.~\eqref{eq:FM_noise_params}] and do not repeat it for the subsequent quantities.

\subsection{The joint Fisher matrix}\label{sec:joint_fisher_matrix}
The Fisher matrix is a useful tool to understand correlations between parameters of interest and as a predictive tool for parameter estimation~\cite{Finn:1992wt,abuse_fisher,Cutler:2007mi,Flanagan:1997kp}. The literature is abundant with Fisher-based analyses of astrophysical parameter estimation so we need not repeat the literature here. However, noise inference is rarely approached from a Fisher perspective and that will be our main focus. In addition, we will also derive the noise and signal fisher-based block and verify that the correlations between signal and noise parameters are essentially zero.  

To compute the fisher matrix, one can use the direct definition 
\begin{equation}
    \mathbb{E}[-\partial_A\partial_B\log\mathcal{L}] = \bGamma_{AB}\,.
\end{equation}
Over the full parameter set $\bTheta = \boldsymbol{\theta} \cup \boldsymbol{\lambda}$, the Fisher matrix carries capital indices, $\bGamma_{AB} = \mathbb{E}[-\partial_A\partial_B\log\mathcal{L}]$. Partitioning each capital index into its signal and noise parts, $A = (a,\alpha)$, casts the Fisher matrix into the $2\times 2$ block form
\begin{equation}\label{eq:joint_fisher_blocks}
    \bGamma_{AB} = \begin{pmatrix}
    \bGamma_{ab} & \bGamma_{a\beta} \\[3pt]
    \bGamma_{\alpha b} & \bGamma_{\alpha\beta}
    \end{pmatrix}\,,
\end{equation}
where Latin indices $a,b$ label the signal parameters and Greek indices $\alpha,\beta$ the noise parameters. The signal--signal block $\bGamma_{ab}$ is the familiar Fisher matrix for astrophysical parameters; the noise--noise block $\bGamma_{\alpha\beta}$ is the noise Fisher matrix; and the cross blocks $\bGamma_{a\beta} = (\bGamma_{\alpha b})^{T}$ encode correlations between the astrophysical signal and noise parameters. 

The usual Fisher matrix for astrophysical parameters is given by a noise-weighted inner product of the signal derivatives. As shown in \cite{Burke:2025bun}, we have that
\begin{equation}
\bGamma_{ab}  = \left(\bm{W} \partial_a \bm{h}\right)^T \bQ^+ \left(\bm{W} \partial_b \bm{h}\right) \,.
\end{equation}

The noise Fisher matrix is slightly more challenging to compute. Using Eq.\eqref{eq:pseudo_inv_derivative} and Eq.\eqref{eq:jacobi_inv_deriv}, one can take a noise parameter derivative of the log-likelihood giving
\begin{equation}
    \partial_{\alpha}\log\mathcal{L} = \frac{1}{2}(\bW\bn)^T\bQ^+  \bQ_{\alpha} \bQ^+ (\bW\bn) \\ - \frac{1}{2}\text{Tr}[\bQ^+ \bQ_\alpha]\,,
\end{equation}
with $\bQ_\alpha = \partial_{\alpha}\bQ$ and $\bQ_{\alpha\beta} = \partial_{\alpha}\partial_{\beta}\bQ$. Defining the matrix quantity $\boldsymbol{\mathcal{Q}}_{\alpha\beta}$ (which is \emph{not} $\partial_{\alpha}\partial_{\beta}\bQ$) via
\begin{widetext}
\begin{equation}\label{eq:calQ_definition}
    \boldsymbol{\mathcal{Q}}_{\alpha\beta} = \bQ^+\bQ_\beta \bQ^+\bQ_\alpha\bQ^+ - \bQ^+\bQ_{\alpha\beta}\bQ^+ +\bQ^+\bQ_{\alpha}\bQ^+\bQ_\beta\bQ^+\,,
\end{equation}
one can then take a second noise parameter derivative of the log-likelihood function and use the identities \eqref{eq:jacobi_inv_deriv}, \eqref{eq:pseudo_inv_derivative} to obtain
\begin{equation}
\partial_{\alpha}\partial_{\beta}\log\mathcal{L} = -\frac{1}{2}(\bW\bn)^T\boldsymbol{\mathcal{Q}}_{\alpha\beta}(\bW\bn) + \frac{1}{2}\text{Tr}(\boldsymbol{\mathcal{Q}}_{\alpha\beta}\bQ) - \frac{1}{2}\text{Tr}(\bQ^+ \bQ_{\alpha}\bQ^+\bQ_{\beta})\,.
\end{equation}
\end{widetext}
Now, our interest is the expectation of the second derivative of the log-likelihood function, which will yield the Fisher information matrix on noise parameters with a non-Whittle based covariance matrix. From equation \eqref{eq:quadratic_form_identity} in appendix \ref{app:identities}, we demonstrate that 
\begin{align}
    \langle (\bW\bn)^T \boldsymbol{\mathcal{Q}}_{\alpha\beta} (\bW\bn)\rangle &= \text{Tr}[\boldsymbol{\mathcal{Q}}_{\alpha\beta} (\bW\langle\bn\bn^T\rangle\bW)] \nonumber \\
    &= \text{Tr}[\boldsymbol{\mathcal{Q}}_{\alpha\beta}\bQ]\,. \label{eq:quadratic_form_identity_Q_main_text}
\end{align}
The expectation of the second derivative of the log-likelihood yields a cancellation of the first two terms, yielding the Fisher information matrix (FM) 
\begin{empheq}[box=\fbox]{equation}\label{eq:FM_noise_params}
    \boldsymbol{\Gamma}_{\alpha\beta} = \langle -\partial_{\alpha}\partial_{\beta}\log\mathcal{L}\rangle = \frac{1}{2}\text{Tr}(\bQ^+ \bQ_{\alpha}\bQ^+\bQ_{\beta})\,.
\end{empheq}
The noise parameter FM result Eq. \eqref{eq:FM_noise_params} is general and is valid for any such process of the form $\bQ = \bW\bSigma\bW$. Equation \eqref{eq:FM_noise_params} can be cast to the frequency domain. One can use the cyclic trace identity
\begin{equation}\label{eq:trace_identity}
    \text{Tr}[\boldsymbol{A}\boldsymbol{B}\boldsymbol{C}] = \text{Tr}[\boldsymbol{C}\boldsymbol{A}\boldsymbol{B}] = \text{Tr}[\boldsymbol{B}\boldsymbol{C}\boldsymbol{A}]
\end{equation}
Giving
\begin{align}
    \boldsymbol{\Gamma}_{\alpha\beta} &= \frac{1}{2}\text{Tr}(\bQ^+ \bQ_{\alpha}\bQ^+\bQ_{\beta})\,, \\
    & = \frac{1}{2}\text{Tr}(\bP^{\dagger}\tilde{\bQ}^+\tilde{\bQ}_{\alpha}\tilde{\bQ}^+\tilde{\bQ}_{\beta}\bP)\,, \\
    & = \frac{1}{2}\text{Tr}(\bP\bP^{\dagger}\tilde{\bQ}^+\tilde{\bQ}_{\alpha}\tilde{\bQ}^+\tilde{\bQ}_{\beta})\,, \\
    & = \frac{1}{2}\text{Tr}(\tilde{\bQ}^+\tilde{\bQ}_{\alpha}\tilde{\bQ}^+\tilde{\bQ}_{\beta})\,\,,\\ 
    & = \tilde{\bGamma}_{\alpha\beta}\,.
\end{align}
With components of $\tilde{\bQ}$ given by \eqref{eq:windowed_frequency_domain_covariance}. 

In appendix \ref{app:cross_block_fisher}, we show that the cross-blocks of the Fisher matrix vanish identically, $\bGamma_{a\beta} = \bGamma_{\alpha b} = 0$. This means that the astrophysical signal parameters and noise parameters are uncorrelated provided the linear-signal approximation is valid (high signal-to-noise ratio (SNR)) and the number of effective samples are large. The cross-block result itself is stronger than this. Its derivation uses only $\langle\bW\bn\rangle = 0$ together with the fact that the signal does not depend on the noise parameters and the covariance does not depend on the signal parameters. It therefore holds for \emph{any} model covariance $\bSigma'$, faithful or not, and for any noise model, including one with free spectral shapes. The joint Fisher matrix is therefore block diagonal and takes the form
\begin{equation}
    \bGamma_{AB} = \begin{pmatrix}
    \bGamma_{ab} & 0 \\[3pt]
    0 & \bGamma_{\alpha\beta}
    \end{pmatrix}\,.
\end{equation}
demonstrating that inference on signal parameters is independent of inference on noise parameters in this regime.

As a test case, we can check that the result Eq. \eqref{eq:FM_noise_params} reduces to the familiar result for stationary and circulant processes (Whittle-based). Let $\bW = \mathbb{I}$ represent a case where no gating function is applied. This means that $\bQ = \bSigma$, giving the usual Whittle-based Fisher matrix
\begin{align}
\boldsymbol{\Gamma}^{\rm Whittle}_{\alpha\beta} &= \frac{1}{2}\text{Tr}[\bSigma^{-1}\bSigma_{\alpha}\bSigma^{-1}\bSigma_{\beta}]\,. \\
&= \frac{1}{2}\text{Tr}[\tilde{\bSigma}^{-1}\tilde{\bSigma}_{\alpha}\tilde{\bSigma}^{-1}\tilde{\bSigma}_{\beta}]\,. 
\end{align}
Notice that we have replaced $\bSigma^+ = \bSigma^{-1}$ since the matrix $\bSigma$ is no longer singular and used the trace identity \eqref{eq:trace_identity}. Here $\bSigma$ is the usual circulant time-domain noise covariance matrix defined by $\mathbb{E}[\bn\bn^T] = \bSigma$. Since $\bSigma$ is circulant, the frequency-domain noise covariance matrix is given by $\tilde{\Sigma}_{ij} = \delta_{ij}S_{n}(f_i)/2\Delta f$ for positive frequency bin components. This results in the expression
\begin{equation}
    \boldsymbol{\Gamma}^{\rm Whittle}_{\alpha\beta} = \sum_{i=1}^{N/2 -1} \frac{\partial_{\alpha}S_{n}(f_i)\partial_{\beta}S_{n}(f_i)}{S_{n}^2(f_i)} 
\end{equation}
or, in the usual continuous representation (see for instance \cite{PhysRevD.109.042001})
\begin{align}\label{eq:FM_whittle}
    \Gamma^{\rm Whittle}_{\alpha\beta} &= T_{\text{obs}} \int_{0}^{\infty} \frac{\partial_{\alpha}S_{n}(f)\partial_{\beta}S_{n}(f)}{S_{n}^2(f)}\,\text{d}f\,. \\ 
    & = T_{\text{obs}} \int_{0}^{\infty} [\partial_{\alpha}\log S_{n}(f)][\partial_{\beta}\log S_{n}(f)]\,\text{d}f\,. \nonumber
\end{align}
The Fisher matrix Eq. \eqref{eq:FM_whittle} demonstrates that the precision on noise parameters scales proportionally to the observation time, or, equivalently, number of effective samples. This is similar in nature to the Fisher matrix on astrophysical parameters where the fisher matrix scales proportionally to the SNR squared. The larger the amplitude, the greater the precision on astrophysical parameters. Analogously, the greater the volume of data points, the better precision measurement on the noise parameters. 
\subsection{The Hessian based likelihood}\label{sec:hessian_based_likelihood}
Returning to the gated noise process, we now want to show that the inverse Fisher matrix yields the covariance matrix on both noise and signal parameters. We will also compute an analytical expression for the maximum likelihood estimates (MLEs) of the parameters and demonstrate that the covariance is, indeed, the inverse Fisher matrix as expected. 

Consider a small perturbation to both the noise and signal parameters, collected into the mixed-index parameter vector $\Theta^{A} = (\theta^{a},\lambda^{\alpha})$, with $A=(a,\alpha)$ denoting signal and noise components respectively. Writing
\begin{equation}
    \Theta^{A}=\Theta^{A}_0+\Delta\Theta^{A}
\end{equation}
we can expand the log-likelihood about $\boldsymbol{\Theta}_0$ in a multivariate Taylor series, assuming $\Delta\Theta^{A}$ is sufficiently small:
\begin{widetext}
\begin{equation}
    \log\mathcal{L}(\boldsymbol{\Theta}) = \log\mathcal{L}(\boldsymbol{\Theta}_0)
    + \Delta\Theta^{A}\,\partial_{A}\log\mathcal{L}(\boldsymbol{\Theta}_0)
    + \frac{1}{2}\Delta\Theta^{A}\Delta\Theta^{B}\,\partial_{A}\partial_{B}\log\mathcal{L}(\boldsymbol{\Theta}_0)
    + \mathcal{O}(\|\Delta\boldsymbol{\Theta}\|^3)\,.
\end{equation}
The linear term includes both the noise and signal contributions, 
\begin{equation}
    \Delta\Theta^{A}\,\partial_{A}\log\mathcal{L}
    = \Delta\theta^{a}\partial_{a}\log\mathcal{L}
    + \Delta\lambda^{\alpha}\partial_{\alpha}\log\mathcal{L}\,.
\end{equation}
Explicitly, the quadratic term contains the noise-noise, signal-signal, and cross contributions,
\begin{equation}
    \Delta\Theta^{A}\Delta\Theta^{B}\,\partial_{A}\partial_{B}\log\mathcal{L}
    = \Delta\theta^{a}\Delta\theta^{b}\partial_{a}\partial_{b}\log\mathcal{L}
    + 2\,\Delta\theta^{a}\Delta\lambda^{\alpha}\partial_{a}\partial_{\alpha}\log\mathcal{L}
    + \Delta\lambda^{\alpha}\Delta\lambda^{\beta}\partial_{\alpha}\partial_{\beta}\log\mathcal{L}\,.
\end{equation}
\end{widetext}
The goal now is to write the above expression as a quadratic form so it can be identified as the form of a multivariate Gaussian distribution, with the parameter covariance matrix given by the inverse Fisher matrix.

To this end we split each block of the Hessian into a deterministic Fisher piece and a mean-zero fluctuation driven by the noise realisation,
\begin{equation}\label{eq:hessian_split_joint}
    \partial_A\partial_B\log\mathcal{L} = -\bGamma_{AB} + \bF_{AB}(\bn)\,,\qquad \langle \bF_{AB}(\bn)\rangle = 0\,,
\end{equation}
and evaluate the signal, cross, and noise blocks in turn. Differentiating the log-likelihood given by \eqref{eq:log_likelihood_Q_TD} and using $\partial_a(\bW\bn) = -\bW\partial_a\bh$ together with the identities \eqref{eq:pseudo_inv_derivative} and \eqref{eq:jacobi_inv_deriv}, gives
\begin{widetext}
\begin{subequations}\label{eq:hessian_blocks}
\begin{align}
    \text{signal--signal:}\quad
    \partial_a\partial_b\log\mathcal{L} &= -\underbrace{(\bW\partial_a\bh)^T\bQ^+(\bW\partial_b\bh)}_{\textstyle \bGamma_{ab}} + \underbrace{(\bW\partial_a\partial_b\bh)^T\bQ^+(\bW\bn)}_{\textstyle \bF_{ab}(\bn)}\,, \label{eq:hessian_ss}\\
    \text{cross:}\quad
    \partial_a\partial_\alpha\log\mathcal{L} &= -\underbrace{\;0\;}_{\textstyle \bGamma_{a\alpha}}\;\underbrace{-\,(\bW\partial_a\bh)^T\bQ^+\bQ_\alpha\bQ^+(\bW\bn)}_{\textstyle \bF_{a\alpha}(\bn)}\,, \label{eq:hessian_cross}\\
    \text{noise--noise:}\quad
    \partial_\alpha\partial_\beta\log\mathcal{L} &= -\underbrace{\tfrac12\text{Tr}(\bQ^+\bQ_\alpha\bQ^+\bQ_\beta)}_{\textstyle \bGamma_{\alpha\beta}} + \underbrace{\Big[{-}\tfrac12(\bW\bn)^T\boldsymbol{\mathcal{Q}}_{\alpha\beta}(\bW\bn) + \tfrac12\text{Tr}(\boldsymbol{\mathcal{Q}}_{\alpha\beta}\bQ)\Big]}_{\textstyle \bF_{\alpha\beta}(\bn)}\,. \label{eq:hessian_nn}
\end{align}
\end{subequations}
\end{widetext}
In equation \eqref{eq:hessian_blocks}, we have identified a deterministic \emph{Fisher-matrix} term $\bGamma$ and a probabilistic fluctuation term $\bF$. Each fluctuation is mean-zero: for the noise block this follows from Eq.~\eqref{eq:quadratic_form_identity_Q_main_text}, and for the signal and cross blocks from $\langle\bW\bn\rangle = 0$. It remains to check that the fluctuations are subdominant. Although each $\bF_{AB}$ is mean-zero, the variance may not be and could dominate the overall Fisher expression. 

For the noise block we show in the appendix that
\begin{align}
    \text{Var}[\bF_{\alpha\beta}] &= \tfrac12\text{Tr}[(\boldsymbol{\mathcal{Q}}_{\alpha\beta}\bQ)^2] \sim N_{\rm eff}\,,
\end{align}
and so scales with the number of effective samples $N_{\rm eff}$, by which we mean the number of frequency bins that carry independent information, $N_{\rm eff} = N_{b}$ per channel here. The linear-in-noise signal and cross fluctuations are also small in comparison to the Fisher matrix relevant for that block. In every block the $1\sigma$ fluctuation scales like $\bF_{AB}\sim N^{1/2}$ whereas the Fisher piece scales like $\bGamma_{AB}\sim N$, so the deterministic curvature dominates in the limit of a large intake of data. This makes sense: both signal and noise parameters become better constrained with higher deterministic curvature with a longer observation. Hence, for large sample sizes,
\begin{equation}
    \partial_A\partial_B\log\mathcal{L} \approx -\bGamma_{AB}\,.
\end{equation}
with
\begin{equation}\label{eq:block_diag_curvature}
    \bGamma_{AB} = \begin{pmatrix} \bGamma_{ab} & 0 \\[3pt] 0 & \bGamma_{\alpha\beta}\end{pmatrix}\,.
\end{equation}

We can thus write the log-likelihood as a joint quadratic form
\begin{widetext}
\begin{equation}\label{eq:joint_quadratic_form}
    \log\mathcal{L} = -\tfrac12(\Delta\Theta^A - \Delta\hat\Theta^A)\,\bGamma_{AB}\,(\Delta\Theta^B - \Delta\hat\Theta^B) + \mathcal{O}(N^{-1/2})\,,
\end{equation}
with joint maximum-likelihood estimate $\Delta\hat\Theta^A = (\bGamma^{-1})_{AB}\,\partial_B\log\mathcal{L}(\bTheta_0)$. Because $\bGamma_{AB}$ is block diagonal [Eq.~\eqref{eq:block_diag_curvature}], the estimates decouple into independent signal and noise sectors,
\begin{subequations}\label{eq:joint_mle}
\begin{align}
    \text{Signal parameter  MLE:} \qquad \Delta\hat\theta^a &= (\bGamma^{-1})_{ab}\,(\bW\partial_b\bh)^T\bQ^+(\bW\bn)\,, \\
    \text{Noise parameter MLE:} \qquad \Delta\hat\lambda^\alpha &= \tfrac12(\bGamma^{-1})_{\alpha\beta}\big\{(\bW\bn)^T\bQ^+\bQ_\beta\bQ^+(\bW\bn) - \text{Tr}[\bQ^+\bQ_\beta]\big\}\,.
\end{align}
\end{subequations}
\end{widetext}
Notice that the overall shift to parameter values are split into a signal piece $\Delta\hat\theta^a$ and a noise piece $\Delta\hat\lambda^\alpha$, each governed by its own Fisher block. The signal MLE is linear in the noise realisation $\bW\bn$, while the noise MLE is quadratic in $\bW\bn$. Both estimators are unbiased, $\langle\Delta\hat\Theta^A\rangle = 0$. A brute-force application of Isserlis' theorem (see \eqref{eq:real_isserlis_theorem} in appendix \ref{app:derivation_scatter_to_width}) for the noise component then shows that the covariance is the inverse of the block-diagonal Fisher matrix,
\begin{equation}\label{eq:joint_covariance}
    \langle\Delta\hat\Theta^A\Delta\hat\Theta^B\rangle = (\bGamma^{-1})_{AB} = \begin{pmatrix} (\bGamma^{-1})_{ab} & 0 \\[3pt] 0 & (\bGamma^{-1})_{\alpha\beta}\end{pmatrix}\,,
\end{equation}
as expected. The signal and noise likelihoods factorise into independent Gaussians, each governed by the inverse of its own Fisher block. In particular the noise--noise block reproduces
\begin{empheq}[box = \fbox]{equation}\label{eq:verify_FM_windowed_noise}
    \langle \Delta\hat{\lambda}^{\alpha} \Delta\hat{\lambda}^{\beta} \rangle = \boldsymbol{\Gamma}^{-1}_{\alpha\beta}\,, \qquad \boldsymbol{\Gamma}_{\alpha\beta} = \frac{1}{2}\text{Tr}(\bQ^+ \bQ_\alpha \bQ^+ \bQ_{\beta})\,,
\end{empheq}
and signal block
\begin{equation}\label{eq:FM_signal_block_correct_model}
\langle \Delta\hat{\theta}^{a} \Delta\hat{\theta}^{b} \rangle = \boldsymbol{\Gamma}^{-1}_{ab}\,, \qquad \boldsymbol{\Gamma}_{ab} = (\bW\partial_{a}\bh)^T\bQ^+ (\bW \partial_{b}\bh)\,,
\end{equation}
consistent with~\cite{Burke:2025bun}.

\subsection{Quantifying noise mismodelling on noise}\label{sec:mismodelling_fisher_matrix}

In ~\cite{Burke:2025bun}, we derived two quantities -- the scatter-to-width ratio $\bUpsilon$ and width-to-width ratios $\bXi$, the first describing the root-mean-square scatter around the truth and the second the degradation in parameter precision due to mis-modelled noise. In this section, we will derive analogous quantities for noise parameter inference. 

The parameter MLEs in the case of mis-modelling is well detailed in ~\cite{Burke:2025bun} so we will not repeat the calculations here. Instead we will focus this section on the noise parameters and quote the results for signal-parameter MLEs later on in the section. 

Consider a model-based log-likelihood of the form
\begin{equation}
    \log \mathcal{L}(\boldsymbol{\lambda}) = -\frac{1}{2}(\bW\bn)^T (\bSigma^\prime)^+(\bW\bn) - \frac{1}{2}\log\det(\bSigma^\prime)\, . \label{eq:noise_model_likelihood}
\end{equation}
where $\bSigma^\prime$ is the noise covariance matrix assumed to fit the noise process. In this prescription, we will consider the impact on recovering the noise parameters when $\bSigma^\prime \neq \bQ = \bW\bSigma\bW$. Many of the calculations in the previous section have the same structure, so we will quote the result. Defining the symmetric matrix $\boldsymbol{\mathcal{C}}_{\alpha\beta}^{\prime}$
\begin{widetext}
\begin{equation}
    \boldsymbol{\mathcal{C}}^\prime_{\alpha\beta} = (\bSigma^\prime)^+\bSigma^\prime_\beta (\bSigma^\prime)^+(\bSigma^\prime)_\alpha(\bSigma^\prime)^+ - (\bSigma^\prime)^+\bSigma^\prime_{\alpha\beta}(\bSigma^\prime)^+ +(\bSigma^\prime)^+\bSigma^\prime_{\alpha}(\bSigma')^+\bSigma^\prime_\beta(\bSigma^\prime)^+\,. \label{eq:calC_definition}
\end{equation}
which is to be read symmetrised in $\alpha$ and $\beta$, exactly as for Eq.~\eqref{eq:calQ_definition}. We can then write down an expression for the first and second noise parameter derivative of the model log-likelihood:
\begin{align}
\partial_{\alpha}\log\mathcal{L} &= \frac{1}{2}(\bW\bn)^T(\bSigma^\prime)^+  \partial_{\alpha}(\bSigma^\prime)(\bSigma^\prime)^+ (\bW\bn) - \frac{1}{2}\text{Tr}[(\bSigma^\prime)^+ \partial_{\alpha}(\bSigma^\prime)]\,, \label{eq:model_noise_score}\\
\partial_{\alpha}\partial_{\beta}\log\mathcal{L} &= -\frac{1}{2}(\bW\bn)^T\boldsymbol{\mathcal{C}}^\prime_{\alpha\beta}(\bW\bn) + \frac{1}{2}\text{Tr}(\boldsymbol{\mathcal{C}}^\prime_{\alpha\beta}\bSigma^\prime) - \frac{1}{2}\text{Tr}((\bSigma^\prime)^+ \bSigma^\prime_{\alpha}(\bSigma^\prime)^+\bSigma^\prime_{\beta})\,,
\\ & = \bF^{\prime}_{\alpha\beta}(\bn) - \boldsymbol{\Gamma}^{\prime}_{\alpha\beta}\,.
\end{align}
\end{widetext}
Here we identified the approximate model (observed) Fisher matrix $\boldsymbol{\Gamma}^{\prime}_{\alpha\beta}$ and the fluctuating piece $\bF^{\prime}_{\alpha\beta}(\bn)$ defined by
\begin{align}
    \boldsymbol{\Gamma}^{\prime}_{\alpha\beta} &= \frac{1}{2}\text{Tr}((\bSigma^\prime)^+ \bSigma^\prime_{\alpha}(\bSigma^\prime)^+\bSigma^\prime_{\beta}) \label{eq:approx_fisher_matrix_windowed}\\
    \bF^{\prime}_{\alpha\beta}(\bn) &= -\frac{1}{2}(\bW\bn)^T\boldsymbol{\mathcal{C}}^\prime_{\alpha\beta}(\bW\bn) + \frac{1}{2}\text{Tr}(\boldsymbol{\mathcal{C}}^\prime_{\alpha\beta}\bSigma^\prime)\,.
\end{align}
Notice that the random variable $\bF^{\prime}_{\alpha\beta}$ is no longer necessarily zero-mean, since 
\begin{equation}
    \langle \bF^{\prime}_{\alpha\beta}(\bn) \rangle = \frac{1}{2}\text{Tr}[\boldsymbol{\mathcal{C}}^\prime_{\alpha\beta}(\bSigma^\prime - \bQ)]\,,
\end{equation}
resulting in a null calculation only if $\bQ = \bW\bSigma\bW = \bSigma^\prime$. This is a very interesting result since this implies that the overall precision on noise parameters is (1) directly related to the model covariance matrix and (2) receives a probabilistic shift to the noise uncertainty as a result of noise mis-modelling. We can write down the observed Fisher information matrix via 
\begin{equation}
    (\boldsymbol{\Gamma}^\prime_{\alpha\beta})^{\text{observed}} = \boldsymbol{\Gamma}^{\prime}_{\alpha\beta} - \bF^{\prime}_{\alpha\beta}(\bn)
\end{equation}
which is a function of the specific noise realisation $\bn$. Averaging over noise realisations, the effective curvature of the log-likelihood is given by 
\begin{equation}
    \langle \partial_\alpha\partial_\beta \log \mathcal{L} \rangle = \langle \bF^{\prime}_{\alpha\beta}\rangle - \boldsymbol{\Gamma}^\prime _{\alpha\beta}
\end{equation}
with Fisher information matrix 
\begin{equation}
    \langle -\partial_\alpha\partial_\beta \log \mathcal{L} \rangle = (\boldsymbol{\Gamma}^{\prime})^{\rm effective}_{\alpha\beta} = \boldsymbol{\Gamma}^\prime _{\alpha\beta} - \langle \bF^{\prime}_{\alpha\beta}\rangle\,. \label{eq:effective_fisher}
\end{equation}
We can therefore write an approximation to the likelihood function using the Hessian-based approach
\begin{equation}
    \log \mathcal{L} = -\frac{1}{2}(\Delta\lambda^{\alpha} - \Delta\hat{\lambda}^{\alpha})(\boldsymbol{\Gamma}^\prime_{\alpha\beta})^{\text{effective}}(\Delta\lambda^{\beta} - \Delta\hat{\lambda}^{\beta})\,,
\end{equation}
with corrections to the likelihood scaling by $\mathcal{O}(N^{-1/2})$\,.
The statistical fluctuations to the recovered noise parameters as a result of noise-mismodelling given by 
\begin{widetext}
\begin{equation}\label{eq:stat_fluc_params_mis-modelling}
    \Delta\hat{\lambda}^{\alpha} = \frac{1}{2}([\boldsymbol{\Gamma}^{\prime}_{\alpha\beta}]^{-1})^{\text{effective}}\left((\bW\bn)^T(\bSigma^\prime)^+  \partial_{\alpha}(\bSigma^\prime)(\bSigma^\prime)^+ (\bW\bn) - \text{Tr}[(\bSigma^\prime)^+ \partial_{\alpha}(\bSigma^\prime)]\right)\,, 
\end{equation}
\end{widetext}
We will now show the expected behaviour of $\Delta\hat{\lambda}^{\alpha}$. First, denoting $\langle \bF'_{\alpha\beta}(\bn)\rangle := \langle \bF'_{\alpha\beta}\rangle$ note that 

\begin{align*}
    \left[(\bGamma^{\prime}_{\alpha\beta})^{\rm effective}\right]^{-1} & = \left[\boldsymbol{\Gamma}'_{\alpha\beta} - \langle\boldsymbol{F}^{\prime}_{\alpha\beta}\rangle\right]^{-1} \\
    &= \left\{\left[\mathbb{I} - (\boldsymbol{\Gamma}')^{-1}_{\alpha\gamma}\langle\boldsymbol{F}^{\prime}_{\gamma\beta}\rangle\right]\boldsymbol{\Gamma}'_{\alpha\beta}\right\}^{-1} \\ 
    &= (\boldsymbol{\Gamma}')^{-1}_{\alpha\beta}\left[\mathbb{I} - (\boldsymbol{\Gamma}')^{-1}_{\alpha\gamma}\langle\boldsymbol{F}^{\prime}_{\gamma\beta}\rangle\right]^{-1}\,.
\end{align*}
We now make the assumption that the effect of mis-modelling is small, i.e., $\bSigma' \approx \bQ$, implying that $\langle\bF'_{\alpha\beta}\rangle \approx 0$, validating the expansion for $\bGamma^{-1}\bF$ small
\begin{equation}
    \left[\boldsymbol{\Gamma}'_{\alpha\beta} - \boldsymbol{F}^{\prime}_{\alpha\beta}\right]^{-1} 
    \approx (\boldsymbol{\Gamma}')^{-1}_{\alpha\beta} + (\boldsymbol{\Gamma}')^{-1}_{\alpha\gamma}\langle\boldsymbol{F}^{\prime}_{\gamma\delta}\rangle(\boldsymbol{\Gamma}')^{-1}_{\delta\beta} + \mathcal{O}(\boldsymbol{F}^2)
\end{equation}
Giving
\begin{equation}\label{eq:derivation_MLE_noise_mismodelling}
    \Delta\hat{\lambda}^\alpha \approx [(\boldsymbol{\Gamma}')^{-1}_{\alpha\beta}\, + (\boldsymbol{\Gamma}')^{-1}_{\alpha\gamma}\,\langle\boldsymbol{F}^{\prime}_{\gamma\delta}\rangle\,(\boldsymbol{\Gamma}')^{-1}_{\delta\beta}]\, \partial_\beta\log\mathcal{L}
\end{equation}
Now, computing $\langle \partial_{\beta}\log\mathcal{L} \rangle$ and noting $\langle (\boldsymbol{Wn})(\boldsymbol{Wn})^T \rangle = \boldsymbol{Q}$ we obtain
\begin{align}\label{eq:score_likelihood_deriv}
    \langle \partial_\beta\log\mathcal{L} \rangle =  \frac{1}{2}\text{Tr}\!\left[(\boldsymbol{\Sigma}')^+\boldsymbol{\Sigma}'_\beta\left((\boldsymbol{\Sigma}')^+\boldsymbol{Q} - \mathbb{I}\right)\right] \,.
\end{align}

The final step involves taking the expectation of \eqref{eq:stat_fluc_params_mis-modelling}, and then substituting in the expression above. The leading order term gives
\begin{empheq}[box = \fbox]{equation}
\langle \Delta\hat{\lambda}^\alpha \rangle \approx \frac{1}{2}(\boldsymbol{\Gamma}')^{-1}_{\alpha\beta}\;\text{Tr}\!\left[(\boldsymbol{\Sigma}')^+\boldsymbol{\Sigma}'_\beta\left((\boldsymbol{\Sigma}')^+\boldsymbol{Q} - \mathbb{I}\right)\right]\,.\label{eq:biased_noise_params_mismodelling}
\end{empheq}
with corrections that scale like $\mathcal{O}(N_{\text{eff}}^{-1/2})$. An equivalent expression for the noise-parameter bias under a mis-specified spectral model was obtained independently in Ref.~\cite{Baghi:2026spec} in the context of time- and frequency-averaged spectra. Notice that \eqref{eq:biased_noise_params_mismodelling} is non-zero under mismodelling where $\bQ \neq \bSigma^\prime$. 

If the windowed noise process matches the modelled covariance matrix $\bQ = \bW\bSigma\bW = \bSigma^\prime$, then we observe that $\langle \Delta\hat{\lambda}^\alpha \rangle = 0$, resulting in an unbiased estimate of the noise parameters. In general, the assumed model covariance matrix $\bSigma^{\prime} \neq \bW\bSigma\bW$ resulting in 
\begin{enumerate}
    \item Altered parameter uncertainties: the quoted parameter uncertainties may not be consistent with the true parameter uncertainties in the case of no mis-modelling errors. 
    \item Actual systematic errors: Noise mis-modelling will incur biased noise parameters.
\end{enumerate}
Notice that this is somewhat analogous to waveform systematics where imperfect 
waveform templates are used to perform search and parameter estimation analyses 
of GW signals. Assuming non-faithful templates results 
in (1) biased parameter estimates and (2) corrections to the model posteriors 
outside of the linear signal approximation~\cite{Antonelli:2021vwg}. Waveform systematics produce a \emph{deterministic} bias that is independent of the 
noise realisation. Noise parameter biases in 
Eq.~\eqref{eq:biased_noise_params_mismodelling} depend on the true noise 
covariance $\boldsymbol{Q}$ through a trace that is itself deterministic. The bias on noise parameters does not average away over noise realisations in either case.

Unfaithful noise models do not bias signal parameters at leading order, Eq.\eqref{eq:MLE_signal_bias_window}, however, the converse does not hold. An unfaithful waveform leaves a coherent residual whose power enters the quadratic noise score and biases the noise parameters. This is a direction we will not pursue here and is left as an area of future work.

From ~\cite{Burke:2025bun}, the signal-signal MLE under noise-mismodelling is given by 
\begin{equation}\label{eq:MLE_signal_bias_window}
\Delta \hat{\theta}^a = \left(\bm{\Gamma}^{\prime}\right)^{-1}_{ab} \left( \bm{W} \partial_b \bm{h} \right)^T \left(\bm{\Sigma}'\right)^{+} \left( \bm{W} \bm{n} \right)
\end{equation}

In the case of mis-modelling the noise process, the signal parameter constraints 
are determined by the specific model covariance matrix of choice and, as shown 
in Ref.~\cite{Burke:2025bun}, the signal parameter estimates remain unbiased 
($\langle\Delta\hat{\theta}^a\rangle = 0$) regardless of the noise 
mis-modelling. However, the \emph{covariance} of those estimates is significantly 
affected. One can compute the corrected covariance of the MLEs as a result of noise mis-modelling of the signal-signal block
\begin{widetext}
\begin{align}\label{eq:derivation_bias_variance}
    \left\langle \Delta \hat{\theta}^a \Delta \hat{\theta}^b  \right\rangle
    &= \left(\bm{\Gamma}^{\prime}\right)^{-1}_{ac} \left(\bm{\Gamma}^{\prime}\right)^{-1}_{bd} \left(\bm{W} \partial_c \bm{h}\right)^T \left(\bm{\Sigma}'\right)^{+} \bQ \left(\bm{\Sigma}'\right)^{+} \left(\bm{W} \partial_d \bm{h}\right) \,,
\end{align}
\end{widetext}
where, only in the limit that $\bSigma' = \bQ = \bW\bSigma\bW$, does the covariance of MLEs collapse to the true Fisher matrix $\bGamma_{ab}$. 

Our goal now is to extend that analysis to account for \emph{both} signal and noise
parameters in the context of noise mis-modelling, providing a unified framework
for assessing the impact of an incorrect noise model on all inferred quantities.

\subsection{Scatter-to-width and width-to-width ratios}\label{subsec:metrics_scatter_to_width}
Consider the \emph{leading order} noise induced statistical fluctuation to the recovered parameters in the case of noise mis-modelling 
\begin{align}\label{eq:noise_param_fluctuations_noise_mismodelling}
    \Delta\hat{\lambda}^{\alpha} &=
     \frac{1}{2}([\boldsymbol{\Gamma}^{\prime}_{\alpha\beta}]^{-1})\big[(\bW\bn)^T(\bSigma^\prime)^+ (\bSigma^\prime_{\alpha})(\bSigma^\prime)^+ (\bW\bn) \nonumber\\ & \qquad \qquad \qquad \qquad \qquad  - \text{Tr}[(\bSigma^\prime)^+ \partial_{\alpha}(\bSigma^\prime)]\big]\,. \\
     \Delta \hat{\theta}^a  &= \left(\bm{\Gamma}^{\prime}\right)^{-1}_{ab} \left( \bm{W} \partial_b \bm{h} \right)^T \left(\bm{\Sigma}'\right)^{+} \left( \bm{W} \bm{n} \right) \label{eq:signal_param_fluctuations_noise_mismodelling}
\end{align}
These quantities depend on the model based Fisher information matrix $\boldsymbol{\Gamma}^{\prime}$ alongside the windowed noise realisations $\bW\bn$.

Following the definitions in Ref.~\cite{Burke:2025bun}, we define the noise-parameter scatter-to-width ratio $\Upsilon_{A}$ and width-to-width ratio $\Xi_{A}$ (no summation over the repeated index $A$, which runs over both signal and noise parameters)

\begin{align}\label{eq:upsilon_matrix_noise}
    \Upsilon_A &= \frac{\sqrt{\langle\Delta\hat{\Theta}^A\Delta\hat{\Theta}^A\rangle}}{\sqrt{(\boldsymbol{\Gamma}')^{-1}_{AA}}} \,, \\
     \Xi_A &= \frac{\sqrt{(\bGamma')^{-1}_{AA}}}{\sqrt{(\bGamma)^{-1}_{AA}}} \label{eq:xi_matrix_noise}
\end{align}

where $\Gamma$ is the Fisher matrix computed with the true covariance (Eq.~(32)). The new calculations here are the noise-noise and signal-noise block. The calculations for both are detailed in appendix \ref{app:derivation_scatter_to_width}, so we will quote the result here for $\bQ = \bW\bSigma\bW$:
\begin{widetext}

\begin{empheq}[box = \fbox]{align}
\underbrace{\langle\Delta\hat{\lambda}^\alpha\Delta\hat{\lambda}^\beta\rangle}_{\text{Noise - Noise}}
&= (\boldsymbol{\Gamma}')^{-1}_{\alpha\rho}(\boldsymbol{\Gamma}')^{-1}_{\beta\sigma}
\begin{aligned}[t]
\Bigg\{ &\frac{1}{2}\text{Tr}\!\left[(\bSigma')^+\bSigma^{\prime}_{\rho}\,(\bSigma')^+\bQ\,(\bSigma')^+\bSigma^{\prime}_{\sigma}\,(\bSigma')^+\bQ\right] \\
&+ \frac{1}{4}\text{Tr}\!\left[(\bSigma')^+\bSigma^{\prime}_{\rho}\big((\bSigma')^+\bQ - \mathbb{I}\big)\right]
   \text{Tr}\!\left[(\bSigma')^+\bSigma^{\prime}_{\sigma}\big((\bSigma')^+\bQ - \mathbb{I}\big)\right]\Bigg\}\,,
\end{aligned} \label{eq:noise_mismodelling_params_covariance}\\
\underbrace{\langle\Delta\hat{\lambda}^\alpha\Delta\hat{\theta}^a\rangle}_{\text{Noise - Signal}} &= 0\,,\\
\underbrace{\left\langle \Delta \hat{\theta}^a \Delta \hat{\theta}^b  \right\rangle}_{\text{Signal - Signal}}
    &= \left(\bm{\Gamma}^{\prime}\right)^{-1}_{ac} \left(\bm{\Gamma}^{\prime}\right)^{-1}_{bd} \left(\bm{W} \partial_c \bm{h}\right)^T \left(\bm{\Sigma}'\right)^{+} \bm{W} \bm{\Sigma} \bm{W} \left(\bm{\Sigma}'\right)^{+} \left(\bm{W} \partial_d \bm{h}\right) \,. \label{eq:noise_mismodelling_signal_params_covariance}
\end{empheq}

With true and model Fisher matrices defined by
\begin{align}
\text{Noise Component:} \ \boldsymbol{\Gamma}_{\alpha\beta} &= \frac{1}{2}\text{Tr}(\bQ^+ \bQ_\alpha \bQ^+ \bQ_\beta)\, &
\boldsymbol{\Gamma}^{\prime}_{\alpha\beta} &= \frac{1}{2}\text{Tr}((\bSigma^\prime)^+ \bSigma^\prime_{\alpha}(\bSigma^\prime)^+\bSigma^\prime_{\beta})  \\ 
\text{Signal Component:} \ \boldsymbol{\Gamma}_{ab} &= (\bW\partial_a \bm{h})^T \bQ^+ (\bW\partial_b \bm{h})\,, &
\boldsymbol{\Gamma}^{\prime}_{ab} &= (\bm{W} \partial_a \bm{h})^T (\bSigma^\prime)^+ (\bm{W} \partial_b \bm{h})
\end{align}
\end{widetext}
constituting the essential ingredients of the scatter-to-width and width-to-width ratios in Eqs.~\eqref{eq:upsilon_matrix_noise} and \eqref{eq:xi_matrix_noise}. The covariance of mis-modelled MLEs reduce to the usual inverse Fisher Matrix when $\bSigma^{\prime} = \bQ$. The overall parameter covariance matrix therefore takes the block form:

\begin{equation}\label{eq:block_diag_scatter}
    \langle\Delta\hat\Theta^A\Delta\hat\Theta^B\rangle =
    \begin{pmatrix}
        \langle\Delta\hat\theta^a\Delta\hat\theta^b\rangle & 0 \\[2pt]
        0 & \langle\Delta\hat\lambda^\alpha\Delta\hat\lambda^\beta\rangle
    \end{pmatrix}\,,
\end{equation}

Demonstrating that, even in the case of mis-modelling, there are no parameter correlations between the signal and noise sectors at leading order. 

Equation \eqref{eq:noise_mismodelling_params_covariance} is the leading order contribution to the covariance of the noise parameters, while the second term is a bias correction due to noise mis-modelling. Notice that the latter two-terms in \eqref{eq:noise_mismodelling_params_covariance}is precisely the square of the bias quantity defined by Eq.\eqref{eq:biased_noise_params_mismodelling}.Formally, Equation \eqref{eq:noise_mismodelling_params_covariance}represents the \emph{mean-square-error} of the noise parameters $\boldsymbol{\lambda}$. Notice that there is no bias term in \eqref{eq:noise_mismodelling_signal_params_covariance} since the signal parameters remain unbiased under noise mis-modelling. 

The scatter-to-width ratio $\Upsilon_A$ quantifies the relative size of the actual scatter of the MLEs to the model-predicted width, while the width-to-width ratio $\Xi_A$ quantifies the relative size of the model-predicted width to the true width. Both ratios are dimensionless and provide a clear diagnostic of noise mis-modelling effects on parameter inference of the underlying noise process.

The scatter-to-width ratio $\Upsilon_A$ is the most 
important diagnostic. A value $\Upsilon_A \simeq 1$ indicates that the 
model's reported uncertainties on signal/noise parameters are 
statistically consistent with the actual scatter of the maximum-likelihood 
estimates across noise realisations. Values $\Upsilon_\alpha > 1$ signal that 
the posteriors are too narrow relative to the true scatter --- for most noise 
realisations, the true noise parameters would be excluded by the reported 
credible intervals. Conversely, $\Upsilon_\alpha < 1$ indicates overly 
conservative posteriors. The width-to-width ratio $\Xi_\alpha$ in Eq.~\eqref{eq:xi_matrix_noise}
separately quantifies information loss: values $\Xi_\alpha > 1$ indicate that 
the model predicts wider posteriors than the correct analysis would yield, typically as a consequence of tapering erasing data near gap edges. An
interactive explorer, which sweeps $\Upsilon$ and $\Xi$ against the posterior
they imply, is available in the companion
documentation~\href{\gaplikedocs/_static/upsilon_xi_explorer.html}{\faGithub\,\texttt{upsilon\_xi\_explorer}}.


We emphasise that these diagnostics can be computed cheaply using only the model likelihood and the ability to simulate noise from the true covariance $\bQ$ describing the windowed noise process. The direct evaluation of $\langle\Delta\hat{\Theta}^A\Delta\hat{\Theta}^B\rangle$ requires computing the full windowed covariance matrix and pseudo inverse of large $N \times N$ matrices, which may be expensive for large data sets. 
However, a Monte Carlo approach provides a practical alternative: one draws many realisations of $\bW\bn$ from the true covariance $\bW\bSigma\bW$, computes the MLE from \eqref{eq:biased_noise_params_mismodelling} for each using only the model likelihood, and estimates 
$\langle\Delta\hat{\Theta}^A\Delta\hat{\Theta}^B\rangle$ from the sample covariance of the MLEs. The only time the full noise covariance matrix is required is when computing the width-to-width ratio $\Xi_{A}$. This 
mirrors the approach discussed in 
Ref.~\cite{Burke:2025bun} for signal parameters, and means that the full 
diagnostic could be obtained (up to monte-carlo error) without ever computing the correct (but expensive) 
gapped covariance. Taken together with the signal-parameter diagnostics of 
Ref.~\cite{Burke:2025bun}, the framework presented here provides a unified and 
efficient toolkit for validating the statistical consistency of LISA data 
analysis pipelines operating on gapped data with simultaneously estimated noise 
and signal parameters.

\subsection{Higher order terms for MLE and covariance}\label{sec:Godambe_White_formalism}
The scatter formulas of Sec.~\ref{subsec:metrics_scatter_to_width} are exact
statements about the \emph{linearized} estimator
[Eq.~\eqref{eq:noise_param_fluctuations_noise_mismodelling}]. For Gaussian
data, the mean and covariance are given \emph{identically} by
Eqs.~\eqref{eq:biased_noise_params_mismodelling} and
\eqref{eq:noise_mismodelling_params_covariance}, with no higher-order
corrections. The estimator actually used in an analysis --- the maximum of
the model likelihood --- is a nonlinear function of the data, and when the
mis-modelling pushes it far from the truth its distribution is instead
governed by the standard theory of mis-specified
M-estimators~\cite{huber1967behavior,white1982maximum}. As derived in Appendix~\ref{app:sandwich}, the estimator is centred
on the Kullback–Leibler (KL) pseudo-true parameters,
\begin{equation}
    \blambda^{*} = \arg\min_{\blambda}\,
    \Big[\ln{\det}^{+}\bSigma'(\blambda)
    + \text{Tr}\big(\bSigma'(\blambda)^{+}\,\bQ\big)\Big]\,.
    \label{eq:pseudo_true}
\end{equation}
Equation~\eqref{eq:pseudo_true}, and the covariance quoted below, are both
controlled by the same object: the \emph{score} of the model log-likelihood
with respect to the noise parameters, $\partial_{\alpha}\log\mathcal{L}$,
which is quadratic in the windowed data
[Eq.~\eqref{eq:model_noise_score}]. On each realization the
maximum-likelihood estimator solves
$\partial_{\alpha}\log\mathcal{L}(\hat{\blambda}) = 0$. The pseudo-true point
is where the \emph{mean} score vanishes,
$\langle\partial_{\alpha}\log\mathcal{L}\rangle\big|_{\blambda^{*}} = 0$
[Eq.~\eqref{eq:pseudo_true_stationarity}]. This is what permits a controlled
expansion of the score about $\blambda^{*}$
[Eq.~\eqref{eq:score_expansion}] rather than about the truth, where the mean
score is large under mis-modelling. Carrying out this expansion
(Appendix~\ref{app:sandwich}) yields the Godambe--White sandwich
covariance~\cite{godambe1960optimum,white1982maximum}
\begin{align}
    \mathrm{Cov}\big[\hat{\blambda}\big] &\simeq H^{-1} J H^{-1}\,,
    \label{eq:sandwich}\\
    H_{\alpha\beta} &=
        \partial_{\alpha}\partial_{\beta}\big\langle -\log\mathcal{L}
        \big\rangle\Big|_{\blambda^{*}}\,,
    \nonumber\\
    J_{\alpha\beta} &=
        \sum_{j,k} A^{*}_{\alpha,j}\,\big|Q_{jk}\big|^{2}\,A^{*}_{\beta,k}\,,
    \nonumber
\end{align}
where $\boldsymbol{A}^{*}_{\alpha} = \mathrm{diag}\big[\bSigma'^{+}
\partial_{\alpha}\bSigma'\,\bSigma'^{+}\big]_{\blambda^{*}}$ for the diagonal
models. The two slices of the sandwich play distinct roles. The ``bread''
$H$ is the mean curvature of the model log-likelihood at the pseudo-true
point; its explicit trace expression is given in
Eq.~\eqref{eq:sandwich_H_explicit}, and by Eq.~\eqref{eq:sandwich_H_effective}
it is nothing but the effective Fisher matrix \eqref{eq:effective_fisher}
evaluated at $\blambda^{*}$ instead of at the truth. The ``meat'' $J$ is the
covariance of the score under the \emph{true} windowed process,
$J_{\alpha\beta} = \mathrm{Cov}\big[\partial_{\alpha}\log\mathcal{L},\,
\partial_{\beta}\log\mathcal{L}\big]_{\blambda^{*}}$. Since the score is a
Gaussian quadratic form, Isserlis' theorem gives the closed form given in the appendix Eq.~\eqref{eq:sandwich_J_general}
\begin{equation}
    J_{\alpha\beta} = \frac{1}{2}\,\text{Tr}\Big[
    \big(\bSigma'^{+}\bSigma'_{\alpha}\bSigma'^{+}\big)\,\bQ\,
    \big(\bSigma'^{+}\bSigma'_{\beta}\bSigma'^{+}\big)\,\bQ
    \Big]\Big|_{\blambda^{*}}\,,
    \label{eq:J_general_maintext}
\end{equation}
This is the first term of the
linearized information matrix \eqref{eq:info_matrix} evaluated at the pseudo-true parameter $\blambda^{*}$ rather than the truth. The second (bias-squared) term of
\eqref{eq:info_matrix} is absent because the score $\partial_{\alpha}\log\mathcal{L}$ is centred there. For the
diagonal-family models, Eq.~\eqref{eq:J_general_maintext} reduces to the sum
quoted in Eq.~\eqref{eq:sandwich} via the complex-normal identity
$\mathrm{Cov}(|r_j|^2, |r_k|^2) = |Q_{jk}|^2$ per channel
[Eq.~\eqref{eq:sandwich_J_diag}]. 
For the correct model,
$\blambda^{*} = \boldsymbol{0}$, the identity is restored, $\boldsymbol{J} = \boldsymbol{H} = \bGamma$,
and Eq.~\eqref{eq:sandwich} reduces to $\bGamma^{-1}$. 

Quantities involving the Godambe-White sandwich covariance and pseudo true point will involve the superscript ``sw'', i.e., $\Upsilon^{\rm sw}$ and $\Delta\hat{\lambda}^{\rm sw}$ to discern them from the linearised Fisher-matrix formalism, $\Upsilon^{\rm lin}$ and $\Delta\hat{\lambda}^{\rm lin}$, presented earlier in this section. 

We make the final remark here that, under \emph{severe} mismodelling where $\bSigma \not \approx \bQ$, then we need to use \eqref{eq:pseudo_true} as the MLE and \eqref{eq:J_general_maintext} alongside \eqref{eq:J_general_maintext} for $\boldsymbol{J}$ and \eqref{eq:sandwich_H_explicit} for $\boldsymbol{H}$, together forming the Godambe-White covariance of the parameter. If, the noise mis-modelling is slight such that $\bSigma' \approx \bQ$, then the MLE is given by \eqref{eq:biased_noise_params_mismodelling} with covariance matrix given by the inverse of Eq.\eqref{eq:approx_fisher_matrix_windowed}. Throughout our analysis, we will use both formalisms and demonstrate two situations where the latter breaks down, but the former is robust.

\section{Approximations to the Windowed FD Covariance}
\label{sec:approximations_windowed_fd_covariance}
\subsection{Full windowed covariance}

Let $n(t)$ be a stationary circulant Gaussian noise process with one-sided power spectral density $S_n(f)$. In the continuous limit, the Wiener-Khintchine theorem states that
\begin{equation}
    \langle \tilde{n}(f)\tilde{n}(f')^* \rangle = \frac{1}{2}\delta(f - f')\,S_n(f)\,,\qquad f, f' > 0\,.
    \label{eq:stationary_fd}
\end{equation}
We introduce a window function $w(t)$ and define the windowed noise process $N(t) = w(t)\,n(t)$. In the Fourier domain, multiplication becomes convolution:
\begin{equation}
    \tilde{N}(f) = \int \tilde{w}(f - u)\,\tilde{n}(u)\,\text{d}u.
    \label{eq:windowed_fd}
\end{equation}

Using Eqs.~\eqref{eq:stationary_fd} and \eqref{eq:windowed_fd}, the FD covariance of the windowed process is
\begin{align}
    \langle \tilde{N}(f)\tilde{N}(f')^* \rangle 
    &= \frac{1}{2}\int \;\tilde{w}(f - u)\,\tilde{w}(f' - u)^*\,S_n(u)\,\text{d}u\,.
    \label{eq:full_fd_cov}
\end{align}
This is the continuous version of the result of \eqref{eq:windowed_frequency_domain_covariance}. This result is fully general and valid for any time-dependent function $w(t)$. Now we will investigate various approximate forms of the windowed FD covariance that could be useful for practical computations. 
\subsection{Leading diagonal approximation to the windowed FD covariance}
Setting $f = f'$ in Eq.~\eqref{eq:full_fd_cov} gives the diagonal elements:
\begin{equation}
    \langle |\tilde{N}(f)|^2 \rangle 
    = \frac{1}{2}\int du\;|\tilde{w}(f - u)|^2\,S_n(u)\,.
    \label{eq:diagonal_exact}
\end{equation}
This is a convolution of the spectral window function $|\tilde{w}|^2$ with the PSD. In the discrete setting with frequency resolution $\Delta f = 1/T$ and cadence $\Delta t$, this becomes
\begin{equation}
    \left[\mathbb{E}[\tilde{\bN}\tilde{\bN}^\dagger]\right]_{ii} 
    = \frac{\Delta f}{2}\sum_{p=0}^{N-1}|\tilde{w}[\overline{i - p}]|^2\,S_n[p]\,.
    \label{eq:diagonal_discrete}
\end{equation}
for $\overline{i - p} \equiv i - p\mod N$. Note that we make the modelling choice to set the off-diagonal elements $\left[\mathbb{E}[\tilde{\bN}\tilde{\bN}^\dagger]\right]_{i\neq j}  = 0$, approximating Eq.~\eqref{eq:windowed_frequency_domain_covariance} by only the leading diagonal. Equation~\eqref{eq:diagonal_discrete} is the \emph{exact} diagonal of the windowed FD covariance matrix. It can be computed in $\mathcal{O}(N\log N)$ operations via FFT convolution, making it a cheap approximation that captures the frequency-dependent power redistribution caused by the window. Similarly, the matrix is invertible with inverse simply the reciprocal of the diagonal elements, which is another $\mathcal{O}(N)$ operations. This is a significant improvement over the $\mathcal{O}(N^3)$ operations required to compute the full windowed FD covariance matrix and its pseudo-inverse.

\subsection{Slowly varying PSD approximation to the windowed FD covariance -- normalisation constant}
If the PSD $S_n(f)$ varies slowly compared to the width of the spectral window $|\tilde{w}|^2$, we can pull $S_n$ outside the convolution:
\begin{align}
    \left[\mathbb{E}[\tilde{\bN}\tilde{\bN}^\dagger]\right]_{ii} 
    &\approx \frac{\Delta f}{2}\,S_n[i]\sum_{p=0}^{N-1}|\tilde{w}(\overline{i - p})|^2 \nonumber \\
    & = \frac{\Delta f}{2}\,S_n[i]\sum_{p=0}^{N-1}|\tilde{w}[p]|^2\,, \\
\end{align}
where in the last step we used the fact that the sum over $p$ is a circular convolution evaluated at zero lag. The remaining task is to evaluate $\sum_p |\tilde{w}[p]|^2$. At this point, we can invoke Parseval's theorem that states 
\begin{equation}
    \sum_{j=0}^{N-1}|\tilde{w}[j]|^2 = N\Delta t^2\sum_{k=0}^{N-1}|w[k]|^2\,,
    \label{eq:parseval}
\end{equation}
for $k$ a time index. 
Assuming a DTFT convention 
\begin{equation}
    \tilde{w}[j] = \Delta t\sum_{k=0}^{N-1}w[k]\,e^{-2\pi i jk/(N\Delta t)}\,,
    \label{eq:dft_convention}
\end{equation}
giving 
\begin{align}
    \left[\mathbb{E}[\tilde{\bN}\tilde{\bN}^\dagger]\right]_{ii} &= \frac{\Delta f}{2}\,S_n[i]\sum_{p=0}^{N-1}|\tilde{w}[p]|^2\,, \\
    &=\frac{1}{2}\Delta f S_{n}[i] N^2\Delta t^2 \left(\frac{1}{N}\sum_{k = 0}^{N-1}|w[k]|^2\right) \\
    &= \frac{1}{2\Delta f}S_{n}[i]W_{c} \\
    & = \left[\mathbb{E}[\tilde{\bn}\tilde{\bn}^\dagger]\right]_{ii} W_c
    \label{eq:diagonal_slow_psd}
\end{align}
For the window normalising constant given by mean-square of the window function provided below
\begin{equation}
    W_{c} = \frac{1}{N}\sum_{k = 0}^{N-1}|w[k]|^2
\end{equation}

Comparing with the unwindowed result $[\mathbb{E}[\tilde{\bn}\tilde{\bn}^\dagger]]_{ij} = \delta_{ij}S_n[i]/(2\Delta f)$, we see that the diagonal of the windowed FD covariance is reduced by precisely the factor $W_c$.

\subsection{Hierarchy of approximations}
\label{eq:hierarchy_of_approximations}

To summarise, we have a hierarchy of increasingly accurate approximations to the diagonal of the windowed FD covariance:
\begin{enumerate}
    \item \textbf{No correction:} $[\mathbb{E}[\tilde{\bN}\tilde{\bN}^\dagger]]_{ii} \approx S_n(f_i)/(2\Delta f)$. This ignores the window entirely and leads to biased noise parameter estimates (noise amplitudes underestimated).
    
    \item \textbf{Scalar $W_c$ correction:} $[\mathbb{E}[\tilde{\bN}\tilde{\bN}^\dagger]]_{ii} \approx W_c\,S_n(f_i)/(2\Delta f)$. This accounts for the overall power loss but assumes the PSD is slowly varying. Valid when the spectral window $|\tilde{w}|^2$ is narrow compared to features in $S_n(f)$.
    
    \item \textbf{Frequency-dependent correction:} $[\mathbb{E}[\tilde{\bN}\tilde{\bN}^\dagger]]_{ii} = \frac{\Delta f}{2}\sum_p |\tilde{w}(f_i - f_p)|^2\,S_n(f_p)$. This is the exact diagonal, capturing the frequency-dependent power but ignoring the off diagonal elements by explicitly setting them to zero. Computable in $\mathcal{O}(N\log N)$ via FFT convolution. Recommended when the PSD has significant structure (steep red noise, zero crossings). Note that this formula is exact bin by bin whatever the PSD does. Spectral structure is what makes it \emph{necessary}. This frequency dependent correction is where the scalar correction of item~(2) departs most from the true variance. Whether the neglected off-diagonal terms also matter is the separate question of item~(4) below, and the zeros of the TDI transfer function are where we find that they do.
    
    \item \textbf{Full covariance $\tilde{\bQ}$:} The complete non-diagonal frequency domain matrix from Eq.~\eqref{eq:full_fd_cov}. Captures all inter-frequency correlations induced by the window. Computable in $\mathcal{O}(N^2\log N)$ but requires pseudo-inverse computation at $\mathcal{O}(N^3)$. Necessary for a fully consistent analysis. For this last item, the cost associated with a full time-domain or frequency-domain analysis is comparable, since dense solves are required using all components of the correlated matrix.
\end{enumerate}
For the noise-parameter Fisher matrix and the associated mismodelling diagnostics $\Upsilon_\alpha$ and $\Xi_\alpha$, using approximation~(3) in place of the Whittle diagonal provides a cheap and significant improvement over the scalar $W_c$ correction.

Every tier above models the covariance of the \emph{windowed} data.
There is a fifth option, which sidesteps the window altogether: one may
discard the gapped samples and work directly with the exact likelihood of the
samples that survive. This is also exact, it is cheaper than the full windowed
covariance, and we defer it to Sec.~\ref{subsec:td_exact} because it is most
naturally described alongside the numerical machinery that makes it
practical. The five options are compared in Table~\ref{tab:gap_scenarios}.

\begin{figure*}[t!]
\centering
\includegraphics[width=\textwidth]{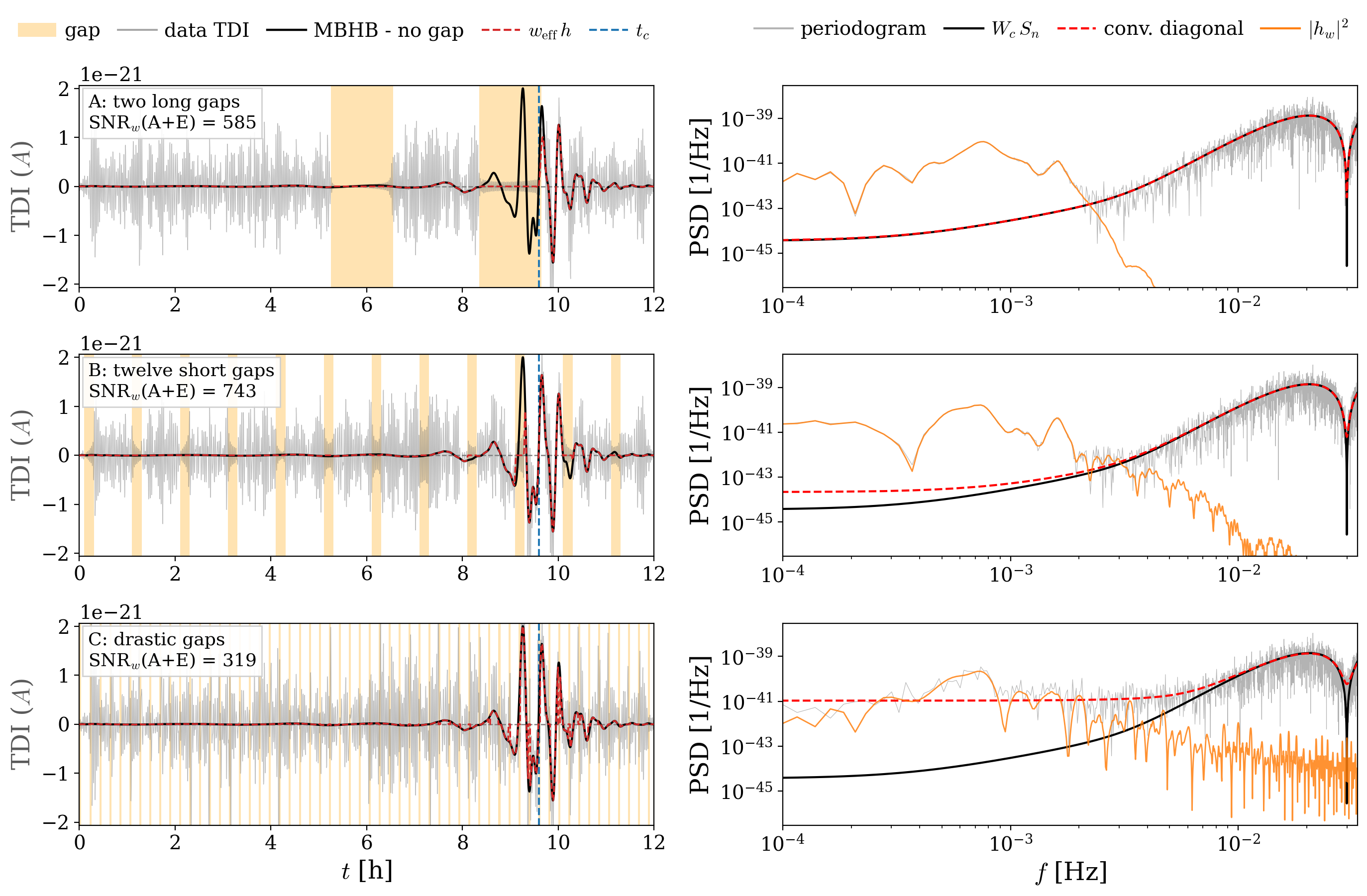}
\caption{The three controlled-gap scenarios, one per row; all rows
share the same axes, so levels can be compared directly across scenarios.
\emph{Left}: the gap pattern (shaded), the windowed TDI-$A$ data (grey),
the MBHB signal without gaps (black), the same signal as seen through the
effective window, $w_{\rm eff}h$ (red dashed), and the merger time
$t_c = 9.6$\,h (vertical dashed); $w_{\rm eff}$ is the segment taper
times the gap gates, defined in Eq.~\eqref{eq:gap_gate} below. The
vertical scale is set by the merger amplitude, so the largest
$\sim\!1\%$ of the displayed noise excursions fall outside it. The grey
trace is low-passed at $10$\,mHz for display only: more than
$99.9\%$ of the time-domain noise variance comes from the loud
OMS/TDI2-transfer region above $10$\,mHz, while the MBHB carries $99\%$
of its power below $1.6$\,mHz, so without that cut the signal would be
buried in the noise and invisible by eye --- every analysis in this work
uses the full band. Scenario A illustrates a ``gap during merger'': the
second gap ends just before $t_c$, its taper has not fully recovered by
then, and the observed merger (red) is therefore suppressed to a small
fraction of its true amplitude (black) --- the two curves separate
precisely there. In scenario C the low-passed noise is
comparable to the merger burst itself, the sub-$10$\,mHz band being
dominated by the aliased OMS floor. \emph{Right}: windowed periodograms
(grey) against the stationary expectation $W_{c}S_{n}$ (black) and the
exact convolved diagonal of Eq.~\eqref{eq:diagonal_discrete} (red
dashed), together with the windowed signal power (orange). The convolved
diagonal departs from $W_{c}S_{n}$ by a factor $\sim\!30$ around the TDI2
transfer zero in scenario B, and by three to five orders of magnitude
across the band in scenario C. Produced by \srclink{make_figures.py}.}
\label{fig:overview}
\end{figure*}
\section{Massive Black Holes, Noise Curves and Gaps}\label{sec:setup_mbh_gaps_simulation}
\subsection{Packages and waveform model}
Waveforms and instrument response are generated with
\texttt{lisabeta}~\cite{Marsat:2020rtl}
using the \texttt{IMRPhenomHM} approximant~\cite{London:2017bcn} with the
$(\ell,m) = (2,2), (2,1), (3,3), (3,2), (4,4), (4,3)$ modes, the full LISA response, and second-generation time-delay interferometry (TDI) observables in
fractional-frequency units. We assume that the LISA satellites are identical, so that the noises acting on each test mass are equal, and likewise for the noise in each optical metrology system. We additionally assume a static configuration of the LISA instrument, where the arm-lengths are both constant and equal ---
facilitating uncorrelated data channels $A$, $E$ and
$T$~\cite{tinto2014time,Tinto:2003vj}; since the $T$ channel is largely
insensitive to GWs at low frequencies, we neglect it from our analysis. Unless stated otherwise, our numerical work is
then carried out in the frequency domain, on the band-restricted
positive-frequency data of the remaining $N_{\rm ch} = 2$ channels $A$ and
$E$, treated as independent zero-mean circularly-symmetric complex Gaussian
vectors with identical covariances. Frequency-domain outputs are conjugated
to match the DFT convention of Eq.~\eqref{eq:DFT}, and time-domain templates
follow the continuous-to-discrete rule
$\bh = \mathrm{irfft}(\tilde{\bh}/\Delta t)$, which makes windowed SNRs
agree with their Whittle integrals in the no-gap limit. The methodologies
presented in this work are applicable to the more general $\{X,Y,Z\}$ basis, though we will not apply it here\footnote{
In such a basis, for less trivial orbital configurations (such as
time-evolving and unequal arm-lengths) a segmented frequency-domain approach
could become viable where each data segment is treated as independent;
otherwise, time-frequency approaches would need to be utilized.}.
Sampling uses the affine-invariant ensemble sampler
\texttt{emcee}~\cite{Foreman-Mackey:2012any}. The analytic machinery of
Secs.~\ref{sec:methodology}--\ref{sec:approximations_windowed_fd_covariance}
is implemented in a small stand-alone Python package
(\texttt{gaplike}); the full implementation, the executed analysis notebook
and every figure and table entry of this and the following sections are
reproducible from the companion repository.
\subsection{Source configuration}
We analyse a single massive black-hole binary in a $T_{\rm obs} = 12$\,h
segment sampled at $\Delta t = 15$\,s ($N = 2880$), in the band $f \in
[10^{-4}, 3.1\times10^{-2}]$\,Hz ($N_b = 1335$ bins per channel). The signal
parameters are $\btheta = (M_{\rm tot}, q, \chi_1, \chi_2, \log_{10} d_L,
\iota, \varphi, \lambda, \beta, \psi, t_c)$, with the merger time entering as
a pure frequency-domain phase $e^{-2\pi i f t_c}$. The injected source has
$M_{\rm tot} = 2.5\times10^{7}\,M_\odot$ ($m_1 = 1.5\times10^{7}$, $m_2 =
10^{7}\,M_\odot$), $q = 1.5$, $\chi_{1,2} = (0.3, 0.1)$, $d_L = 40$\,Gpc,
$\iota = 0.6$, and angles (defined in the LISA frame) $(\varphi, \lambda, \beta, \psi) = (1.0,
1.0, 0.4, 0.3)$; the merger sits at $t_c = 0.8\,T_{\rm obs} = 9.6$\,h,
deliberately late in the segment. A smooth spectral turn-on below
$2.45\times10^{-4}$\,Hz prevents the (formally infinite) early inspiral from
wrapping around the finite segment. With the segment window only, the (ungapped) network
SNR in $A+E$ is $902$. 
\subsection{Noise curves, TDI2 and its zeros}
\begin{figure*}
\centering
\includegraphics[width=\textwidth]{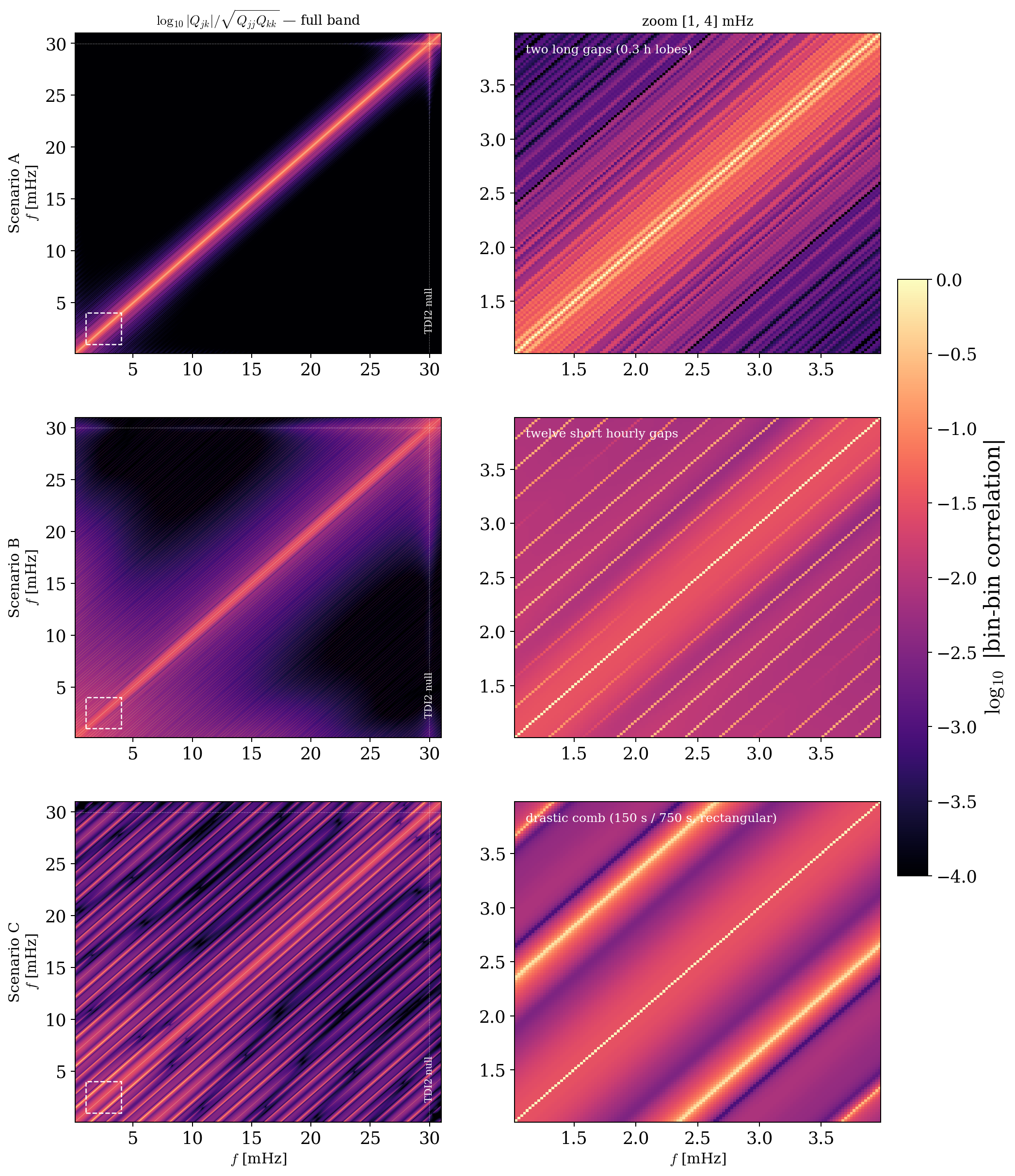}
\caption{Correlation structure of the windowed frequency-domain covariance
[Eq.~\eqref{eq:windowed_frequency_domain_covariance}],
$\log_{10}|Q_{jk}|/\sqrt{Q_{jj}Q_{kk}}$, for scenarios A (top), B
(middle) and C (bottom); right panels zoom into $[1,4]$\,mHz. Scenario A
produces a smooth
near-diagonal leakage halo of width $\sim\!1/(1\,{\rm h})$ with beating from
the two-gap separation; scenario B produces the comb of sideband diagonals at
multiples of $0.28$\,mHz and a bright cross at the TDI2 transfer zero
($\simeq 30$\,mHz, dotted): the bins around the zero contain leaked power
only, and are therefore strongly correlated with the rest of the band.
The rectangular comb of scenario C produces the densest structure:
alias diagonals at multiples of $1/750\,{\rm s} = 1.33$\,mHz spanning the
whole band. These are the bin--bin correlations that the exact solvers of
Secs.~\ref{subsec:fullcov_trick} and \ref{subsec:td_exact} unmix. Produced by \srclink{fig_cov_colormap.py}.}
\label{fig:cov_cmap}
\end{figure*}

We assume that the primary noise sources, such as laser
frequency noise, are suppressed to negligible levels by the post-processed
TDI variables. What remains are the so-called secondary noises. Specifically,
we consider the contributions from the acceleration noise of each test mass (TM) and the optical metrology system (OMS) noise, whose PSDs are assumed equal for all of the LISA satellites and given by
\begin{align}
    S_{\rm tm}(f) &= \frac{A_{\rm tm}^{2}}{(2\pi f c)^{2}}
        \left[1 + \left(\frac{f_1}{f}\right)^{2}\right]
        \left[1 + \left(\frac{f}{f_2}\right)^{4}\right],
    \label{eq:psd_tm}\\
    S_{\rm oms}(f) &= A_{\rm oms}^{2}
        \left(\frac{2\pi f}{c}\right)^{2}
        \left[1 + \left(\frac{f_3}{f}\right)^{4}\right],
    \label{eq:psd_oms} 
\end{align}
with $A_{\rm tm} = 3\times10^{-15}\,{\rm m\,s^{-2}\,Hz^{-1/2}}$, $A_{\rm oms}
= 15\,{\rm pm\,Hz^{-1/2}}$ and $(f_1, f_2, f_3) = (0.4, 8, 2)$\,mHz.
Propagating both terms through the second-generation TDI $A/E$ combinations 
gives \cite{2023PhRvD.108h2004Q}
\begin{align}
    S^{\rm TM}_{A}(f) &= 4\,\mathcal{R}(f)\,
        \big(3 + 2\cos x + \cos 2x\big)\, S_{\rm tm}(f)\,,
    \label{eq:psd_AE_tm}\\
    S^{\rm OMS}_{A}(f) &= 2\,\mathcal{R}(f)\,\big(2 + \cos x\big)\,
        S_{\rm oms}(f)\,,
    \label{eq:psd_AE_oms}\\
    \mathcal{R}(f) &= 16\,\sin^{2}\!x\,\sin^{2}\!2x\,,
    \label{eq:tdi2_transfer}
\end{align}
with $x \equiv 2\pi f L/c$ and armlength $L = 2.5$\,Gm. The same expression also applies to channel $E$. For parameter estimation, the overall noise PSD is modeled as
\begin{equation}
    S_{A}(f;\blambda)
    = 10^{\lambda_{\rm tm}}\, S^{\rm TM}_{A}(f)
    + 10^{\lambda_{\rm oms}}\, S^{\rm OMS}_{A}(f)\,.
    \label{eq:psd_family}
\end{equation}
The noise parameters are the two dimensionless $\log_{10}$ deviations of TM and OMS
noise powers from their reference values, with truth
$\blambda = \boldsymbol{0}$. All inference below is joint in the $11$ signal  and $2$ noise parameters.
The transfer function \eqref{eq:tdi2_transfer} vanishes at $f = k\,c/(2L)$
and, through its second (second-generation) factor, at $f = k\,c/(4L)$; for
$L = 2.5$\,Gm the first zero of the latter family, $f_{\varnothing} \simeq
30$\,mHz, lies \emph{inside} our analysis band. At such a zero the stationary
PSD drops by orders of magnitude: the instrument is locally quiet, and any
power found there in the windowed data must have leaked from neighbouring
frequencies. The zeros of the TDI transfer therefore act as natural
amplifiers of window mis-modelling --- a point that drives the results of
Sec.~\ref{sec:results} and is visible directly in the correlation structure
of $\bQ$ (Fig.~\ref{fig:cov_cmap}): for the gapped windows the bins around
$f_{\varnothing}$ are strongly correlated with the rest of the band, because
their content is entirely leakage. The same structure explains why the
hierarchy of Sec.~\ref{eq:hierarchy_of_approximations} jumps from
the leading diagonal directly to the full covariance, with no banded
(multi-diagonal) tier in between: hard truncation of $\tilde{\bQ}$ to a band
of diagonals does not in general remain positive semi-definite, and a valid
banded model would require smoothly tapering the off-diagonals to zero, which we do not pursue here.

\subsection{Gap families}
The effective window is the product of a fixed segment taper and a gap gate,
\begin{equation}
    w_{\rm eff}(t) = w_{\rm seg}(t)\,\prod_{i} g_i(t)\,,
    \label{eq:gap_gate}
\end{equation}
\begin{equation}
    g_i(t) =
    \begin{cases}
        0\,, & |t-t_i| \le d_i/2\,,\\[2pt]
        \sigma\!\big(\tfrac{|t-t_i| - d_i/2}{\tau}\big)\,,
            & d_i/2 < |t-t_i| < d_i/2 + \tau\,,\\[2pt]
        1\,, & \text{otherwise}\,,
    \end{cases}
    \label{eq:gap_notch}
\end{equation}
where $w_{\rm seg}$ is a Tukey window with taper fraction $\alpha = 0.05$,
$t_i$ and $d_i$ are the gap centres and durations, and
\begin{equation}
    \sigma(u) = \Big[\,1 +
    \exp\!\Big(\frac{1}{u} - \frac{1}{1-u}\Big)\Big]^{-1}\,,
    \qquad u \in (0,1)\,,
    \label{eq:planck_ramp}
\end{equation}
is the $C^{\infty}$ Planck ramp~\cite{McKechan:2010kp}. The segment window tapers the data at the start and end of the observation period, while the gap windows set the data to zero in the gaps, with tapering over a duration $\tau$ at the edges. 
We consider three controlled families, chosen to separate the ways gaps interact with the analysis (Fig.~\ref{fig:overview}):
\subsubsection*{Scenario A: Two long gaps}
Two one-hour gaps ($d_i = 1$\,h) centred
at $t_i = 5.9$ and $9.0$\,h, generously tapered ($\tau = 0.3$\,h). The first
interrupts the inspiral; the second ends close enough to the merger that its
recovery lobe overlaps it ($w_{\rm eff}(t_c) \simeq 0.18$), i.e.,\ this
scenario includes a ``gap during merger''. This family has similar characteristics to long,
scheduled interruptions, for example for antenna repointing or planned
maintenance, where the gap times are known in advance and the signals are well separated, so that tapering can be easily applied by the user.
\subsubsection*{Scenario B: twelve short gaps.} Twelve nine-minute gaps ($d_i =
0.15$\,h), one per hour ($t_i = (0.2 + k)$\,hr, $k = 0,\dots,11$), with short
tapers ($\tau = 0.05$\,h $= 3$\,min). This family has similar characteristics to frequent,
unscheduled interruptions of the science data stream --- for example due to onboard glitches or
safe-mode events temporarily disabling the measurement or its downlink, or
short losses of the communication link --- where long tapers would consume
an unacceptable fraction of the surviving data. The hourly repetition
produces a spectral comb: sidebands at multiples of $1/(1\,\mathrm{h})
\simeq 0.28$\,mHz around every frequency, with the broad leakage floor set
by the sharp $3$-minute edges.
\subsubsection*{Scenario C: drastic gaps.} Ten consecutive samples ($150$\,s) are
removed at the start of every $50$-sample ($750$\,s) block, with \emph{no}
taper: the window degenerates into a bare rectangular mask. This family has the characteristics of a (hopefully unlikely) scenario in which interruptions are so frequent that tapering is
no longer a meaningful option --- any taper long enough to smooth the edges
would consume the surviving data --- so the window carries no design freedom
at all. The periodic mask turns the window into a fine comb: aliasing sidebands
appear at every multiple of $1/750\,{\rm s} = 1.33$\,mHz with slowly
decaying harmonics, and the loud OMS-dominated region of the spectrum folds
down onto the quiet signal band --- the convolved diagonal at $0.3$\,mHz
exceeds $W_{c}S_{n}$ by a factor $\sim\!1.8\times10^{3}$ (up to
$\sim\!2\times10^{5}$ across the band). The
$150$\,s-block case, although dramatic and potentially unrealistic, keeps the source detectable and provides a stress test of our methodology. The corresponding key-parameter corner plot for this scenario is shown in
Fig.~\ref{fig:corner_key_C} (Sec.~\ref{subsec:results_C}).

All quoted SNRs are computed with the \emph{full windowed covariance},
\begin{equation}\label{eq:SNR_full_windowed_cov}
\mathrm{SNR}^{2} = 2\sum_{\rm ch}\mathrm{Re}\big[(\bW\bh)^{\dagger}
\bQ^{+}(\bW\bh)\big]
\end{equation}
with $\bW\bh$ the windowed, band-restricted template and $\bQ^+ = (\bW\bSigma\bW)^+$ the pseudo-inverse of the covariance matrix of the full windowed noise. 
By using $\bQ$, the only loss of information related to signal statistics (SNRs, likelihoods) is during the period of missing data, not the tapering process. The pseudo-inverse is formed by discarding the singular values annihilated by the gaps, that is, those directions of the covariance corresponding to the periods of missing data. 

With no gaps, the Whittle and windowed covariance matrix SNRs agree with SNR $\sim 902$. The window normalising constants are $W_{c} = 0.968$ (no gap; the residual
$\sim 3\%$ is the always-applied segment-edge Tukey taper), $0.742$ (A), $0.767$
(B) and $0.775$ (C); the retained singular values of the gated covariance
are $r/N_{b} = 1104/1335$ (A), $1178/1335$ (B) and $1331/1335$ (C). The
windowed SNRs are $585$ (A), $743$ (B) and $319$ (C). Scenario A loses substantially more than the duty-cycle estimate
$\sqrt{W_{c}^{\rm A}/W_{c}^{\rm nogap}}$ would suggest, because the tapered 
merger phase carries a significant fraction of the SNR. Scenario B sits close to its
duty-cycle estimate; and scenario C, at essentially the same duty cycle as A
and B, loses a further factor $\sim\!2$ to alias leakage alone. Throughout,
the three scenarios share a single noise realization, so that model
comparisons are paired. Table~\ref{tab:gap_scenarios} collects our scenarios, together with the conclusions of
Secs.~\ref{sec:verification} and \ref{sec:results} on which covariance
treatments remain viable for each.
\begin{table*}[t]
\centering
\caption{The three gap scenarios and the suitability of each covariance
treatment. Columns: number and duration of the (fully zeroed) gap cores; the
Planck-ramp taper length $\tau$ applied to each gap edge
[Eq.~\eqref{eq:planck_ramp}]; the duty cycle, i.e.\ the fraction of samples
with nonzero window (taper regions count as observed); the window
normalising constant $W_c$ [Eq.~\eqref{eq:diagonal_slow_psd}]; and the
windowed full-covariance SNR. The first row is the gap-free reference: the
always-applied segment taper ($\alpha = 0.05$ Tukey) is why $W_c = 0.968
\neq 1$ there. The five treatments are the hierarchy of
Sec.~\ref{sec:approximations_windowed_fd_covariance} plus the exact
time-domain solver: the raw stationary PSD (Whittle), the $W_c$-rescaled PSD,
the convolved diagonal [Eq.~\eqref{eq:diagonal_discrete}], the
restricted-stationary time-domain solver (Sec.~\ref{subsec:td_exact}), and
the full windowed covariance [Eq.~\eqref{eq:full_fd_cov}]. Verdicts
aggregate the signal and noise sectors (Table~\ref{tab:upsxi},
Sec.~\ref{sec:results}): \checkmark\ = unbiased and calibrated with
near-optimal widths; $(\checkmark)$ = usable with the stated caveat;
$\times$ = unusable (biases or mis-calibration far exceeding the quoted
uncertainties). Scenario definitions, duty cycles, $W_c$ and windowed SNRs
are produced by \srclink{scenarios.py} and \srclink{scenC.py}.}
\label{tab:gap_scenarios}
\begin{ruledtabular}
\begin{tabular}{lccccccccccc}
scenario & gaps & length & taper $\tau$ & duty cycle & $W_c$ & SNR &
PSD & $W_c\,$PSD & conv.\ diag.\ & TD solver & full cov. \\
\hline
--- (no gaps) & 0 & --- & --- & $100\%$ & $0.968$ & $902$ &
\checkmark & \checkmark & \checkmark & \checkmark & \checkmark \\
A: two long gaps (planned) & 2 & $1$\,h & $18$\,min & $83.3\%$ & $0.742$ & $585$ &
$(\checkmark)$\footnote{Usable with caution: the pseudo-true biases are
small in absolute terms ($|\lambda^{*}| \lesssim 0.30$) but amount to $4$--$8$ posterior widths in $\lambda_{\rm tm}$ and
$19$--$33$ in $\lambda_{\rm oms}$, so the (extremely narrow) noise
posteriors exclude the truth (Sec.~\ref{subsec:results_A}).} &
$(\checkmark)$\footnotemark[1] & \checkmark & \checkmark & \checkmark \\
B: twelve short gaps (unplanned) & 12 & $9$\,min & $3$\,min & $85.0\%$ & $0.767$ & $743$ &
$\times$ & $\times$ & \checkmark & \checkmark & \checkmark \\
C: drastic gaps (rectangular) & 58 & $150$\,s & --- & $80.0\%$ & $0.775$ & $319$ &
$\times$ & $\times$ &
$(\checkmark)$\footnote{Calibrated but inefficient: unbiased with honest
scatter ($\Upsilon \le 1.23$), yet the quoted widths exceed what the data in principle allow by up to an order of magnitude ($\Xi_{s} = 8.6$--$10.2$,
$\Xi_{\rm tm} = 13.7$; Sec.~\ref{subsec:results_C}).} &
\checkmark & \checkmark \\
\end{tabular}
\end{ruledtabular}
\end{table*}

\section{Practical Implementation: Frequency Domain}\label{sec:numerics}

\subsection{Computing the full windowed covariance}
\label{subsec:computing_full_cov}

The dense covariance is evaluated directly from
Eq.~\eqref{eq:windowed_frequency_domain_covariance}, restricted to the
analysis band. Writing $\tilde{w}$ for the DFT of the effective window and
$S_2[p]$ for the two-sided PSD on the DFT grid, we precompute the rectangular
matrix
\begin{equation}
    D_{jp} = \tilde{w}\big[\overline{j-p}\big]\,\sqrt{S_2[p]}\,,
    \label{eq:cov_factor_matrix}
\end{equation}
with $j$ running over the $N_b$ band bins only and $p$ over the full grid, so
that (up to the overall unit normalization, which cancels in every
dimensionless quantity used below, though not in the SNR of
Eq.~\eqref{eq:SNR_full_windowed_cov}, which is evaluated with the normalised
$\bQ$)
\begin{equation}
    \bQ \propto \boldsymbol{D} \boldsymbol{D}^{\dagger}\,.
    \label{eq:cov_outer_product}
\end{equation}
This is the band-restricted version of Algorithm~1 of
Ref.~\cite{Burke:2025bun}: the same circulant contraction of the spectral
window with the two-sided PSD, but with the rows and columns kept only for
the $N_b$ positive-frequency bins inside the analysis band, rather than for
the full two-sided grid $f \in [-f_{\rm nyq}, f_{\rm nyq}]$. Two remarks on
this restriction. First, the inner sum over $p$ still runs over the
\emph{entire} grid, so leakage into the band from frequencies outside it,
including from negative frequencies, is retained. We discard the covariance \emph{among} out-of-band bins, which the analysis
never uses. Second, working with the positive-frequency block of a real
process treats the retained bins as circularly-symmetric complex variables. The improper $\pm f$-coupling terms generated by window leakage across DC
and Nyquist are dropped. For the tapered windows of scenarios A and B these
couplings are negligible (the agreement of Sec.~\ref{sec:verification} is an
a posteriori bound). They do become visible using the
rectangular mask of scenario C, as quantified in Sec.~\ref{subsec:results_C},
which is also the regime where the exact time-domain formulation of
Sec.~\ref{subsec:td_exact} or fully windowed covariance matrix must take over.

A single matrix product of cost $\mathcal{O}(N_b^2 N)$ therefore yields the
band-restricted covariance without ever forming an $N\times N$ object; for
the configuration of Sec.~\ref{sec:setup_mbh_gaps_simulation} ($N = 2880$,
$N_b = 1335$) this takes seconds on a laptop. 

From the noise models \eqref{eq:psd_AE_oms}, the covariance is computed
\emph{per noise component} at the reference amplitudes, $\bC_{\rm TM}$ and
$\bC_{\rm OMS}$, so that the modelled covariance is linear in two fixed
matrices, 
\begin{equation}
    \bSigma'(\blambda)
    = 10^{\lambda_{\rm tm}}\,\bC_{\rm TM}
    + 10^{\lambda_{\rm oms}}\,\bC_{\rm OMS}\,,
    \label{eq:sigma_two_component}
\end{equation}
a structure exploited heavily below. The per-component split relies only on
the PSD being \emph{linear} in its parameters at fixed spectral shapes. What does \emph{not} extend is the
two-matrix simultaneous diagonalization of
Sec.~\ref{subsec:fullcov_trick}, which is specific to a two-component
family. With more components, or with parameters entering the shapes
nonlinearly (knee frequencies, spectral tilts), each likelihood evaluation
would require a fresh factorization\footnote{We elect for computational convenience rather than a more flexible model, indicating a clear direction for future work -- noise mis-modelling with flexible approximations, e.g., based on splines.}. The exact leading diagonal
[Eq.~\eqref{eq:diagonal_discrete}] never requires the dense object: it is a
circular convolution of $|\tilde w|^2$ with the PSD and costs
$\mathcal{O}(N\log N)$ by FFT.

One can assess the validity of our numerical recipe for $\bQ = \bW\bSigma\bW$ via brute force, by windowing and transforming an
ensemble of simulated stationary noise realizations. Their averaged
periodograms reproduce the convolved diagonal of
Eq.~\eqref{eq:diagonal_discrete} to $0.06\%$ (median over the band), and
their averaged outer products $\tilde{\bn}\tilde{\bn}^{\dagger}$
reproduce every entry of the band-restricted $\bQ$ of
Eq.~\eqref{eq:cov_outer_product} within its Monte-Carlo error: the covariance builders, not just their diagonals,
match the statistics of simulated windowed data.

\subsection{Diagonalisation of a two-parameter noise matrix}
\label{subsec:fullcov_trick}

A na\"ive implementation of the model likelihood
[Eq.~\eqref{eq:noise_model_likelihood}] with the dense $\bSigma'(\blambda)$
would require an $\mathcal{O}(N N_b^2)$ factorization at every point visited by
the sampler. Because the parameter dependence in
Eq.~\eqref{eq:sigma_two_component} is confined to two scalar amplitudes, the
entire $\blambda$-dependence can instead be diagonalized \emph{once}. We first restrict to the observed subspace. Writing the eigendecomposition
\begin{equation}
  \bC_{\rm TM} + \bC_{\rm OMS}
    = \boldsymbol{U}\,\boldsymbol{\Lambda}\,\boldsymbol{U}^{\dagger},
  \qquad
  \boldsymbol{\Lambda} = \mathrm{diag}(\Lambda_1 \ge \cdots \ge \Lambda_{N_b})\,,
  \label{eq:gated_eigendecomposition}
\end{equation}
--- which, the matrix being Hermitian and positive semi-definite, is also
its singular value decomposition --- we retain the $r$ directions with
$\Lambda_i > 10^{-8}\,\Lambda_1$ and collect the corresponding
eigenvectors in the $N_b \times r$ matrix $\boldsymbol{U}_r$, discarding
the directions annihilated by the gaps, in the same spirit as the
pseudo-inverse treatment of Ref.~\cite{Burke:2025bun}. The choice of tolerance is not critical: the
spectrum of the gated covariance shows a clean separation of many orders of
magnitude between the retained singular values and the near-null ones, and
we verified that every quantity reported below is unchanged when the
tolerance is varied across $10^{-6}$--$10^{-10}$ (see Fig.9 of ~\cite{Burke:2025bun}). 

Our analysis assumes a linear noise model in the overall amplitudes of  TM and OMS noises. Instead of forming the dense noise covariance matrices at every instance of a likelihood calculation, we find it computationally convenient to diagonalise the matrices to recover a cost comparable to a Whittle-based evaluation.

To see this, observe that the constant in $\blambda$ matrices $\bC_{\rm OMS}$ and $\bC_{\rm TM}$ can be eigendecomposed into $\boldsymbol{A} = \boldsymbol{U}_{r}^{\dagger}\bC_{\rm TM}\boldsymbol{U}_{r}$ and $\boldsymbol{B} = \boldsymbol{U}_{r}^{\dagger}\bC_{\rm OMS}\boldsymbol{U}_{r}$. By removing the singular values corresponding to the gap, $\boldsymbol{B}$ is positive definite allowing for a Cholesky decomposition of the form $\boldsymbol{B} = \boldsymbol{L}\boldsymbol{L}^{\dagger}$. Using $\boldsymbol{L}$, one can effectively whiten both the $\boldsymbol{A}$ and $\boldsymbol{B}$ matrices via
\begin{align}
\boldsymbol{\hat{B}} &= \boldsymbol{L}^{-1}\boldsymbol{B}\boldsymbol{L}^{-\dagger} = \boldsymbol{L}^{-1}\boldsymbol{L}\boldsymbol{L}^{\dagger}\boldsymbol{L}^{-\dagger} = \mathbb{I} \\
\boldsymbol{\hat{A}} &= \boldsymbol{L}^{-1}\boldsymbol{A}\boldsymbol{L}^{-\dagger}\,. 
\end{align}

Now, diagonalising the matrix $\boldsymbol{\hat{A}} = \boldsymbol{V}\text{diag}(\mu_j)\boldsymbol{V}^{\dagger}$ for eigenvalues $\mu_j$, one can set the operator $\boldsymbol{T} = \boldsymbol{V}^{\dagger}\boldsymbol{L}^{-1}\boldsymbol{U}_{r}^{\dagger}$ to obtain the two equations
\begin{align}
\boldsymbol{T}\bC_{\rm OMS}\boldsymbol{T}^{\dagger} &= \mathbb{I} \\
\boldsymbol{T}\bC_{\rm TM}\boldsymbol{T}^{\dagger} &= \boldsymbol{V}^{\dagger}\boldsymbol{\hat{A}}\boldsymbol{V} = \text{diag}(\mu_j)
\end{align}
Implying that 
\begin{equation}
\boldsymbol{T}\bSigma^\prime (\blambda)\boldsymbol{T}^{\dagger} = 10^{\lambda_{\rm tm}}\text{diag}(\mu_j) + 10^{\lambda_{\rm oms}}\,\mathbb{I} = \text{diag}(v_j)\,,
\end{equation}
hence the matrix $\boldsymbol{T}$ diagonalises the matrix $\boldsymbol{\Sigma}^\prime$ and this diagonalisation trick is only required \emph{once}. This is the most expensive step in the overall likelihood cost. 

The quadratic piece in the likelihood simplifies
\begin{align}
\boldsymbol{r}^T \bSigma^{\prime}(\blambda)^+ \boldsymbol{r} &= (\boldsymbol{T}\boldsymbol{r})^{T} \big(\boldsymbol{T}\boldsymbol{\Sigma}^{\prime}(\blambda)\boldsymbol{T}^{\dagger}\big)^{-1}(\boldsymbol{T}\boldsymbol{r}) \\
& = \boldsymbol{z}^{T}\,\text{diag}\big(v_j(\blambda)\big)^{-1} \boldsymbol{z} = \sum_{j=1}^{r}\frac{|z_j|^2}{v_j(\blambda)} \label{eq:trick_quadform}
\end{align}
for $\boldsymbol{z} = \boldsymbol{T}\boldsymbol{r}$ and $v_j(\boldsymbol{\lambda}) = 10^{\lambda_{\rm tm}}\mu_j + 10^{\lambda_{\rm oms}}$.
The determinant piece reduces to 
\begin{equation}
    \ln{\det}^{+}\,\bSigma'(\blambda)
    = \sum_{j=1}^{r}\ln v_j(\blambda) + \text{const}\,.
    \label{eq:trick_logdet}
\end{equation}

Each likelihood call then costs one $r\times N_b$ projection of the residual
(which changes with the signal parameters) plus $\mathcal{O}(r)$ work for any
value of $\blambda$ --- about three orders of magnitude cheaper than repeated
dense solves, and cheap enough that the full-covariance model can be sampled
jointly in all $13$ parameters with an off-the-shelf ensemble sampler. Our calculations have used the model covariance matrix in the likelihood $\bSigma^{\prime}$, but any covariance matrix (provided it follows the same noise family) is valid in our computations. 

We finish this section with two observations. Firstly, this structure means that finding the exact
\emph{profile} maximum-likelihood estimate of the noise parameters at fixed
signal --- the nonlinear completion of
Eq.~\eqref{eq:noise_param_fluctuations_noise_mismodelling} --- reduces to a
two-dimensional numerical maximization of
Eqs.~\eqref{eq:trick_quadform}--\eqref{eq:trick_logdet} and runs in seconds;
the same holds for the Kullback--Leibler (KL) pseudo-true parameters and the
Godambe--White sandwich covariance introduced in
Sec.~\ref{subsec:ensemble}. Secondly, we note that  the removal of singular values corresponding to the gaps defining the
pseudo-inverse [Eq.~\eqref{eq:pseudo_inv_def_1}] should be applied with the
\emph{same} tolerance in the sampled likelihood and in every analytic
quantity; with near-singular gated covariances an inconsistent cut amounts to
comparing two slightly different models, which becomes visible in these data as a
spurious offset between predicted and sampled noise posteriors.

\section{Practical Implementation: Time Domain}
\label{subsec:td_exact}
\
The frequency-domain hierarchy of
Sec.~\ref{sec:approximations_windowed_fd_covariance} presumes that one
windows the data and then models the windowed covariance. The
alternative, explored in Ref.~\cite{Burke:2025bun}, is to discard the
gapped samples altogether and work with the exact likelihood of those
that survive. Such a time-domain analysis is often said to be infeasible
at LISA data lengths, on the grounds that it requires an $N \times N$
covariance matrix. That objection is about \emph{storing and factorising}
that matrix, not about the analysis domain. What the likelihood actually
needs is the action of the covariance on a vector, and for a stationary
process that action costs $\mathcal{O}(N\log N)$ with no matrix formed at
all. The Fourier transform enters here as the means of applying an
operator, not as a change of analysis domain: the estimator remains the
exact time-domain likelihood of the observed samples, correlations across
the gaps included, which is precisely what a diagonal frequency-domain
model cannot represent. 
This section builds that evaluation. A stationary covariance is diagonalised by the Fourier transform, so its action on a vector costs a pair of transforms and a pointwise multiplication (Sec.~\ref{subsec:td_cg}). Deleting the gapped samples turns this into the action of a principal submatrix --- zero-embed, transform, restrict --- and the linear system it defines is then solved iteratively rather than by factorisation (Sec.~\ref{subsec:td_missing}). What is left is the determinant, the one ingredient a matrix--vector product does not deliver; Sec.~\ref{subsec:td_determinant} shows that it is often not needed at all, and gives an exact and a stochastic route for the cases where it is.
\subsection{Circulant covariances and the fast matrix–vector product}
\label{subsec:td_cg} 
Given that a circulant matrix $\bSigma$ is diagonalisable by the DFT matrix with eigenvalues proportional to the PSD of the process $S_n$, then knowledge of the $S_n$ encapsulates the full structure of the underlying circulant matrix. The time-domain matrix is given via 
\begin{equation}
    \gamma(\tau) = \frac{1}{2\Delta t}\,\mathrm{ifft}(S_{n})[\tau], \quad \tau = 0, 1, \ldots, N-1
\end{equation}
where $S_{n}$ denotes the one-sided PSD evaluated on the full two-sided DFT
grid, $S_{n}[k] = S_{n}(|f_{k}|)$, and $\mathrm{ifft}$ the $1/N$-normalised
inverse transform. The same convention is used for the preconditioner in
Eq.~\eqref{eq:preconditioner}. With the circulant property
\begin{equation}
\Sigma_{jk} = \gamma[(j - k) \mod N]\,.
\end{equation}
Since multiplying by a circulant matrix is the same as applying a convolution $(\bSigma \bv)_i = \sum_{k}\gamma[(i - k)\mod N]v_k$, one can then compute the matrix-vector product
\begin{equation}\label{eq:cheap_TD_full_matvec}
\bSigma \bv = \text{ifft}(\text{fft}[\gamma] \cdot \rm \text{fft}(v))
\end{equation}
where 
\begin{equation}
\text{fft}[\gamma] = \frac{1}{2\Delta t}S_{n}\,.
\end{equation}
Here ``$\cdot$'' stands for pointwise multiplication between the two vectors. One has been able to compute the vector $\bSigma v$ without having to build the matrix $\bSigma$. The cost is still $\mathcal{O}(N\log N)$, applied three times. 

\begin{figure*}[t]
  \includegraphics[width=\textwidth]{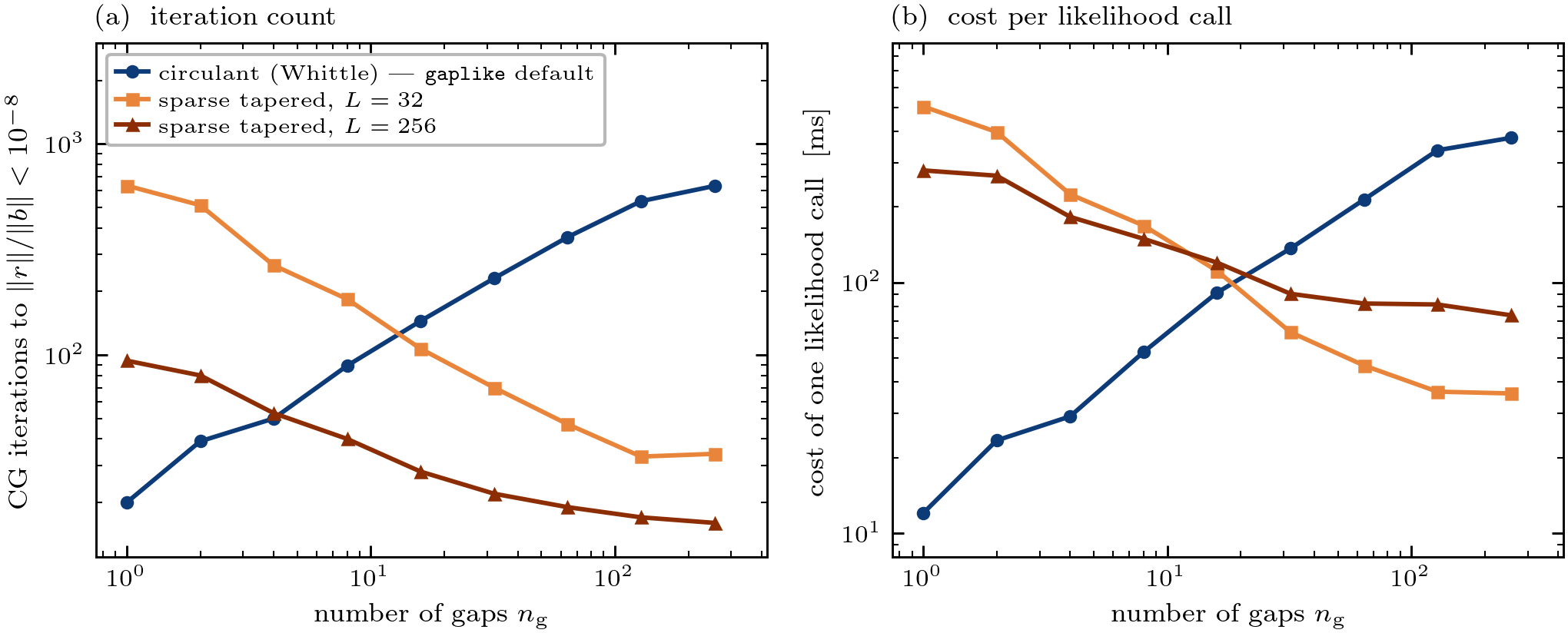}
  \caption{Two preconditioners for the matrix-free time-domain
    likelihood, as the same missing time is chopped ever more finely.
    $N = 8192$ at $\Delta t = 15\,$s, $15\%$ of the data removed at
    every $n_{\rm g}$, conjugate gradients run to
    $\|r\|/\|b\| < 10^{-8}$.  \emph{(a)} Iteration count.  The circulant
    (Whittle) preconditioner is exact between the gaps, so it degrades
    with their number; the sparse tapered preconditioner of
    Ref.~\cite{Baghi:2016myd} discards correlations beyond lag $L$, which
    only hurts while the surviving segments are longer than $L$, so it
    improves with their number.  \emph{(b)} The cost of one likelihood
    call at a new $\lambda$, which for the sparse method includes the cost of computing the banded Cholesky decomposition.  The crossover sits near $n_{\rm g} \simeq 20$;
    scenarios A and C of this paper sit either side of it.
    \srclink{fig_precond_compare.py}}
  \label{fig:precond_compare}
\end{figure*}
\subsection{Accounting for missing data}
\label{subsec:td_missing}
Now, the same principle applies to gated data. Let $O \subset \{0, 1,
\ldots, N-1\}$ denote the set of indices at which data are observed, with $m
= |O| < N$ the number of surviving samples, and let $\bR \in \{0,
1\}^{m\times N}$ be the associated selection matrix whose $a$th row is the
unit vector picking out the $a$th observed index. It is important to keep
three distinct objects apart. The first is the full data stream $\bx \in
\mathbb{R}^{N}$, whose gapped entries are simply never measured. The second
is the \emph{compressed} vector
\begin{equation}\label{eq:compressed_data}
\bx_{O} = \bR \bx \in \mathbb{R}^{m}\,,
\end{equation}
which is what the analyst actually holds: the surviving samples with the
gaps closed up. The third is its \emph{zero-embedding} $\bR^{T}\bx_{O} \in
\mathbb{R}^{N}$, which restores the original time stamps and writes zeros
into the gaps. In other words, $\bR$ deletes and $\bR^{T}$ zero-embeds. These
two operations satisfy
\begin{equation}\label{eq:selection_identities}
\bR\bR^{T} = \mathbb{I}_{m}\,, \qquad
\bR^{T}\bR = \boldsymbol{\Pi}_{\rm obs}\,,
\end{equation}
with $\boldsymbol{\Pi}_{\rm obs}$ the projector of
Eq.~\eqref{eq:pseudo_inv_gate}; for a rectangular (untapered) gate
$\boldsymbol{\Pi}_{\rm obs}$ is precisely the window matrix $\bW$. The first
identity states that embedding and then deleting returns the original
vector, so that $\bR^{T}$ is a right inverse of $\bR$. The second states
that deleting and then embedding merely zeros the gaps. We stress that
$\bR^{T}$ is the pseudo-inverse of $\bR$ and \emph{not} its inverse: since
$\bR^{T}\bR\bx = \boldsymbol{\Pi}_{\rm obs}\bx \neq \bx$, the gapped samples
are irretrievably lost, exactly as they should be.

Taking expectations, the covariance of the compressed data is the principal
submatrix
\begin{align}\label{eq:restricted_cov}
\bSigma_{OO} &= \langle \bx_{O} \bx_{O}^T\rangle = \bR \bSigma \bR^{T} \in
\mathbb{R}^{m \times m}\,, \\
(\bSigma_{OO})_{ab} &= \gamma\big[(t_{a} - t_{b}) \mod N\big]\,,
\end{align}
for $t_{a}, t_{b} \in O$. Notice that $\bSigma_{OO}$ is not circulant, since the
observed indices are irregularly spaced and the entries are no longer
constant along diagonals, but it is a principal submatrix of a circulant
matrix. Its entries remain \emph{exact} values of the autocovariance,
including those lags that straddle the gaps and thus encode the noise
correlations across missing data. The action of $\bSigma_{OO}$ then follows by
associativity,
\begin{align}\label{eq:zero_embed_matvec}
\bSigma_{OO}\bv_{O} = \bR \bSigma \bR^T \bv_O = \bR (\bSigma (\bR^T \bv_{O}))\,.
\end{align}
Unpacking this expression from right to left, one takes the compressed
vector $\bv_{O}$ and \emph{zero-embeds} it, $\bv_{O} \mapsto \bR^{T}\bv_{O}$,
such that it lives inside an $N$ dimensional vector space with zeros imputed
in place of the gaps. The operation $\bSigma(\bR^T \bv_{O})$ is then computed
via circular convolutions using Eq.~\eqref{eq:cheap_TD_full_matvec}, and the
result restricted to the $m$ dimensional subspace of observed data, namely
$\bSigma_{OO}\bv_{O} \in \mathbb{R}^{m}$. The imputed zeros annihilate
precisely those columns of $\bSigma$ that deletion removed, and the final
restriction retains precisely those rows that it kept, so
Eq.~\eqref{eq:zero_embed_matvec} reproduces the principal submatrix
\eqref{eq:restricted_cov} \emph{exactly}. We stress that circularity of
$\bSigma$ is a convenience rather than a requirement. For a stationary but
non-circular process the covariance is Toeplitz, and any $N \times N$
Toeplitz matrix embeds in a circulant matrix of size at least $2N \times
2N$~\cite{gray2006toeplitz}. Zero-padding the vector to that length,
applying the enlarged circulant by FFT, and discarding the padded entries
returns the exact Toeplitz product, at the cost of transforms twice as long.
Everything that follows therefore applies unchanged to a Toeplitz
$\bSigma$. Notice that this operation scales
similarly to before, only requiring a few fft computations, scaling with
$\mathcal{O}(N\log N)$.

Returning to the (rectangular) windowed log-likelihood in the time-domain
\begin{equation}
\log \mathcal{L} \propto -\frac{1}{2}(\bW\br)^T(\bW \bSigma\bW)^+(\bW\br) - \frac{1}{2}\log\det^{+}(\bW\bSigma\bW) 
\end{equation}
for $\br = (\bd - \bh(\boldsymbol{\theta}))$ then, noting that $\bW = \bR^T\bR$ and $\bR^T = \bR^+$ giving $\bR\bR^+ = \mathbb{I}_m$, we obtain an identical likelihood in terms of observed data with gaps marginalised out
\begin{equation}
\begin{split}
    \log\mathcal{L} = \log\mathcal{L}_O 
    \propto {}& -\frac{1}{2}\br_{O}^{\,T}\,\bSigma_{OO}(\blambda)^{-1}\,\br_{O} \\
    &  - \frac{1}{2}\log\det\bSigma_{OO}(\blambda)\,,
\end{split}
    \label{eq:td_restricted_likelihood}
\end{equation}
with $\br_{O} = (\bd - \bh(\btheta))_{O}$. This formalism was explored in Appendix C of ~\cite{Burke:2025bun}: samples within the gaps are discarded, and only those unaffected by the gaps are retained. Equation \eqref{eq:td_restricted_likelihood} is \emph{exact}: it is the estimator that every tier of the windowed hierarchy approximates. By construction there is no mis-modelling error, with $\Upsilon = \Xi = 1$. It is also, we note, precisely the marginal likelihood that Bayesian data augmentation targets by sampling the missing samples as auxiliary parameters~\cite{Baghi:2019eqo}; here we evaluate the same object directly, without augmentation.

In order to compute the likelihood \eqref{eq:td_restricted_likelihood}, we need to solve expressions of the form
\begin{equation} \label{eq:setup_CG_method}
\bSigma_{OO} \boldsymbol{u} = \br_{O}\,,
\end{equation}
which can be computed via Conjugate Gradient (CG) methods. Since the LISA noise spectrum may span many orders of magnitudes, particularly when the nulls of the PSD are included, the matrix $\bSigma_{OO}$ is likely to be ill conditioned. One can then precondition \eqref{eq:setup_CG_method} using a matrix $\boldsymbol{M} \approx \bSigma_{OO}$ in order to cluster the eigenvalues around unity to aid the CG method. The natural choice for $\bM$ is the gap-free (Whittle) covariance restricted to the observed subspace
\begin{equation}
\bM = \bR\bSigma_{\rm whittle}\bR \in \mathbb{R}^{m\times m}
\end{equation}
whose inverse is applied by the same pair of FFTs used in \eqref{eq:zero_embed_matvec}. Define
\begin{align}\label{eq:preconditioner}
\boldsymbol{M}^{-1}\bv_{O} &= \bR\big(\bSigma^{-1}_{\rm whittle}(\bR^{T}\bv_{O})\big)\,, \\
\bSigma_{\rm whittle}^{-1}\boldsymbol{w} &= \text{ifft}\!\left(\frac{2\Delta t}{S_{n}}\cdot \text{fft}(\boldsymbol{w})\right)\,,
\end{align}
with the eigenvalues $S_{n}$ floored near the nulls of the TDI transfer function, where $1/S_{n}$ would otherwise diverge. We therefore are interested in solving the equation
\begin{equation} \label{eq:precond_CG_method}
\boldsymbol{M}^{-1}\bSigma_{OO} \boldsymbol{u} = \boldsymbol{M}^{-1}\br_{O}\,,
\end{equation}
where, using \eqref{eq:zero_embed_matvec}, every application of the operator on the left-hand side is a sequence of zero-embeddings, FFTs and restrictions,
\begin{equation}
\bSigma_{OO}\boldsymbol{u} = \bR (\bSigma (\bR^T \boldsymbol{u})) = \br_{O}\,,
\end{equation}
so the goal here is to iteratively solve for the optimal value of $\boldsymbol{u}$ that satisfies the equation above.

The Conjugate Gradient (CG) algorithm is an excellent tool for this task. It solves symmetric
positive-definite systems using only matrix--vector products, the single
operation that Eq.~\eqref{eq:zero_embed_matvec} provides, without ever
requiring access to the matrix itself. This means that quadratic forms of likelihoods can be computed using a handful of FFTs without storing $\bSigma_{OO}$ in memory. The number of iterations is governed
by the spectrum of the (preconditioned) operator rather than by its
dimension: the more tightly the eigenvalues cluster around unity, the
fewer iterations are needed, whatever the size of $\bSigma_{OO}$. This is determined by the choice of preconditioner. The matrix $\bSigma_{OO}$ \emph{is} the stationary operator
everywhere except at the gaps, so $\boldsymbol{M}^{-1}\bSigma_{OO}$ is
the identity plus a correction localised at the gap edges. The spectrum clusters around unity, and the iteration count falls accordingly. This is the gapped-data analogue of the classical circulant preconditioners for
Toeplitz systems~\cite{strang1986proposal, chan1988optimal}. 

Two further
refinements come for free. First, the iteration must be initialised
somewhere, and the natural starting guess is the Whittle solution
$\boldsymbol{u}_0 = \boldsymbol{M}^{-1}\br_{O}$. The CG method then
only computes the correction induced by the gaps. Second, within a
stochastic sampler the accepted solution of the previous step is a better
starting point still, since a small move in the parameters barely changes
$\boldsymbol{u}$. The iteration is terminated at a fixed residual tolerance, so the resulting likelihood evaluation is fully deterministic.

The preconditioner we use by default is the inverse of the gap-free
(Whittle) covariance.  It is exact wherever the data is
uninterrupted, so what it leaves for the iteration to resolve is the
geometry of the gaps alone, and its cost grows with their number.
Reference~\cite{Baghi:2016myd} proposes a second option: taper the
time-domain autocovariance to zero beyond a lag $L$, restrict it to the
observed samples, and factorise the resulting banded matrix.  That
preconditioner is built from the true restricted covariance --- it knows
where the holes are --- but throws away every correlation beyond lag $L$.
The approximation becomes invalid 
when a surviving segment is longer than
$L$, and when the gaps are closely spaced none of them are.
 
Figure~\ref{fig:precond_compare} compares the cost of the two preconditioners under a controlled
scenario: a fixed $15\%$ of the data missing, chopped into $n_{\rm g}$
equal gaps.  The two preconditioners are complementary rather than competing: adding gaps
degrades the circulant one and improves the sparse one, because
fragmenting the data destroys exactly the long-range correlations that
the taper throws away. What decides between them is the cost of a full likelihood call rather than the iteration count, since the sparse routine must refactorise at every new $\blambda$; that places the crossover at a few tens of gaps. The
scenarios of this paper fall on either side of it --- the circulant
preconditioner is an order of magnitude cheaper for scenario A, the
sparse one by a similar factor for scenario C, with scenario B close to
the crossover --- and both return the same quadratic form to a few parts
in $10^{15}$.  We keep the circulant preconditioner as the default, since
it has no tuning parameter to set, and recommend the sparse alternative
whenever the data is heavily fragmented.

\subsection{The determinant}
\label{subsec:td_determinant}
\begin{figure*}[t!]
\centering
\includegraphics[width=\textwidth]{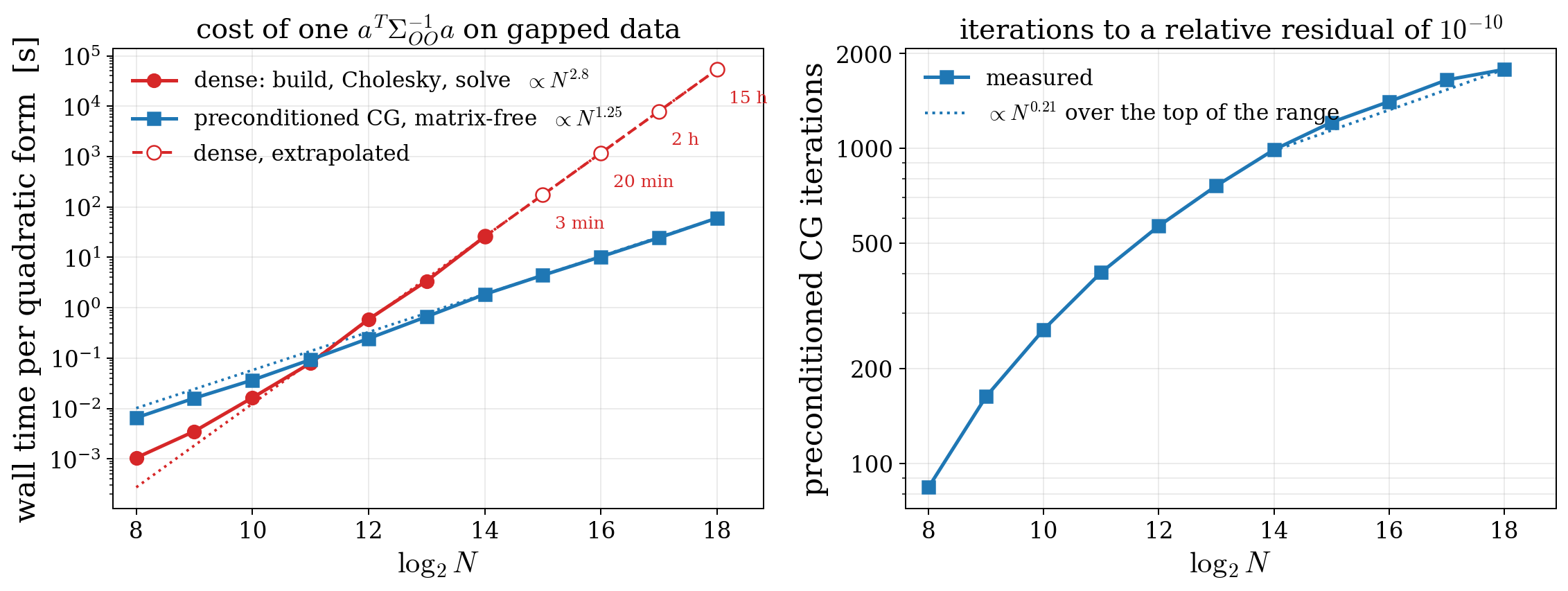}
\caption{Cost of one quadratic form
$\bn_{O}^{T}\bSigma_{OO}^{-1}\bn_{O}$ on gapped data as a function of data length. We use the gap scenario of C where we drop 10 of every 50 samples. We use the TDI2 $A$ channel spectrum and a fixed cadence $\Delta t = 15\,$seconds. \emph{Left}: minimum wall time over
repeated solves, for the dense route (assemble $\bSigma_{OO}$, Cholesky,
triangular solve; filled red) and for the matrix-free preconditioned
conjugate gradient (blue); dotted lines are the measured (empirical) power laws. \emph{Right}: the number of iterations required, which is
what determines the total cost from the $N\log N$ of a single iteration. The
growth decelerates as $N$ is increased logarithmically. The dotted line is the $N^{0.21}$ fitted
over the upper half of the range and used in the text. Produced by \srclink{fig_cg_scaling.py}.}
\label{fig:cg_scaling}
\end{figure*}

\begin{figure}[t!]
\centering
\includegraphics[width=\columnwidth]{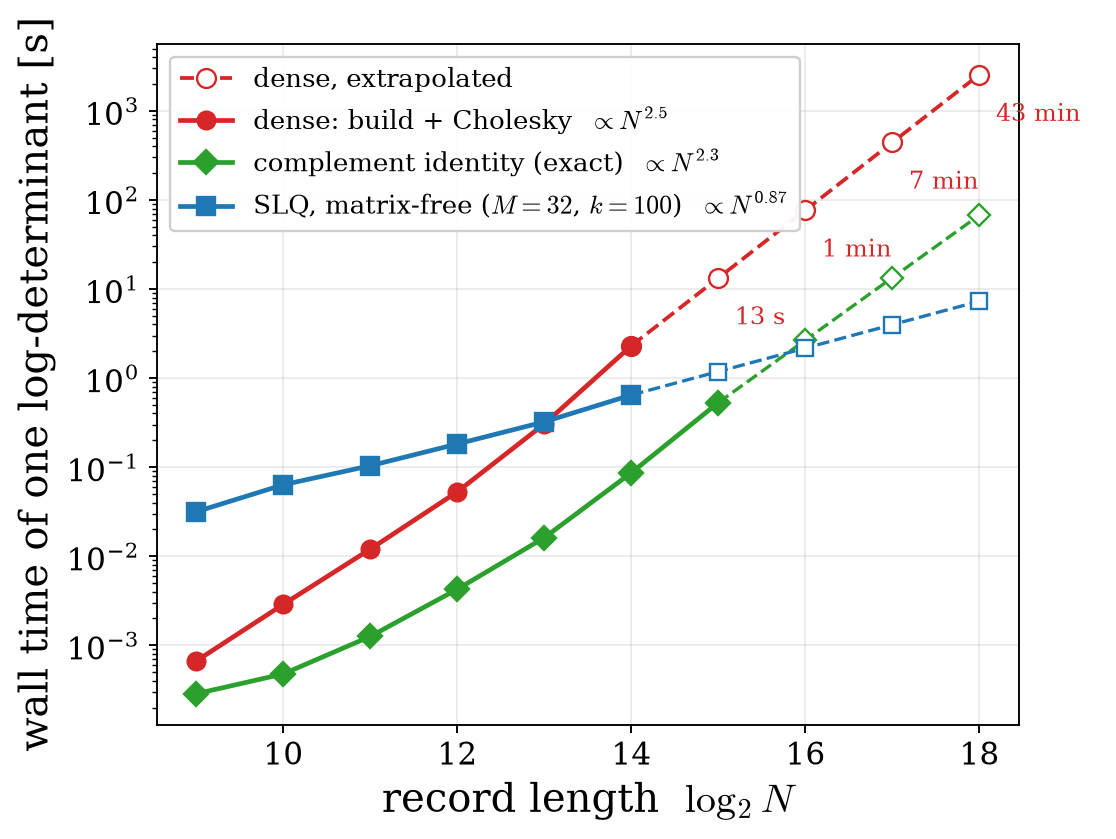}
\caption{\small Cost of one log-determinant $\log\det\bSigma_{OO}$ on the
scenario-C comb. Filled markers are measured, open markers extrapolated
along the fitted power law. The dense route (red) assembles $\bSigma_{OO}$
and factorises it. The complement identity \eqref{eq:complement_identity}
(green) is exact to machine precision but pays for the $g$ missing samples
rather than the $m$ retained ones. Stochastic Lanczos quadrature (blue)
carries a fixed overhead set by the probe and Lanczos counts, and wins once
the data is long enough. Produced by \srclink{fig_det_scaling.py}.}
\label{fig:det_scaling}
\end{figure}
The quadratic form is only half of the likelihood
\eqref{eq:td_restricted_likelihood}. The determinant term
$\log\det\bSigma_{OO}$ is the more challenging half, since it depends on the
whole spectrum of the matrix rather than on its action on a single vector.
Four observations render it harmless, and which one applies depends on the
noise model and on how much of the data is missing.

\emph{The determinant is often not needed at all.} The term
$\log\det\bSigma_{OO}(\blambda)$ is independent of the signal parameters
$\btheta$. Any move at fixed noise parameters cancels it within an MCMC
acceptance ratio, so signal-only parameter estimation requires no
determinant at any scale.

\emph{For a two-amplitude model it is available in closed form.} The matrix
$\bSigma_{OO}$ inherits the structure of
Eq.~\eqref{eq:sigma_two_component},
$\bSigma_{OO}(\blambda) = 10^{\lambda_{\rm tm}}\bC^{\rm tm}_{OO}
+ 10^{\lambda_{\rm oms}}\bC^{\rm oms}_{OO}$, so the simultaneous
diagonalisation of Sec.~\ref{subsec:fullcov_trick} applies verbatim. A
single $\mathcal{O}(m^{3})$ decomposition yields
$\log\det\bSigma_{OO}(\blambda) = \sum_{j}\log v_j(\blambda) + \text{const}$,
an $\mathcal{O}(m)$ sum of scalar logarithms for \emph{every} value of the
noise parameters. At the scale of this study ($m \simeq 2300$) this direct
route is all that is required, and it is the one used in
Sec.~\ref{subsec:results_C}. Its limitation is the one noted in
Sec.~\ref{subsec:fullcov_trick}: it is specific to two amplitudes entering
linearly through fixed spectral shapes, and we do not expect it to hold for 
more general noise models, such as splines or parametrized noise models modelling the spectral shapes (in frequency).

\emph{A complement identity gives the exact determinant at the cost of the
gaps rather than the data.} Order the full grid so that the $m = |O|$
observed samples come first and the $g = |G| = N - m$ gapped samples second,
and partition the covariance conformally,
\begin{equation}
    \bSigma = \begin{pmatrix} \bSigma_{OO} & \bSigma_{OG} \\
                              \bSigma_{GO} & \bSigma_{GG} \end{pmatrix}\,.
    \label{eq:block_partition}
\end{equation}
As shown in Appendix~\ref{app:schur_det}, the determinant over the observed
block obeys the exact identity
\begin{empheq}[box = \fbox]{equation}
    \log\det\bSigma_{OO} = \log\det\bSigma
    + \log\det\big[\bSigma^{-1}\big]_{GG}\,.
    \label{eq:complement_identity}
\end{empheq}

where $\bSigma_{OO}$ is the time-domain covariance matrix over the observed samples and $\big[\bSigma^{-1}\big]_{GG}$ is the gap block of the \emph{inverse} covariance. The determinant over the $m$ samples we kept has been traded for one over
the $g$ samples we lost. Both terms on the right are cheap because $\bSigma$
is circulant. For the first, the DFT matrix, $\bP$, defined in Eq.~(\ref{eq:def_P_jk_matrix}), is unitary,
$\bP\bP^{\dagger} = \bP^{\dagger}\bP = \mathbb{I}$, so conjugation by $\bP$
leaves the determinant unchanged and it may be evaluated in the frequency
domain, where $\tilde{\bSigma}$ is diagonal
[Eq.~\eqref{eq:diagonal_cov_circulant}]. With the normalisation of
Eq.~\eqref{eq:defFDcov},
\begin{align}
    \log\det\bSigma &= \log\det\tilde{\bSigma} - N\log\!\big(N\Delta t^{2}\big) \nonumber \\
    & = \sum_{k}\log\frac{S_{n}[k]}{2\,\Delta f\,N\Delta t^{2}} \nonumber \\
    &= \sum_{k}\log\frac{S_{n}[k]}{2\Delta t}\,,
    \label{eq:logdet_circulant}
\end{align}
where the second equality distributes the constant across the $N$ terms and
the third uses $\Delta f = 1/(N\Delta t)$, recovering the circulant
eigenvalues $\lambda_{k} = S_{n}[k]/(2\Delta t)$ of
Sec.~\ref{subsec:td_cg}. This is an $\mathcal{O}(N)$ sum with no
factorisation at all. For the second, the
inverse of a circulant matrix is again circulant, with autocovariance
$\tilde{\gamma} = \mathrm{ifft}(2\Delta t/S_{n})$, so the entries of
$\big[\bSigma^{-1}\big]_{GG}$ are read directly off $\tilde{\gamma}$ at the
gapped lags. Building that $g \times g$ block costs $\mathcal{O}(g^{2})$ and
its Cholesky factorisation $\mathcal{O}(g^{3})$.

Two properties make Eq.~\eqref{eq:complement_identity} worth having. It is
\emph{exact} and deterministic, so it may be used inside an acceptance ratio
without further thought. And it assumes nothing about how $\bSigma$ is
parametrised, so unlike the closed form above it is valid for more general noise models, for example when there are more than two noise
components, a free spectral index, or a spline model. Its cost is governed by
the number of \emph{missing} samples rather than by the length of the
data, which is the opposite of the usual situation. How far that carries
depends on the duty cycle. A realistic four-year LISA masking simulation
gives a median duty cycle of $84.3\%$, of which the scheduled interruptions
account for only $3.9\%$ of the lost data and the unplanned ones for the
remainder~\cite{burke_tdi_gaps}. Keeping the cost of a $g \times g$ factorisation below
one second per likelihood evaluation places a limit of $N \sim 2\times10^{5}$ if
only the planned gaps are gated, and $N \sim 5\times10^{4}$ at the full
$15.7\%$ loss. In terms of data length these are segments of days to
weeks rather than years, which is precisely the regime in which the stream
must be partitioned anyway for the noise to be treated as stationary. The
identity is therefore the natural tool for the determinant of an individual
block, with the block determinants summing over the partition. Beyond that 
stochastic estimation (described below) becomes more efficient, visible as the cross-over in
Fig.~\ref{fig:det_scaling}.

\emph{At scale the determinant may be estimated stochastically and is matrix free.} Where even
a single $\mathcal{O}(m^{3})$ factorisation is unavailable and the gaps are
too numerous for Eq.~\eqref{eq:complement_identity}, one may partition the
stream into locally-stationary shorter-duration blocks, or estimate the
determinant matrix-free. Consider the identity $\log\det\bSigma_{OO} =
\text{Tr}\log\bSigma_{OO}$. The trace may be written as an expectation over
random probes, $\text{Tr}[\boldsymbol{A}] =
\mathbb{E}[\boldsymbol{z}^{T}\boldsymbol{A}\boldsymbol{z}]$ for any probe
vector with $\mathbb{E}[\boldsymbol{z}\boldsymbol{z}^{T}] = \mathbb{I}$, so
that
\begin{align}
\log \det \bSigma_{OO} = \mathbb{E}[\boldsymbol{z}^T \log\bSigma_{OO}\boldsymbol{z}]\,.
\end{align}
The obstacle is that we have products with $\bSigma_{OO}$, not with its
logarithm. Lanczos quadrature supplies it: $k$ matrix--vector products
build a $k \times k$ tridiagonal surrogate whose eigenvalues $\theta_i$
stand in for the spectrum of $\bSigma_{OO}$, so that
\begin{equation}
    \boldsymbol{z}^{T}\log(\bSigma_{OO})\,\boldsymbol{z}
    \simeq \|\boldsymbol{z}\|^{2}\sum_{i=1}^{k}
    \big[(\boldsymbol{u}_i)_1\big]^{2}\,\log\theta_i\,,
    \label{eq:lanczos_quadrature}
\end{equation}
with $(\boldsymbol{u}_i)_1$ the first components of its eigenvectors. The
logarithm is thus taken of $k$ scalars, never of the operator; a few tens
of steps and of probes suffice here, and
Refs.~\cite{hutchinson1989stochastic,ubaru2017fast} give the error
analysis.
The same probing supplies the noise-sector Fisher
matrix in the time domain,
\begin{align}
\bGamma_{\alpha\beta} &=
\tfrac{1}{2}\text{Tr}[\bSigma_{OO}^{-1}\partial_{\alpha}
\bSigma_{OO}\,\bSigma_{OO}^{-1}\partial_{\beta}\bSigma_{OO}]\,, \\
& = \frac{1}{2}\mathbb{E}[\boldsymbol{z}^T\bSigma_{OO}^{-1}\partial_{\alpha}
\bSigma_{OO}\,\bSigma_{OO}^{-1}\partial_{\beta}\bSigma_{OO}\boldsymbol{z}]\,,
\end{align}
at the cost of a handful of CG solves per probe, with accuracy improving as
$1/\sqrt{n_{\rm probe}}$. We emphasise that none of this stochasticity
touches the sampled likelihood, which requires only the deterministic CG
solve.

\subsection{Computational cost}
We will now provide some quantitative numbers for the quadratic forms computed for Scenario C. To remind the reader, here there are $N = 2880$ samples with $m = 2300$ observed (not gapped). Solving Eq.~\eqref{eq:setup_CG_method} to a
relative residual of $10^{-10}$ takes $583$ iterations unpreconditioned and
$343$ with a Whittle-based preconditioner (see Eq.~\eqref{eq:preconditioner}), falling to $277$
when the previously accepted solution seeds the next iteration. Over a full
sampling run the mean is $342$ and the maximum $420$. Building the preconditioner away
from the truth results in a marginal change -- across the prior box the cost of a solve
varies by less than a factor of two. Evaluated this way the log-likelihood
reproduces the efficient diagonalised covariance defined by Eq.\eqref{eq:trick_quadform} to a relative error of $10^{-12}$. At this size the closed form is the cheaper option per evaluation, costing  $4.0$\,ms per evaluation against $208$\,ms for two
conjugate-gradient solves. 

The left most panel of  Figure~\ref{fig:cg_scaling}  gives the single-threaded wall-time comparing the CG iterative solver (blue) and dense solves (red) when computing a single quadratic form $\bn_{O}^{T}\bSigma_{OO}^{-1}\bn_{O}$ as the data length grows. The dense solve accounts for forming $\bSigma_{OO}$ from the
autocovariance and solving using a via Cholesky decomposition. 
We compare with the dense solve as the noise models that will most liklely be employed in future analyses will require more flexibility than the two parameter model used here, for example flexible spline based models. 

The costs of the two approaches cross between $N = 2^{11}$ and
$2^{12}$. By $N = 2^{14}$ the dense route takes $26$\,s and requires 5.5GB of memory, compared to the matrix-free CG method that requires $1.9$\,s on a single thread. For $N = 2^{18} \approx 2\times10^{5}$ samples, the iterative solution requires a minute whereas we estimate the dense solve would take 15 hours. Not only that, but the dense solver would require approximately 350GB of memory. The measured computational costs scale as $N^{2.8}$ for the dense (Cholesky) solve  and $N^{1.25}$ for the CG iterative solve. Each iteration
costs four Fourier transforms of length $N$, two for the matrix--vector product and
two for the preconditioner. The iteration count does grow, but slowly,
empirically as $N^{0.21}$ as demonstrated in the right panel of Fig.~\ref{fig:cg_scaling}.

Throughout, the two approaches agree on the value of the quadratic form to between $10^{-14}$ and $10^{-12}$ relative error, two orders of magnitude inside the requested tolerance.

We remark that neither approach has been optimised. At $N = 2^{18}$ the four transforms account for
over $90\%$ of an iteration, so the cost of the method is the cost of an FFT.
The transforms timed in Fig.~\ref{fig:cg_scaling} are single-threaded. A
threaded transform recovers a substantial part of the gap on the same
processor, and it is against that threaded baseline, rather than against the
curve plotted here, that any accelerator should be compared.

The case for a GPU does not rest on the bandwidth ratio alone. A
double-precision transform of this length is limited by memory bandwidth
rather than by arithmetic, so a device with several times the bandwidth helps
directly. Our likelihood needs one solve per
TDI channel, and an ensemble sampler proposes many walkers at once, so the
transforms arrive in batches of tens to hundreds of independent problems of
identical length. This is precisely the regime in which a batched FFT reaches
its peak throughput. This would be easily parallelisable on a GPU. The iteration is matrix-free, so the $350$\,GB matrix is never required in memory.

We conclude this section with two practical remarks. The residual tolerance used here lies below single-precision resolution, so the transforms must be carried in double.
That argues for a scientific accelerator rather than a consumer one, although
a mixed-precision iteration, with the solve carried in single and refined in
double, would relax the requirement. Second, the iteration count is the
quantity worth attacking. It varies by a factor of a few with the treatment
of the transfer nulls in the preconditioner, and a scheme that deflects the
near-null modes would buy a speed-up comparable to the hardware, with the two
gains composing. For a low-latency analysis on segments of a few weeks the
combination is already comfortable, and a year of data is best handled by
partitioning it into such segments, which the slow drift of the instrumental
noise requires in any case.

We close with the conceptual point of this subsection. The estimator is the
exact likelihood \eqref{eq:td_restricted_likelihood} throughout: the
numerical machinery above introduces only \emph{numerical} error, controlled
by a tolerance and an iteration count chosen by the analyst. This is in
sharp contrast to the windowed hierarchy of
Sec.~\ref{eq:hierarchy_of_approximations}, whose
\emph{modelling} error is fixed by the window and the spectrum and cannot be
dialled away by adjusting tolerances. 

\section{Verification}\label{sec:verification}
Before using the analytic diagnostics of Sec.~\ref{subsec:metrics_scatter_to_width} as our main tool, we check them against sampled posteriors. Two descriptions are under test: the linearised statistics of Sec.~\ref{subsec:metrics_scatter_to_width}, and the pseudo-true point with the Godambe--White covariance of Sec.~\ref{sec:Godambe_White_formalism}. Our interest is in whether they reproduce the posteriors that an MCMC actually returns, and in which of the two should be quoted for a given covariance model.

\begin{figure}[t]
  \includegraphics[width=\columnwidth]{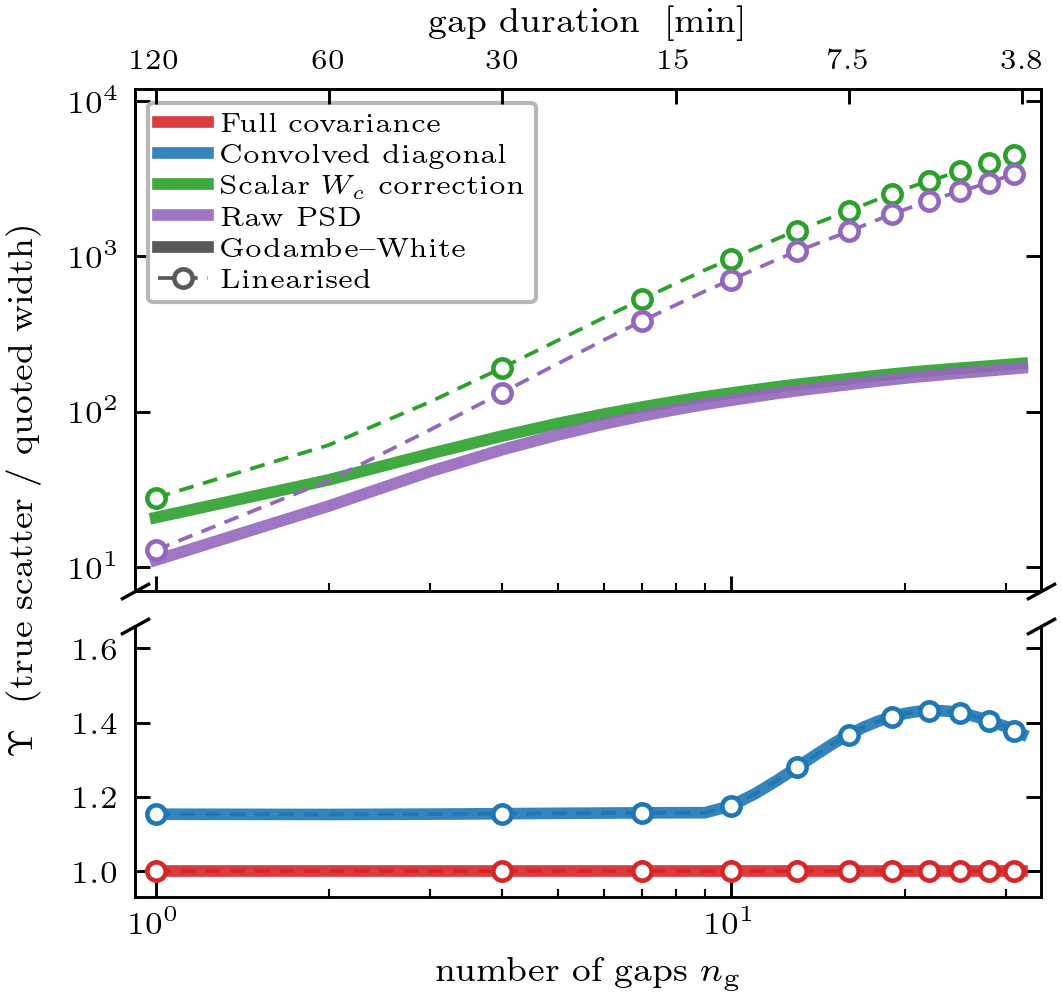}
  \caption{The scatter-to-width diagnostic $\Upsilon$ computed both ways,
    for the four covariance models, as a fixed $2\,$h of missing time is
    chopped into $n_{\rm g}$ gaps of decreasing duration (upper axis).
    Thick lines are the Godambe/White sandwich, thin dashed lines with open circles the linearised formalism.  For the full
    covariance and the convolved diagonal the pseudo-true point does not
    move from the truth and the two formalisms coincide identically ---
    the circles sit centred on the thick lines throughout the lower
    panel, which is on a linear scale below the axis break.  For the raw
    PSD and the scalar $W_c$ correction the pseudo-true point does move,
    and the linearised formalism overstates $\Upsilon$. \srclink{fig_upsilon_formalisms.py}}
  \label{fig:upsilon_formalisms}
\end{figure}

We perform three checks. We first compare both predictions against individual sampled posteriors, drawn as dashed curves on the corner plots of Sec.~\ref{sec:results}. We then show, in Fig.~\ref{fig:upsilon_formalisms}, that the criterion for choosing between them is a property of the covariance model rather than of the gap pattern. Finally, we repeat the exercise over $300$ noise realizations and compare the ensemble statistics with both predictions in Table~\ref{tab:ensemble}. We stress that this section validates the machinery only; the scenario-by-scenario discussion of which models are usable is left to Sec.~\ref{sec:results}.

\emph{Setup.} Scenarios A and B are run under all four models: full windowed
covariance, exact convolved diagonal, scalar $W_c$ correction and PSD with no
correction. For scenario C we run parameter estimation only under the full
covariance, the convolved diagonal and the exact conjugate-gradient
time-domain solver of Sec.~\ref{subsec:td_exact}; the PSD and $W_c$
correction produced chains that failed to converge in every parameter. For
each scenario and each covariance model of the hierarchy of
Sec.~\ref{eq:hierarchy_of_approximations} we sample the joint $13$-parameter
posterior on a common data realization with $48$ walkers, $1800$ burn-in
$+\,4200$ steps, with
the walkers started in a small ball around the truth. The priors are uniform
boxes centred on the truth, identical for all models of a given scenario.
Each box is wide enough to contain every model's forecast and its
mis-modelling displacement, but capped so that it stays inside the
locally-linear regime of the Hessian-based description: at most $\pm 900$\,s
in $t_c$ and $\pm 1.8$ in each $\log_{10}$-power noise parameter.
The boxes are further clipped to the physically admissible range of each
parameter --- bounded spins, $q \ge 1$, $\iota \in [0, \pi]$, and the
ecliptic latitude $\beta \in [-\pi/2, \pi/2]$, each implemented with a
small standoff from the exact endpoint --- though for the injection
analysed here no such bound is active. The ecliptic longitude  $\lambda$,
the reference phase $\varphi$ and the polarisation angle $\psi$ are
periodic and need none.
 
\begin{figure*}[t!]
\centering
\includegraphics[width=0.92\textwidth]{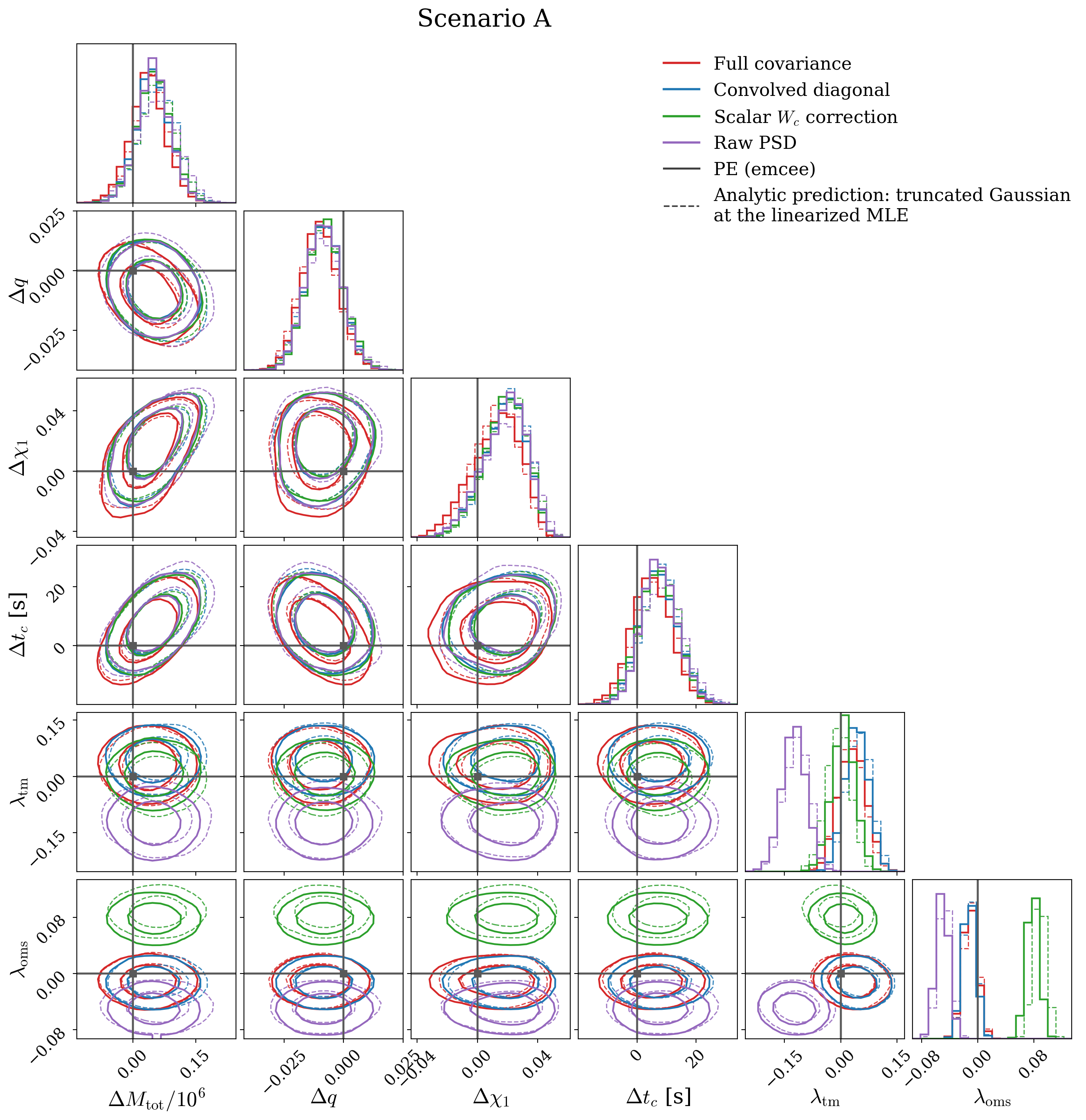}
\caption{Scenario A: joint posterior for a representative subset of the $13$
sampled parameters ($M_{\rm tot}, q, \chi_1, t_c$ and the two noise
amplitudes) under the four covariance models of the hierarchy of
Sec.~\ref{eq:hierarchy_of_approximations} (colours). Solid: MCMC; dashed: the
truncated Gaussian implied by the model Fisher matrix centred on the
linearized MLEs
[Eqs.~\eqref{eq:signal_param_fluctuations_noise_mismodelling} and
\eqref{eq:noise_param_fluctuations_noise_mismodelling}]. For the full covariance and the convolved diagonal the pseudo-true
displacement vanishes, so the linearised and nonlinear predictions coincide.
Note that this figure tests the predicted centre and width on a single
realization; the scatter-to-width ratios of Table~\ref{tab:upsxi} are
ensemble statements, and are verified separately in
Table~\ref{tab:ensemble}. Black lines mark the truth. The remaining signal
parameters are in the companion repository. Produced by \srclink{make_figures.py}.}
\label{fig:corner_key_A}
\end{figure*}

\emph{Prediction.} The analytic prediction is shown as a dashed curve on
every corner plot of Sec.~\ref{sec:results}
(Figs.~\ref{fig:corner_key_A}, \ref{fig:corner_key_B} and
\ref{fig:corner_key_C}). In the signal block it is always the truncated
Gaussian implied by the model Fisher matrix, centred on the linearised MLE
[Eq.~\eqref{eq:signal_param_fluctuations_noise_mismodelling}]. In the noise
block the choice between the two descriptions is set by the pseudo-true
displacement $\blambda^{*}$, not by the scenario. When $\blambda^{*} =
\boldsymbol{0}$ the expansion point is the truth itself and the two coincide
exactly; this is the case for the full covariance and the convolved diagonal
in all three scenarios, and Table~\ref{tab:upsxi} shows $\Upsilon^{\rm lin} =
\Upsilon^{\rm sw}$ for those rows. The descriptions separate as
$|\blambda^{*}|$ grows: for the non-convolved models the ratio
$\Upsilon^{\rm lin}/\Upsilon^{\rm sw}$ is $\simeq 1.7$ in scenario A, where
$\lambda^{*}_{\rm oms} = 0.30$, and $\simeq 8.5$ in scenario B, where
$\lambda^{*}_{\rm oms} = 1.30$. We therefore draw the linearised prediction
when $\blambda^{*}$ vanishes, and the exact profile MLE with its
Godambe--White curvature
[Eqs.~\eqref{eq:pseudo_true}--\eqref{eq:sandwich}] otherwise; the legend of
each figure states which is shown. The rest of this section establishes that
this rule is the right one, and that both descriptions do what they claim.
 
\emph{Beyond the three scenarios.}\label{subsec:formalism_sweep} That the
choice is determined by $\blambda^{*}$ and not otherwise by the gap pattern
is a claim about the covariance models, which cannot be fully assessed with
only three scenarios. Figure~\ref{fig:upsilon_formalisms} tests it on a
one-parameter family that contains them: a fixed amount of missing time
chopped into $n_{\rm g}$ equal gaps with proportional tapers, from a single
long interruption to the comb of scenario C. Scenario A is the $n_{\rm g} =
2$ member of this family exactly, and B and C sit beside members of it, so
the two ratios just quoted are points on a continuous curve rather than three
isolated measurements. Along that curve $\blambda^{*}$ stays at the truth for
the full covariance and the convolved diagonal at every pattern, while the
non-convolved models never recover it: the segment-edge taper alone is enough
to displace their pseudo-true point, whatever the gaps do.
 
\emph{Ensemble.}\label{subsec:ensemble} We repeat the exercise on $300$
independent noise realizations per scenario, computing for each realization
and each model both the linearised statistic and the exact profile noise MLE.
The results are summarised in Table~\ref{tab:ensemble}. First, the linearised
statistic follows its analytic distribution
[Eqs.~\eqref{eq:biased_noise_params_mismodelling} and
\eqref{eq:noise_mismodelling_params_covariance}] at Monte-Carlo precision
everywhere, including where the displacement it predicts is far too large to
be meaningful. Second, the exact profile MLEs behave as the nonlinear
Godambe--White formalism requires when a mis-specified model is fit to many
realizations: they cluster around the pseudo-true point $\blambda^{*}$
[Eq.~\eqref{eq:pseudo_true}] rather than the truth, with a
realization-to-realization spread $\sigma_{\rm sw}$
[Eq.~\eqref{eq:sandwich}] rather than the error bar the model itself quotes.
Table~\ref{tab:ensemble} contains eleven (scenario, model) rows and hence
$22$ (model, parameter) combinations, of which $21$ are comparable once the
flagged entry is set aside. In every one of those $21$ the ensemble mean
lands within $0.5\,\sigma_{\rm sw}$ of $\blambda^{*}$, and the measured
scatter $s$ matches $\sigma_{\rm sw}$ to $15\%$ or better in $19$ cases. The
two exceptions are the measurements of the TM noise in the two non-convolved
models in scenario B, where $s$ exceeds $\sigma_{\rm sw}$ by $50\%$: there the
OMS amplitude has moved to $\lambda^{*}_{\rm oms} \simeq 1.2$--$1.3$ to
absorb the leakage, and around a point displaced by so much the profile
likelihood is no longer quadratic in the TM direction, so a width read off
its curvature can only be indicative.
 
\emph{Outcome.} Both analytic descriptions are therefore verified, and the
rule stated above for choosing between them is justified. The exact profile MLE is reproduced by the
pseudo-true point of Eq.~\eqref{eq:pseudo_true} together with the
Godambe--White covariance of Eq.~\eqref{eq:sandwich}, in mean and in spread,
across every model and every scenario tested. The cheaper linearised
statistics reproduce it as well wherever $\blambda^{*}$ has not moved far,
and quantitatively so: expanding Eq.~\eqref{eq:sandwich} about
$\blambda^{\rm true}$ gives a relative error of order $\tfrac{1}{2}\ln
10\,|\Delta\blambda^{*}|$. In the signal block that condition is met in every
case considered, and the linearised Fisher formalism suffices throughout; in
the noise block it is not, and severe mis-modelling requires the non-local
Godambe--White formalism of Sec.~\ref{sec:Godambe_White_formalism}. Having
verified the regimes in which our analytic estimates are consistent with
Bayesian inference, we now investigate each scenario in turn, with the aim of
establishing which model covariance matrices, if any, are suitable for
precise and accurate inference of signal and noise parameters in the presence
of gaps.

\begingroup
\squeezetable
\begin{table}[t]
\centering
\caption{Ensemble verification of the noise-sector scatter ($300$
noise realizations per scenario; each entry lists $\lambda_{\rm tm}$,
$\lambda_{\rm oms}$.
$\langle\hat{\blambda}\rangle \pm s$ are the ensemble mean and
standard deviation of the \emph{exact} profile MLEs, to be compared
with the pseudo-true values $\blambda^{*} \pm \sigma_{\rm sw}$ of
Eqs.~\eqref{eq:pseudo_true}--\eqref{eq:sandwich}. The linearized
statistic (not shown) matches Eqs.~\eqref{eq:biased_noise_params_mismodelling}
and \eqref{eq:noise_mismodelling_params_covariance} at Monte-Carlo precision
in every entry. Models as in the hierarchy of
Sec.~\ref{eq:hierarchy_of_approximations}: full windowed covariance,
convolved diagonal (diag), scalar $W_c$ correction, raw PSD; scenario C
additionally lists the exact time-domain solver of
Sec.~\ref{subsec:td_exact} (TD), whose model is correct by construction, so
that the ensemble measures $\Upsilon = \Xi = 1$ rather than assuming it.
$\dagger$: in scenario C the convolved diagonal carries almost no
information on $\lambda_{\rm tm}$, its profile likelihood is flat in that
direction and the exact MLE has no finite mean or variance. Produced by
\srclink{ensemble_noise.py}.}
\label{tab:ensemble}
\begin{ruledtabular}
\begin{tabular}{llcccc}
& model & $\langle\hat{\blambda}\rangle$ & $s$ & $\blambda^{*}$ & $\sigma_{\rm sw}$ \\
\hline
A & full  & $-0.00,\,+0.00$ & $0.038,\,0.010$ & $0,\,0$         & $0.036,\,0.010$ \\
A & diag  & $-0.00,\,+0.00$ & $0.040,\,0.012$ & $0,\,0$         & $0.038,\,0.010$ \\
A & $W_c$ & $-0.14,\,+0.30$ & $0.086,\,0.128$ & $-0.14,\,+0.30$ & $0.082,\,0.148$ \\
A & psd   & $-0.28,\,+0.16$ & $0.086,\,0.128$ & $-0.28,\,+0.18$ & $0.082,\,0.148$ \\
\hline
B & full  & $-0.00,\,-0.00$ & $0.038,\,0.010$ & $0,\,0$         & $0.038,\,0.010$ \\
B & diag  & $-0.00,\,+0.00$ & $0.068,\,0.012$ & $0,\,0$         & $0.062,\,0.010$ \\
B & $W_c$ & $+0.00,\,+1.20$ & $0.210,\,0.294$ & $+0.04,\,+1.30$ & $0.140,\,0.274$ \\
B & psd   & $-0.12,\,+1.10$ & $0.210,\,0.294$ & $-0.08,\,+1.18$ & $0.140,\,0.274$ \\
\hline
C & full  & $-0.01,\,+0.00$   & $0.058,\,0.011$   & $0,\,0$ & $0.060,\,0.009$ \\
C & diag  & $\dagger,\,-0.01$ & $\dagger,\,0.021$ & $0,\,0$ & $1.008,\,0.025$ \\
C & TD    & $-0.01,\,+0.00$   & $0.050,\,0.010$   & $0,\,0$ & $0.050,\,0.010$ \\
\end{tabular}
\end{ruledtabular}
\end{table}
\endgroup

\section{Results}\label{sec:results}
Gaps affect parameter estimation through two channels: they remove
information, and they invalidate the stationary covariance model.
Appendix~\ref{app:abc_corner} (Fig.~\ref{fig:corner_full_ABC}) isolates the
first effect, comparing the exactly calibrated posteriors of the \emph{correct}
(full covariance) analysis in the three scenarios: their differences arise from different amounts of information being lost under each gap scenario. 
This section will concentrate on the second effect: how far
each approximate model falls from those reference posteriors.
Table~\ref{tab:upsxi} collects the diagnostics and Fig.~\ref{fig:upsxi}
resolves them per parameter.

\subsection{Scenario A: isolated gaps through inspiral and merger}
\label{subsec:results_A}
With long, well-tapered gaps the windowed covariance is benign. The leakage is
confined to a smooth near-diagonal halo of width $\sim\!1/(1\,{\rm h})$
(Fig.~\ref{fig:cov_cmap}, top), and every level of the hierarchy performs
close to its stationary expectation. Focusing on the top panel of Fig.~\ref{fig:upsxi} the full model is exactly calibrated
($\Upsilon = \Xi = 1$) and the convolved diagonal has a mild
scatter excess from the neglected bin--bin correlations
($\Upsilon = 1.14,\,1.15$; Table~\ref{tab:upsxi}). The convolved diagonal approximation would therefore be suitable for signal and noise parameter inference.

The two non-convolved models are a different matter. Their pseudo-true
offsets are small in absolute terms ($\blambda^{*}$, Table~\ref{tab:upsxi})
but, measured against the width each model itself quotes, they are
$\Upsilon^{\rm sw} = 5$--$36$, so the reported credible intervals exclude the
truth in essentially every realization. We therefore regard both tiers as
unusable for noise estimation, in this and in the remaining scenarios.

We note that whether or not the linearised description is accurate is a statement about the distance of $\blambda^{}$ from the true parameters and hence whether Eq.~\eqref{eq:noise_mismodelling_params_covariance} can be used; whether the model is usable is a statement about whether the noise-induced scatter in the MLE about $\blambda^{}$ is comparable to that model's own quoted uncertainties. In scenario A the linearised description is adequate, but the two non-convolved models are nonetheless unusable: their average displacements are small in absolute terms but the MLEs fluctuate by tens of posterior widths.

The signal sector over all four models, by contrast has
scatter-to-width and width-to-width ratios within $\sim\!20\%$ of unity. The average quantities  $\bar{\Upsilon}_{s}$ and $\bar{\Xi}_{s}$ are given in  Table~\ref{tab:upsxi}, and per-parameter results can be found in Fig.~\ref{fig:upsxi}). For isolated, well-tapered
gaps the choice of noise approximation is, for the signal parameters, almost
irrelevant. This is true even though the second gap clips the merger. This is
largely consistent with the findings of Ref.~\cite{Burke:2025bun}. It is, of
course, only applicable if a taper has been applied to control the spectral
leakage, which renders the Whittle likelihood reasonable for inference over
the signal parameters alone. The signal parameters would have larger scatter-to-width ratios if the window was rectangular due to the excess spectral leakage. 

\subsection{Scenario B: repeated short gaps}
\label{subsec:results_B}
\begin{figure*}[t!]
\centering
\includegraphics[width=0.92\textwidth]{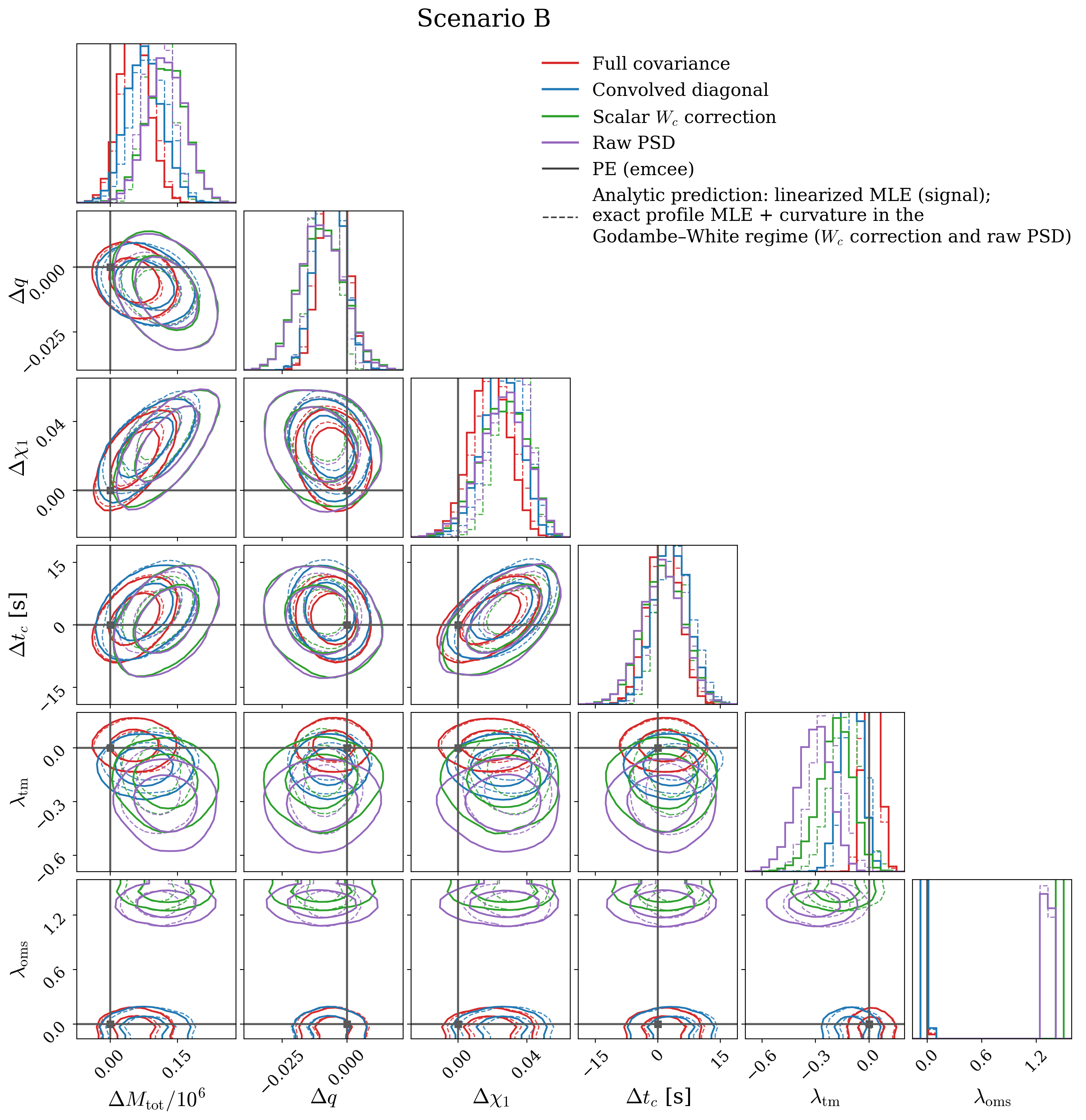}
\caption{Scenario B: joint posterior for the key parameters under the four
covariance models, in the same format as Fig.~\ref{fig:corner_key_A} except
for the noise-sector overlay, as stated in the legend. The signal sector
stays centred on the truth for every model, but the two non-convolved models
($W_c$ correction, green; raw PSD, purple) are displaced in
$\lambda_{\rm oms}$ by more than an order of magnitude in noise power --- over
a hundred of their own posterior widths, matching the pseudo-true values of
Table~\ref{tab:upsxi}. For these two models the dashed noise-sector
prediction is the exact profile MLE with its Godambe--White curvature
[Eqs.~\eqref{eq:pseudo_true}--\eqref{eq:sandwich}]; the linearized prediction
would overshoot by an order of magnitude. The full and convolved-diagonal
posteriors remain centred on the truth in every direction. Produced by \srclink{make_figures.py}.}
\label{fig:corner_key_B}
\end{figure*}

The hourly comb of sharp gaps behaves very differently. The $3$-minute tapers
spread leakage over a $\sim\!5\,$mHz-wide floor, and around the TDI2 zero
$f_{\varnothing} \simeq 30\,$mHz. The TDI2 zero crossing is where the stationary model expects
essentially no power. On top of this, the exact windowed variance exceeds $W_c\,S_n$ by up to
a factor $\sim\!30$ (see the middle rows of Figs.~\ref{fig:overview} and
\ref{fig:cov_cmap}). A model without the convolution has only one way to
account for that excess, which is to inflate the high-frequency (OMS) amplitude. The two non-convolved
models displace $\lambda_{\rm oms}$ by more than an order of magnitude in
noise power, over a hundred of their own posterior widths --- and the sampled
posteriors land on these displaced values, not near the truth
(Fig.~\ref{fig:corner_key_B}; all values in Table~\ref{tab:upsxi}).

The \emph{leading-order} bias formula
[Eq.~\eqref{eq:biased_noise_params_mismodelling}] correctly diagnoses the
failure but grossly overshoots its size.  The parameter displacements this large lie far
outside the linearization's domain of validity. Only the nonlinear
completions [Eqs.~\eqref{eq:pseudo_true}--\eqref{eq:sandwich}] describe them
quantitatively. In Table~\ref{tab:upsxi}, $\Upsilon^{\rm lin}$ overstates
$\Upsilon^{\rm sw}$ by an order of magnitude (per-parameter detail in
Fig.~\ref{fig:upsxi}). Second, the signal sector remains \emph{unbiased} even
here, as guaranteed by
Eq.~\eqref{eq:signal_param_fluctuations_noise_mismodelling}, but its error
bars are no longer reliable: under the non-convolved models the true scatter
exceeds the quoted widths by tens of per cent ($\bar{\Upsilon}_{s}$,
Table~\ref{tab:upsxi}), and the $M_{\rm tot}$ estimate
sits several posterior widths from the truth
(Fig.~\ref{fig:corner_key_B}). The convolved diagonal, by contrast, is
unbiased in \emph{both} sectors, its scatter is consistent with its quoted
widths to within $\sim\!20\%$, and the posterior width is 
at most $\sim\!30\%$ broader than what can be reached with the true likelihood (diag rows of Table~\ref{tab:upsxi}).  

\subsection{Scenario C: drastic gaps and the exact time-domain solver}
\label{subsec:results_C}
\begin{figure*}[t!]
\centering
\includegraphics[width=0.92\textwidth]{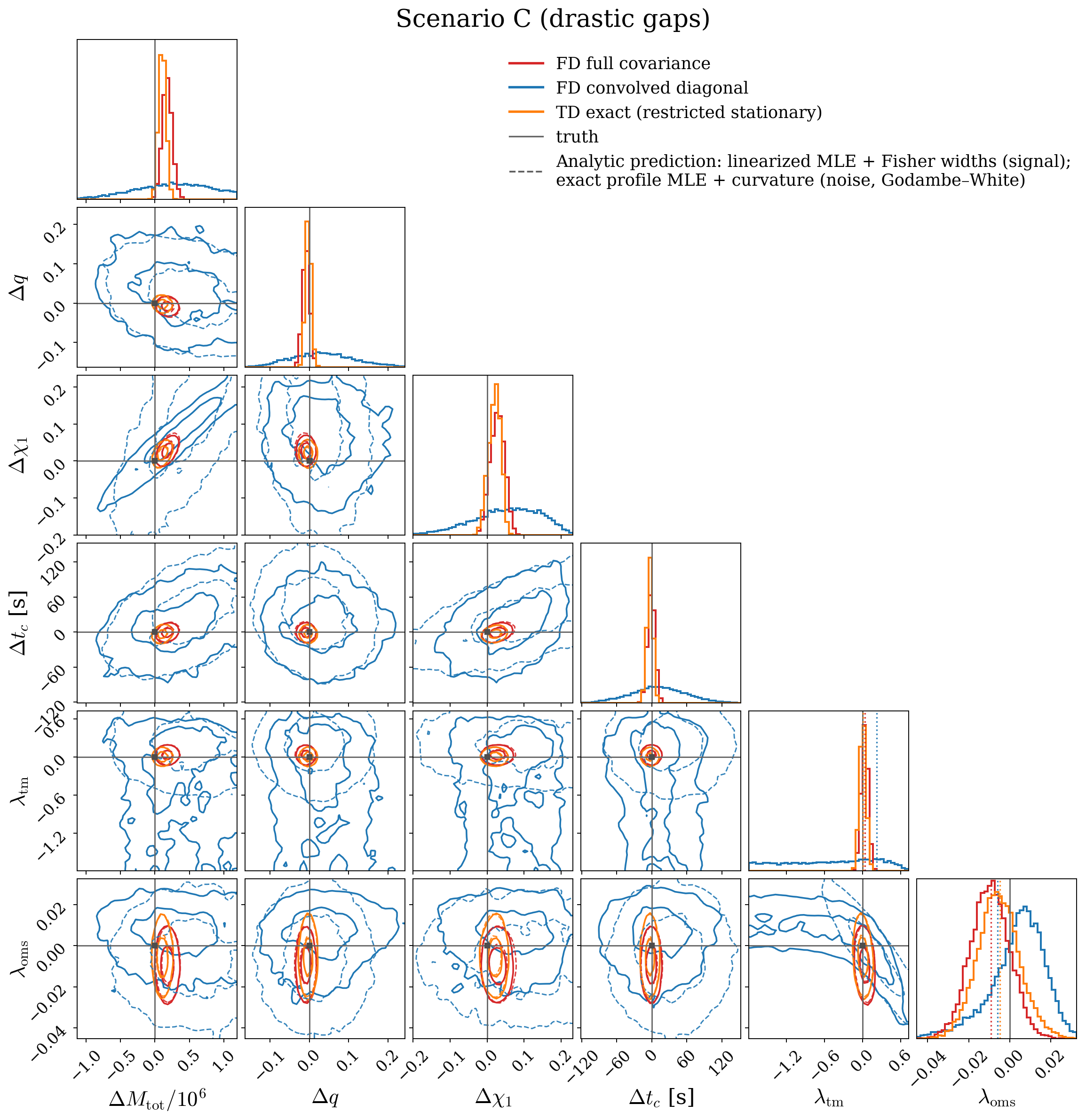}
\caption{Scenario C (drastic gaps): joint posterior for the key parameters
under the frequency-domain full covariance (red), the frequency-domain
convolved diagonal (blue), and the exact time-domain restricted-stationary
likelihood of Sec.~\ref{subsec:td_exact} (orange). All three are centred on
the truth; the diagonal model is calibrated but an order of magnitude wider
in every signal direction ($\Xi_{s} = 8.6$--$10.2$), while the time-domain
solver --- exact by construction, $\Upsilon = \Xi = 1$ --- is marginally
tighter than the band-restricted frequency-domain full covariance,
whose band restriction marginalises over the out-of-band bins and so
gives up part of the information. Dashed curves show the full analytic
prediction of each model in \emph{both} sectors, as stated in the legend: in
the signal block, the linearized MLE
[Eq.~\eqref{eq:signal_param_fluctuations_noise_mismodelling}] with the
model's Fisher widths; in the noise block, the exact (nonlinear) profile MLE
--- marked by dotted vertical lines --- with its curvature, which for the two
exact tiers coincides with their Fisher forecast. Produced by \srclink{plot_scenC.py}.}
\label{fig:corner_key_C}
\end{figure*}

\begin{figure*}[t!]
\centering
\includegraphics[width=\textwidth]{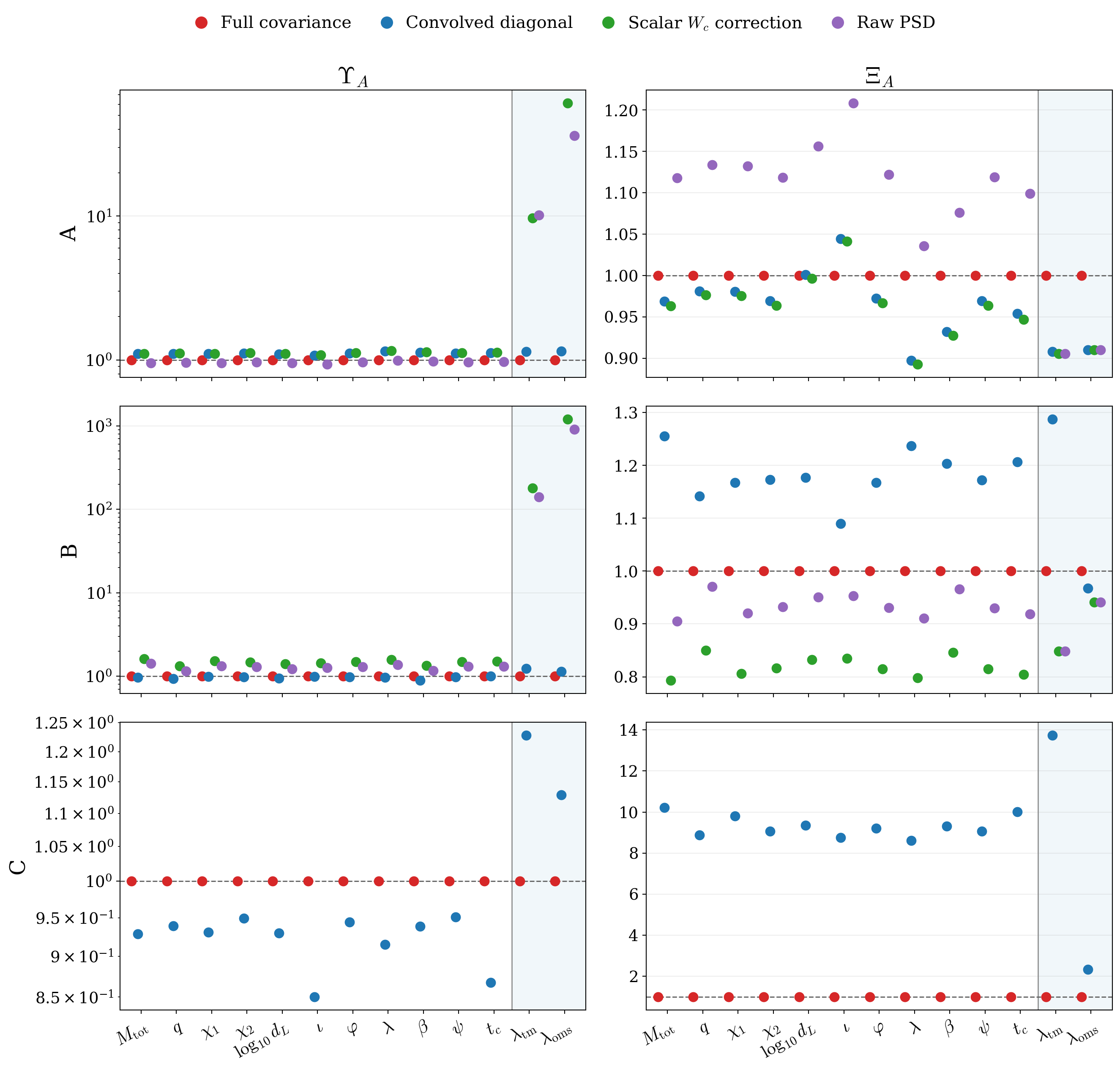}
\caption{Scatter-to-width ($\Upsilon_A$, left) and width-to-width ($\Xi_A$,
right) diagnostics [Eqs.~\eqref{eq:upsilon_matrix_noise} and
\eqref{eq:xi_matrix_noise}] for every sampled parameter (signal and noise
sectors), the covariance models (colours) and the three gap scenarios
(rows), at leading order. Scenario C carries only the full covariance and
the convolved diagonal: the two non-convolved tiers are already ruled out by
scenario B and are not run there, and the exact time-domain solver has
$\Upsilon = \Xi = 1$ by construction. For the non-convolved models in
scenario B the noise-sector $\Upsilon$ leaves the axis: the leading-order
values ($10^2$--$10^3$) overstate the nonlinear sandwich completion
($\simeq 130$--$140$, Table~\ref{tab:upsxi}), both far above any acceptable
calibration error. Scenario C shows the opposite failure. The convolved
diagonal stays calibrated in $\Upsilon$ there ($1.23,\,1.13$ on the noise
amplitudes, $0.85$--$0.95$ across the signal parameters), but its $\Xi$
reaches $13.7$ on $\lambda_{\rm tm}$ and $8.6$--$10.2$ on the signal
parameters. Produced by \srclink{make_figures.py}.}
\label{fig:upsxi}
\end{figure*}

In scenario C the window carries no design freedom, there is nothing to
taper, and the comb structure puts the approximation hierarchy under maximal
stress. The exact convolved diagonal remains \emph{accurate}: its pseudo-true
parameters vanish identically, its noise-sector scatter-to-width ratios are
$\Upsilon = (1.23,\,1.13)$, and its signal sector is calibrated to
$\Upsilon_{s} = 0.85$--$0.95$. But because the diagonal model must treat the
huge aliased variance in every bin as independent noise, while the exact
analyses exploit the strong bin--bin correlations of the comb to unmix it,
its quoted widths are far from what the data allow. The width-to-width ratio has an excess of
$\Xi_{s} = 8.6$--$10.2$ across the signal parameters and
$\Xi = (13.7,\,2.3)$ for the noise amplitudes
(Fig.~\ref{fig:corner_key_C}). In this regime the convolved diagonal is not enough,
and one of the two exact treatments is required: the full windowed
covariance, or the time-domain CG solver that treats the missing data as
marginalised observations [Eq.~\eqref{eq:td_restricted_likelihood}].

As Fig.~\ref{fig:corner_key_C} shows, the time-domain posterior coincides
with the frequency-domain full covariance on every parameter and is exactly
calibrated --- $\Upsilon = \Xi = 1$ by construction, with the exact profile
MLE of the noise parameters landing at
$(\lambda_{\rm tm}, \lambda_{\rm oms}) = (+0.003, -0.005)$--- at a per-evaluation cost equal to that of the
\emph{approximate} diagonal tiers.

The time-domain solver quotes tighter signal widths than the
frequency-domain full covariance. This needs care, because $\Xi$ compares
each model against the best width attainable within \emph{its own} analysis,
and the two exact analyses do not share a reference. The frequency-domain
rows are referred to the band-restricted covariance, the time-domain row to
the full band. Both therefore report $\Xi = 1$, but using different models.

To place them on a common scale, the entries in parentheses in
Table~\ref{tab:upsxi} refer every ratio to the same denominator, the
time-domain Fisher matrix. On that scale the band-restricted full covariance
has $\bar{\Xi}_{s} = 1.27$ and the convolved diagonal $11.8$, in place of the
$1.00$ and $9.3$ each analysis quotes against itself. Across the $11$ signal
parameters the width ratio runs from $1.12$ to $1.33$.

Most of this $27\%$ is the band restriction. Restricting to a band
is a marginalisation rather than an approximation: the retained bins keep
their exact covariance, including the leakage from outside the band, but the
discarded bins are integrated out, and a marginal never carries more Fisher
information than the joint. In scenario C the comb scatters signal power well
beyond the band edge, so the discarded bins are not empty and the loss is
measurable.

In terms of wall-time, each time-domain likelihood evaluation costs about
twice one on the convolved diagonal ($\sim\!8$\,ms against $\sim\!4$\,ms),
and in both the waveform generation dominates, so at this gap density the
exact treatment is almost free. The conjugate-gradient route is only
mildly sensitive to the noise parameters at which the preconditioner is
built: across the prior box the iteration count varies by less than a factor
of two (Sec.~\ref{subsec:td_exact}), so no special handling is required as
the sampler explores $\blambda$.

\begin{table*}[t]
\centering
\caption{Noise-sector diagnostics for the three gap scenarios (each entry
lists $\lambda_{\rm tm}$, $\lambda_{\rm oms}$). The KL pseudo-true offsets
$\blambda^{*}$ [Eq.~\eqref{eq:pseudo_true}] are the population-level biases;
$\Upsilon$ [Eq.~\eqref{eq:upsilon_matrix_noise}] is quoted both at leading
order and from the nonlinear sandwich completion [Eq.~\eqref{eq:sandwich}];
$\Xi$ from Eq.~\eqref{eq:xi_matrix_noise}. The correct model (full) is
exactly calibrated by construction; the convolved diagonal has
$\blambda^{*} = \boldsymbol{0}$ identically because its variance reproduces
$\mathrm{diag}\,\bQ$ exactly. In scenario C only the full and
convolved-diagonal models are run (the non-convolved tiers are already ruled
out by scenario B), together with the exact time-domain solver of
Sec.~\ref{subsec:td_exact}, for which $\Upsilon = \Xi = 1$ by construction.
Note that $\Xi$ is defined against the best width attainable
within each model's \emph{own} analysis. In scenario C the frequency-domain
and time-domain rows therefore use different references, which is why both
report $\Xi = 1$. The parenthetical entries repeat every ratio against a
single common reference, the time-domain Fisher matrix; the difference
between the two is discussed in Sec.~\ref{subsec:results_C}.
The last two columns condense the signal sector into single numbers, in the
spirit of Ref.~\cite{Burke:2025bun}: the diagnostics averaged over the
$d = 11$ signal parameters, $\bar{\Upsilon}_{s} =
\frac{1}{d}\sum_{a}\Upsilon_{a}$ and $\bar{\Xi}_{s} =
\frac{1}{d}\sum_{a}\Xi_{a}$ (the signal MLE is linear in the data, so no
sandwich completion arises there). Produced by \srclink{dump_upsxi.py}.}
\label{tab:upsxi}
\begin{ruledtabular}
\begin{tabular}{llcccccc}
& model & $\blambda^{*}$ & $\Upsilon^{\rm lin}$ & $\Upsilon^{\rm sw}$ & $\Xi$ & $\bar{\Upsilon}_{s}$ & $\bar{\Xi}_{s}$ \\
\hline
A & full  & $0,\,0$         & $1.00,\,1.00$ & $1.00,\,1.00$ & $1.00,\,1.00$ & $1.00$ & $1.00$ \\
A & diag  & $0,\,0$         & $1.14,\,1.15$ & $1.14,\,1.15$ & $0.91,\,0.91$ & $1.11$ & $0.97$ \\
A & $W_c$ & $-0.14,\,+0.30$ & $9.7,\,61$    & $5.0,\,36$    & $0.91,\,0.91$ & $1.12$ & $0.96$ \\
A & psd   & $-0.28,\,+0.18$ & $10,\,36$     & $8.7,\,25$    & $0.91,\,0.91$ & $0.96$ & $1.12$ \\
\hline
B & full  & $0,\,0$         & $1.00,\,1.00$ & $1.00,\,1.00$ & $1.00,\,1.00$ & $1.00$ & $1.00$ \\
B & diag  & $0,\,0$         & $1.23,\,1.14$ & $1.23,\,1.14$ & $1.29,\,0.97$ & $0.96$ & $1.18$ \\
B & $W_c$ & $+0.04,\,+1.30$ & $179,\,1200$  & $4.4,\,141$   & $0.85,\,0.94$ & $1.47$ & $0.82$ \\
B & psd   & $-0.08,\,+1.18$ & $140,\,911$   & $5.0,\,129$   & $0.85,\,0.94$ & $1.28$ & $0.94$ \\
\hline
C & full  & $0,\,0$         & $1.00,\,1.00$ & $1.00,\,1.00$ & $1.00\,(1.19),\,1.00\,(0.94)$ & $1.00$ & $1.00\,(1.27)$ \\
C & diag  & $0,\,0$         & $1.23,\,1.13$ & $1.23,\,1.13$ & $13.7\,(16.4),\,2.3\,(2.2)$   & $0.92$ & $9.3\,(11.8)$  \\
C & TD    & $0,\,0$         & $1.00,\,1.00$ & $1.00,\,1.00$ & $1.00,\,1.00$                 & $1.00$ & $1.00$ \\
\end{tabular}
\end{ruledtabular}
\end{table*}

\section{Conclusions}\label{sec:conclusions}
\subsection{Summary}
In this work we have demonstrated and numerically verified both a Fisher-matrix-based formalism and a non-local Hessian based approximation (Godambe-White) that can cheaply evaluate the impact of noise mis-modelling on both noise and signal parameter estimation. We have extended the results of ~\cite{Burke:2025bun} to include noise-parameter inference in the simple case of a two-parameter noise model, demonstrating
\begin{itemize}
\item Inference using a Whittle-based likelihood after a suitable taper is applied to the gap edges correctly recovers the uncertainty of signal parameters but can grossly mis-estimate the noise parameters. The same conclusions follow when the diagonal likelihood is corrected by simply applying a normalising constant.
\item We found significant improvement on both signal and noise parameter inference when a diagonal approximation was used to the full windowed covariance matrix. Convolving the PSD with the window function in the frequency domain correctly weights the noise spectrum with the window across frequencies of interest, giving correct parameter uncertainties for mild and moderate gap families. 
\item The diagonal approximation fails in the limit of many short-duration gaps tightly spaced together where leakage effects are predominant. This was true for our drastic gap scenario, where $150$\,s of data are lost every $750$\,s and tapering is no longer a viable option. The signal/noise parameters are consistent with the truth, but this comes at the cost of significantly greater uncertainties when compared to what is achievable with the correct model-likelihood. A remedy for this is an efficient (time-domain) solver, utilising conjugate gradient methods to iteratively solve principal submatrices of circulant matrices. This time-domain solver will only suffice provided the PSD does not change from gap to gap, otherwise a segmented likelihood must be applied. 
\end{itemize}
In summary, as a first-order (relatively robust) approximation one should approximate the model-covariance matrix as diagonal with elements given via a weighted sum between the frequency domain window and noise PSD
\begin{equation}\label{eq:best_cheap_approximation_covariance}
\tilde{\bSigma}(f;\blambda) = \frac{1}{2}\int |\tilde{w}(f - u)|^2 S_{n}(u;\blambda)\,\text{d}u
\end{equation}
and using a window function, $w$, that exhibits mild tapering pre and post gap. Approximating the model-covariance this way ensures (1) a fast diagonal based likelihood adding extra complexity on the order of only $\mathcal{O}(N\log N)$ and (2) consistent and accurately measured parameter estimates for both the noise and signal elements.
\subsection{Future work}
This manuscript has focused primarily on a frequency domain analysis to treat missing data in the context of parameter estimation. Aside from the time-domain motivated conjugate gradient iterative Toeplitz solver, many of the toolkits here have been implemented in the frequency domain. Time-frequency approaches within GW astronomy are becoming ever more popular, due to their apparent advantages for handling mild non-stationary features for noise inference. Similar techniques to those presented here can be applied in the time-frequency domain and a similar analysis could be conducted to understand if better approximations to \eqref{eq:best_cheap_approximation_covariance} could be developed. 

We have also focused on only one type of signal -- massive black holes, which are relatively compact in time but spread in frequency. It would be interesting to extend this analysis to galactic binaries (compact in frequency, spread in time) or, more challenging, extreme mass-ratio inspirals -- compact in neither frequency nor time. For such signals, the length of the data becomes challenging to model the full covariance matrix, however, one could easily monte-carlo over the noise and signal parameter estimates \eqref{eq:joint_mle} to gain an intuitive understanding of the impact on parameter estimation of the noise mis-modelling. Based on this work, we believe that applying a small taper and using the approximation \eqref{eq:best_cheap_approximation_covariance} may be sufficient to treat data gaps provided the gaps are not too frequent such that a tapering scheme does not reduce the duty cycle of the data stream too significantly. 

A word of caution on how far the decoupling should be pushed. The
vanishing of the cross blocks is a statement about the Fisher matrix and the
scatter matrix at leading order, and it holds for quadratic likelihoods of the form considered here.  
This does not mean that signal and noise parameters may safely be estimated
independently in every setting. Two limitations matter. First, our noise
model carries fixed spectral shapes and two free amplitudes. A flexible model
with free knees, tilts or spline knots can absorb coherent signal power into
the noise estimate, which is a failure of identifiability rather than of
Eq.~\eqref{eq:block_diag_fisher}. Second, in a global fit the unresolved
foreground is simultaneously signal and noise, and the clean factorisation we
demonstrate for a single loud massive black hole binary does not carry over
without further work. The measured correlation bound of
$|\mathrm{corr}| \le 0.047$ quoted above applies to the configuration
analysed here and should not be read as a general result.

Finally, within the LISA Science Ground Segment,  
teams are developing Global Fit pipelines to develop software that can simultaneously infer the parameters of the many overlapping GWs in the presence of (mildly) non-stationary noise. Later data challenges being released by the end of this year will contain data gaps where many of the conclusions presented here would apply and they should be tested in that more realistic setting.  
Additionally, the techniques described here do not just apply to space-based interferometers, but also ground-based detectors. Data challenges for the ET detector will feature gaps, and similarly the LVK collaboration regularly treats data streams with window functions and in most current analyses, normalising constants are not subsequently applied to the data stream. 
It would also be interesting to use the approximation given by \eqref{eq:best_cheap_approximation_covariance} for inference in that context. 

\section{Data Availability}
All data sets analysed are
simulated and can be regenerated in full from the companion code.
The
analysis code --- \texttt{gaplike}, a pip-installable stand-alone
\texttt{Python} package implementing arbitrary gap-pattern generation, the
windowed covariances of
Sec.~\ref{sec:approximations_windowed_fd_covariance} and the full
likelihood hierarchy including the exact time-domain solver, together with
the paper pipeline, the executed notebooks and every figure and table entry
of this manuscript --- is openly available at
\href{\gaplikerepo}{\faGithub\,\texttt{\detokenize{github.com/gaplike/gaplike}}};
the version used for this manuscript is release~\texttt{\gapliketag}, to
which every link in this article is pinned.
It is installable with \texttt{pip install gaplike}, and that release is
archived on Zenodo at
\href{https://doi.org/\gaplikedoi}{\texttt{\gaplikedoi}}.
Documentation, including a derivation-level walkthrough of the likelihood
hierarchy, the $\Upsilon$ and $\Xi$ diagnostics and instructions for
reproducing each figure of this manuscript from the cached posterior chains
and benchmark outputs,
is hosted at \href{\gaplikedocs}{\texttt{\detokenize{gaplike.github.io/gaplike}}}.
Each figure caption names, and links to, the script that produces it.
Waveform and instrument-response generation uses the publicly available
\texttt{lisabeta}~\cite{Marsat:2020rtl}, sampling uses
\texttt{emcee}~\cite{Foreman-Mackey:2012any}. The analysis further relies
on the open-source scientific Python stack:
\texttt{numpy}~\cite{harris2020array}, \texttt{scipy}~\cite{2020SciPy-NMeth},
\texttt{matplotlib}~\cite{Hunter:2007} and \texttt{corner}~\cite{corner}.



\section{Acknowledgements}
O.Burke and F.Pozzoli contributed equally in this work. MM and JG contributed to discussions in the early stages of formulation of the project, and contributed to the editing and finalisation of the manuscript. We thank Quentin Baghi for both detailed drafts on the report and suggesting alternative pre-conditioners for our time-domain analysis. O.B.~thanks John Veitch, Lorenzo Speri, Christian Chapman-Bird, Joseph Bayley and Graham Woan for the many fruitful discussions that helped scope this work. F.P. ~thanks Riccardo Buscicchio, Diganta Bandopadhyay and Alessandro Santini.
O.B.~acknowledges financial support from the Grant UKRI972 awarded via the UK Space Agency. 
F.P. is supported by MPG. M.M. gratefully acknowledge the support of the German Space Agency, DLR. The work is supported by the Federal Ministry for Economic Affairs and Climate Action based on a resolution of the German Bundestag (Project Ref. No. FKZ 50 OQ 2301). We~thank the noise subgroup of the coordination unit L2D for many useful discussions concerning data gaps. 
\appendix
\begin{widetext}
\section{Identities}\label{app:identities}
From ~\cite{Burke:2025bun}, the quantity $\bQ = (\bW\bSigma\bW)$ is not directly invertible so we use the Moore-Penrose pseudo inverse with identities:
\begin{subequations}
\begin{align}
	 \boldsymbol{A}^{+} \boldsymbol{A} \boldsymbol{A}^{+} &= \boldsymbol{A}^{+} \,, \label{eq:pseudo_inv_def_1}\\
	 \boldsymbol{A} \boldsymbol{A}^{+} \boldsymbol{A} &= \boldsymbol{A} \,, \\
	 \left(\boldsymbol{A} \boldsymbol{A}^{+} \right)^{\dagger} &= \boldsymbol{A}\boldsymbol{A}^{+} \,, \\
	 \left(\boldsymbol{A}^{+} \boldsymbol{A}\right)^{\dagger} &= \boldsymbol{A}^{+} \boldsymbol{A} \,.
\end{align}
	\label{eq:defpseudoinverse}
\end{subequations}
From Eq. \eqref{eq:pseudo_inv_def_1} and Eq.\eqref{eq:pseudo_inv_gate}, one can show over the subspace of observed samples that 
\begin{equation}
\partial_{\alpha}\bQ^+ = - \bQ^+ \partial_{\alpha}\bQ\bQ^+ \label{eq:pseudo_inv_derivative}
\end{equation}
Similarly, the Jacobi identity states that 
\begin{equation}
\partial_{\alpha}\log\det \bQ = \text{Tr}[\bQ^+ \partial_{\alpha}\bQ]\,.\label{eq:jacobi_inv_deriv} 
\end{equation}
We now derive an identity for quadratic forms that will be used extensively throughout this work. Let $\boldsymbol{v} \in \mathbb{R}^{N}$ be a random vector with covariance $\boldsymbol{V} = \mathbb{E}[\boldsymbol{v}\boldsymbol{v}^T]$, and let $\boldsymbol{M} \in \mathbb{R}^{N\times N}$ be a deterministic and symmetric matrix. Then
\begin{equation}
    \mathbb{E}[\boldsymbol{v}^T\boldsymbol{M}\boldsymbol{v}] 
    = \sum_{ij} M_{ij}\,\mathbb{E}[v_i\,v_j] 
    = \sum_{ij} M_{ij}\,V_{ij} 
    = \mathrm{Tr}[\boldsymbol{M}\boldsymbol{V}]\,,
    \label{eq:quadratic_form_identity}
\end{equation}
where the last equality follows from the Frobenius inner product $\sum_{ij}A_{ij}B_{ij} = \mathrm{Tr}[\boldsymbol{A}^T\boldsymbol{B}]$ and the symmetry of $\boldsymbol{V}$. We will also make use Isserlis' theorem ~\cite{isserlis_1918_1431593}, also known as Wick's theorem~\cite{wickstheorem} to compute even moments of a noise process $\bn \sim N(\boldsymbol{0},\bSigma)$. For the fourth moment, we have 
\begin{align}\label{eq:real_isserlis_theorem}
\mathbb{E}[\bn_i \bn_j \bn_k \bn_p] & = \mathbb{E}[\bn_i \bn_j]\,\mathbb{E}[\bn_k \bn_p] + \mathbb{E}[\bn_i \bn_k]\,\mathbb{E}[\bn_j \bn_p] + \mathbb{E}[\bn_i \bn_p]\,\mathbb{E}[\bn_j \bn_k] \\ 
& = (\bSigma)_{ij}(\bSigma)_{kp} + (\bSigma)_{ik}(\bSigma)_{jp} + (\bSigma)_{ip}(\bSigma)_{jk}\,,
\end{align}
We remark here that this should not be confused with the complex version of Isserlis' theorem, which although functionally different would provide the same result. For zero-mean multivariate Gaussian distributions, any odd numbered moment vanishes. Take for instance the third moment: since $\bn \sim N(0,\bSigma)$ is a zero-mean Gaussian vector with symmetric distribution around the mean, the symmetry $\bn \mapsto -\bn$ is a symmetry of the distribution. Hence
\begin{equation}\label{eq:third_moment_vanishes}
\langle \bn_i \bn_j \bn_k \rangle = \langle (-\bn_i)(-\bn_j)(-\bn_k) \rangle = -\langle \bn_i \bn_j \bn_k \rangle\end{equation}
which implies that $\langle \bn_i \bn_j \bn_k \rangle = 0$. Using the same rationale, it is trivial to prove by induction that higher order (odd) moments must vanish.

\section{Vanishing of the Signal--Noise Cross Block}\label{app:cross_block}

\subsection{Signal and noise Fisher block}\label{app:cross_block_fisher}

For the data model $\bd = \bh(\btheta) + \bn$ of
Eq.~\eqref{eq:data_model}, we write the ($\btheta$-dependent) residual
$\bn = \bn(\btheta) = \bd - \bh(\btheta)$, which coincides with the true noise
realisation at the true parameters. The windowed model log-likelihood over the
observed subspace then reads
\begin{equation}\label{eq:joint_likelihood}
    \log\mathcal{L}(\btheta,\blambda)
    = -\frac{1}{2}\,(\bW\bn)^T(\bSigma')^+(\bW\bn)
      - \frac{1}{2}\log\det(\bSigma')\,,
\end{equation}
with model covariance $\bSigma' = \bSigma'(\blambda)$; setting $\bh = 0$
recovers the noise-only likelihood~\eqref{eq:noise_model_likelihood}. We remark
here that the model covariance matrix could be that of any model,
including $\bSigma^\prime = \bW\bSigma\bW = \bQ$ as defined in
Sec.~\ref{sec:joint_fisher_matrix}. Since the signal and noise parameters are disjoint, we must have that
\begin{equation}\label{eq:disjoint_channels}
    \partial_\alpha \bh = 0\,, \qquad \partial_a \bSigma' = 0\,.
\end{equation}
At the true parameters the windowed residual (noise term) reduces to the windowed noise
$\bW\bn$, a zero-mean Gaussian vector with $\langle\bW\bn\rangle = 0$ and
$\langle(\bW\bn)(\bW\bn)^T\rangle = \bW\bSigma\bW = \bQ$.

Using $\partial_a(\bW\bn) = -\bW\partial_a\bh$, the symmetry of
$(\bSigma')^+$, and Eq.~\eqref{eq:disjoint_channels}, the signal parameter derivative of the log-likelihood is linear in the noise,
\begin{equation}\label{eq:signal_score}
    \partial_a\log\mathcal{L} = (\bW\partial_a\bh)^T(\bSigma')^+(\bW\bn)\,,
\end{equation}
whereas the noise parameter derivative, reproduced from Eq.~\eqref{eq:model_noise_score}, is
quadratic.
\begin{equation}\label{eq:noise_score_recap}
    \partial_\alpha\log\mathcal{L}
    = \frac{1}{2}(\bW\bn)^T(\bSigma')^+\bSigma'_\alpha(\bSigma')^+(\bW\bn)
      - \frac{1}{2}\text{Tr}[(\bSigma')^+\bSigma'_\alpha]\,.
\end{equation}

Differentiating the signal score~\eqref{eq:signal_score} with respect to a
noise parameter, using the pseudo-inverse identity
\eqref{eq:pseudo_inv_derivative},
$\partial_\alpha(\bSigma')^+ = -(\bSigma')^+\bSigma'_\alpha(\bSigma')^+$,
together with $\partial_\alpha(\bW\bn) = 0$, gives
\begin{equation}\label{eq:cross_second_deriv}
    \partial_\alpha\partial_a\log\mathcal{L}
    = -(\bW\partial_a\bh)^T(\bSigma')^+\bSigma'_\alpha(\bSigma')^+(\bW\bn)\,.
\end{equation}
Taking the expectation over noise realisations and using
$\langle\bW\bn\rangle = 0$,
\begin{equation}\label{eq:cross_fisher_zero}
    \bGamma'_{a\alpha}
    = \langle -\partial_\alpha\partial_a\log\mathcal{L}\rangle
    = (\bW\partial_a\bh)^T(\bSigma')^+\bSigma'_\alpha(\bSigma')^+
      \langle\bW\bn\rangle
    = 0\,.
\end{equation}
The same result follows by differentiating the noise
score~\eqref{eq:noise_score_recap} with respect to $\theta^a$, confirming the
symmetry $\bGamma'_{a\alpha} = \bGamma'_{\alpha a}$. Only
$\langle\bW\bn\rangle = 0$ was used, so
Eq.~\eqref{eq:cross_fisher_zero} holds for \emph{any} model covariance
$\bSigma'$, whether or not it matches the true windowed covariance $\bQ$. The
Fisher matrix is therefore block diagonal,
\begin{equation}\label{eq:block_diag_fisher}
    \bGamma'_{AB} =
    \begin{pmatrix} \bGamma'_{ab} & 0 \\[2pt] 0 & \bGamma'_{\alpha\beta}\end{pmatrix}\,,
    \qquad
    \bGamma'_{ab} = (\bW\partial_a\bh)^T(\bSigma')^+(\bW\partial_b\bh)\,,
\end{equation}
with the noise block $\bGamma'_{\alpha\beta}$ given by
Eq.~\eqref{eq:approx_fisher_matrix_windowed}.

\subsection{Scatter-to-width cross block}\label{app:subsec:scatter_to_width_cross_block}
Since the Fisher matrix~\eqref{eq:block_diag_fisher} is block diagonal, so is
its inverse, and the leading-order maximum-likelihood fluctuations
$\Delta\hat\Theta^A = (\bGamma')^{-1}_{AB}\,\partial_B\log\mathcal{L}$ do not
mix the two sectors,
\begin{equation}\label{eq:decoupled_mle}
    \Delta\hat\theta^a = (\bGamma')^{-1}_{ab}\,\partial_b\log\mathcal{L}\,,
    \qquad
    \Delta\hat\lambda^\alpha = (\bGamma')^{-1}_{\alpha\beta}\,\partial_\beta\log\mathcal{L}\,,
\end{equation}
the latter coinciding with
Eq.~\eqref{eq:noise_param_fluctuations_noise_mismodelling}. The cross block of
the scatter matrix is therefore
\begin{equation}\label{eq:cross_scatter_general}
    \langle\Delta\hat\theta^a\,\Delta\hat\lambda^\alpha\rangle
    = (\bGamma')^{-1}_{ab}(\bGamma')^{-1}_{\alpha\beta}\,
      \big\langle\,\partial_b\log\mathcal{L}\;\partial_\beta\log\mathcal{L}\,\big\rangle\,,
\end{equation}
so everything hinges on the score cross-covariance. Writing
$\bN := \bW\bn$, $u_i := [(\bSigma')^+\bW\partial_b\bh]_i$ and
$(\mathcal{A}_\beta)_{jk} := [(\bSigma')^+\bSigma'_\beta(\bSigma')^+]_{jk}$,
Eqs.~\eqref{eq:signal_score} and \eqref{eq:noise_score_recap} give
\begin{equation}\label{eq:score_cross_index}
    \big\langle\,\partial_b\log\mathcal{L}\;\partial_\beta\log\mathcal{L}\,\big\rangle
    = \frac{1}{2}\,u_i(\mathcal{A}_\beta)_{jk}\,\langle\bN_i\bN_j\bN_k\rangle
      - \frac{1}{2}\text{Tr}[(\bSigma')^+\bSigma'_\beta]\,u_i\langle\bN_i\rangle\,.
\end{equation}
Both terms are odd moments of the zero-mean Gaussian vector
$\bN\sim\mathcal{N}(0,\bQ)$, and so vanish given Eq.\eqref{eq:third_moment_vanishes}, giving 
\begin{equation}\label{eq:cross_scatter_zero}
    \langle\Delta\hat\theta^a\,\Delta\hat\lambda^\alpha\rangle = 0\,.
\end{equation}


Both the Fisher matrix and the scatter matrix are block diagonal in
$(\btheta,\blambda)$,
\begin{equation}
    \langle\Delta\hat\Theta^A\Delta\hat\Theta^B\rangle =
    \begin{pmatrix}
        \langle\Delta\hat\theta^a\Delta\hat\theta^b\rangle & 0 \\[2pt]
        0 & \langle\Delta\hat\lambda^\alpha\Delta\hat\lambda^\beta\rangle
    \end{pmatrix}\,,
\end{equation}
so the scatter-to-width matrix $\bUpsilon$ of
Eq.~\eqref{eq:upsilon_matrix_noise} inherits the same block structure. The
signal-parameter diagnostics of Ref.~\cite{Burke:2025bun} and the
noise-parameter diagnostics developed here are therefore independent and may be
computed separately, even under noise mis-modelling.

\section{Variance of the Fluctuating Pieces $F_{AB}[\bn]$}\label{app:variance_fluctuating}

The observed curvature of the log-likelihood decomposes block-wise into a
deterministic Fisher part and a fluctuating part,
$-\partial^2\log\mathcal{L} = \bGamma - \bF(\bn)$
[Eq.~\eqref{eq:hessian_blocks}]. This appendix shows that each fluctuating
block is subdominant: the signal--signal piece by the linear-signal
approximation, the noise--noise piece by
$\mathcal{O}(N_{\text{eff}}^{-1/2})$ self-averaging, and the noise--signal
cross piece --- whose mean vanishes identically --- by
$\mathcal{O}(N_{\text{eff}}^{-1/2})$ in the induced parameter correlation.

For the signal-signal block, recall that the fluctuating piece takes the form
\begin{equation}
\bF_{ab} = (\bW\partial_a\partial_b\bh)^T\bQ^+(\bW\bn)
\end{equation}
The variance of this quantity takes the form
\begin{align}
\text{Var}[\bF_{ab}(n)] = \mathbb{E}[(\bF_{ab}(n))^2] - \mathbb{E}[\bF_{ab}(n)]^2
\end{align}

Since the quantity $\bF_{ab}$ is linear in the noise term $\bn$, we necessarily have $\mathbb{E}[\bF_{ab}[n]] = 0$. The second term takes the form
\begin{equation}
\mathbb{E}[(\bF_{ab})^2] = (\bW\partial_a\partial_b\bh)^T\bQ^+(\bW\partial_a\partial_b\bh)\,.
\end{equation}
Drawing comparisons to $\Gamma_{ab} = (\bW\partial_{a}\bh)^T\bQ^+(\bW\partial_b\bh)$, we see that they share the same $\bQ^+$ structure, with the only difference being the number of parameter derivatives applied to the waveform template. Via the Linear Signal Approximation, we implicitly assume that corrections $|\partial_{a}\bh| \gg |\partial_{a}\partial_{b}\bh|$, so the variance of this term is negligible compared to the signal-signal Fisher matrix. In the limit of high SNR, this term is therefore negligible.

Consider now the noise--signal cross block,
$\bF_{a\alpha} = -(\bW\partial_a\bh)^T\bQ^+\bQ_{\alpha}\bQ^+(\bW\bn)$
[Eq.~\eqref{eq:cross_second_deriv}]. Being linear in $\bn$, it has
$\mathbb{E}[\bF_{a\alpha}] = 0$ exactly, cf.~Eq.~\eqref{eq:cross_fisher_zero}.
Writing $\partial_{\alpha}\log\bQ \equiv \bQ^{+}\bQ_{\alpha}$ and using
$\bQ^{+}\bQ\bQ^{+} = \bQ^{+}$,
\begin{align}
\text{Var}[\bF_{a\alpha}]
&= (\bW\partial_a\bh)^T\bQ^+\bQ_{\alpha}\bQ^+\bQ_{\alpha}\bQ^+(\bW\partial_a\bh) \\
&= (\bW\partial_a\bh)^T(\partial_{\alpha}\log\bQ)\,\bQ^{+}\,
   (\partial_{\alpha}\log\bQ)^{T}(\bW\partial_a\bh)
   \;\sim\; \bGamma_{aa}\,,
\end{align}
since $\partial_{\alpha}\log\bQ$ is essentially an $\mathcal{O}(1)$ contribution. The same entries, traced, give
$\bGamma_{\alpha\alpha} = \tfrac{1}{2}\text{Tr}[(\partial_{\alpha}\log\bQ)^{2}]
= \mathcal{O}(N_{\text{eff}})$. 
Given that $\Gamma_{\alpha a} = \Gamma_{a \alpha} = 0$, the observed curvature matrix (via the Hessian) is given by 
\begin{equation}
\boldsymbol{H} = \begin{pmatrix} \bGamma_{aa} & \bF_{a\alpha} \\
\bF_{a\alpha} & \bGamma_{\alpha\alpha} \end{pmatrix}\,, \qquad \boldsymbol{H}^{-1} \approx \frac{1}{\bGamma_{a a}\bGamma_{\alpha\alpha}}\begin{pmatrix} \bGamma_{\alpha\alpha} & -\bF_{a\alpha} \\
-\bF_{\alpha a} & \bGamma_{aa} \end{pmatrix}\,.  
\end{equation} 
where we have simplified $\det{H} = \bGamma_{aa}\bGamma_{\alpha\alpha} - \bF_{a\alpha}\bF_{\alpha a} \approx \bGamma_{a a}\bGamma_{\alpha\alpha}$ since 
\begin{equation}\bGamma_{aa}\bGamma_{\alpha\alpha} \sim \mathcal{O}(\text{SNR}^2 \cdot N_{\rm eff}) \gg \bF_{a\alpha}\bF_{\alpha a} \sim \mathcal{O}(\text{SNR}^2).  
\end{equation}
with parameter correlation given by the cross covariance normalised by the $1\sigma$ deviations predicted by the diagonal components:
$-\bF_{a\alpha}/\sqrt{\bGamma_{aa}\bGamma_{\alpha\alpha}}$. Finally, since
$\bF_{a\alpha}$ is zero-mean, its typical size is
$\sqrt{\text{Var}[\bF_{a\alpha}]}$, hence
\begin{equation}
    \rho_{a\alpha}
    = \frac{\sqrt{\text{Var}[\bF_{a\alpha}]}}
           {\sqrt{\bGamma_{aa}\,\bGamma_{\alpha\alpha}}}
    \;\sim\; \bGamma_{\alpha\alpha}^{-1/2}
    = \mathcal{O}\big(N_{\text{eff}}^{-1/2}\big)\,,
    \label{eq:cross_block_suppression}
\end{equation}
independent of the SNR, and entering the marginal widths only at
$\rho_{a\alpha}^{2} =
\text{Var}[\bF_{a\alpha}]/(\bGamma_{aa}\bGamma_{\alpha\alpha}) =
\mathcal{O}(N_{\text{eff}}^{-1})$. This is the same order as the
fluctuations neglected throughout
Sec.~\ref{sec:hessian_based_likelihood}; empirically, the joint chains of
Sec.~\ref{sec:verification} measure $|\mathrm{corr}| \le 0.047$. It is therefore safe to neglect this cross-block term since it carries negligible weight to (1) the overall precision measurement in parameters and (2) predicted values for the MLEs. 

Now focus on the noise-noise block. The variance of the fluctuating pieces $F_{\alpha\beta}$ in \eqref{eq:hessian_nn}  
\begin{align}
    \text{Var}[\bF_{\alpha\beta}] &= \langle \bF_{\alpha\beta}^2 \rangle - \langle \bF_{\alpha\beta}\rangle^2 \\
    & = \frac{1}{4}\text{Var}[(\bW \bn)^T \boldsymbol{\mathcal{Q}}_{\alpha\beta}(\bW \bn)] \\
    & = \frac{1}{4}\left(\langle\left[(\bW \bn)^T \boldsymbol{\mathcal{Q}}_{\alpha\beta}(\bW \bn)\right]^2\rangle - \langle\left[(\bW \bn)^T \boldsymbol{\mathcal{Q}}_{\alpha\beta}(\bW \bn)\right]\rangle^2\right)\label{eq:latter_term_variance_calc}
\end{align}
To make progress we can use index calculus. Let the vector $(\bW\bn)_i = \bN_i$; focusing on the first term,
\begin{equation}
    \frac{1}{4}\langle\left[(\bW \bn)^T \boldsymbol{\mathcal{Q}}_{\alpha\beta}(\bW \bn)\right]^2\rangle = \frac{1}{4}\boldsymbol{\mathcal{Q}}_{\alpha\beta,ij}\boldsymbol{\mathcal{Q}}_{\alpha\beta,kp}\langle \bN_i\bN_j\bN_k\bN_p\rangle 
\end{equation}
Using Isserlis' theorem \eqref{eq:real_isserlis_theorem},
\begin{align}\label{eq:isserlis_theorem}
\mathbb{E}[\bN_i \bN_j \bN_k \bN_p] = (\bQ)_{ij}(\bQ)_{kp} + (\bQ)_{ik}(\bQ)_{jp} + (\bQ)_{ip}(\bQ)_{jk}\,,
\end{align}
Notice that the first term in this expression
\begin{equation} \boldsymbol{\mathcal{Q}}_{\alpha\beta,ij}\boldsymbol{\mathcal{Q}}_{\alpha\beta,kp}\bQ_{ij}\bQ_{kp} = \text{Tr}[\boldsymbol{\mathcal{Q}}_{\alpha\beta}\bQ]^2 = \langle\left[(\bW \bn)^T \boldsymbol{\mathcal{Q}}_{\alpha\beta}(\bW \bn)\right]\rangle^2
\end{equation}
gives a neat cancellation of the second term in \eqref{eq:latter_term_variance_calc}. The non-zero terms are thus 
\begin{align}
\text{Var}[\bF_{\alpha\beta}] & = \frac{1}{4} \boldsymbol{\mathcal{Q}}_{\alpha\beta,ij}\boldsymbol{\mathcal{Q}}_{\alpha\beta,kp}\left((\bQ)_{ik}(\bQ)_{jp} + (\bQ)_{ip}(\bQ)_{jk}\right)\,.
\end{align}
noting that both $\boldsymbol{\mathcal{Q}}^T = \boldsymbol{\mathcal{Q}}$ and $\bQ = \bQ^T$, we identify each of these quantities as a trace 
\begin{equation}
\text{Var}[\bF_{\alpha\beta}] = \frac{1}{2} \text{Tr}[(\boldsymbol{\mathcal{Q}}_{\alpha\beta}\bQ)^2]
\end{equation}
Since $\text{Tr}[(\boldsymbol{\mathcal{Q}}_{\alpha\beta}\bQ)^{2}] =
\mathcal{O}(N_{\text{eff}})$ while $\bGamma_{\alpha\beta} =
\mathcal{O}(N_{\text{eff}})$, the relative size of the fluctuation is
$\sqrt{\text{Var}[\bF_{\alpha\beta}]}/\bGamma_{\alpha\beta} =
\mathcal{O}(N_{\text{eff}}^{-1/2})$: the noise-sector curvature
self-averages, consistent with the expansion of
Sec.~\ref{sec:mismodelling_fisher_matrix}. 

\section{Derivation of Scatter-to-Width and Width-to-Width Ratios}\label{app:derivation_scatter_to_width}

For ease of notation, let $\boldsymbol{\mathcal{A}}_{\alpha} = (\bSigma^\prime)^+  \partial_{\alpha}(\bSigma^\prime)(\bSigma^\prime)^+$, $\boldsymbol{\mathcal{T}}_{\alpha} = \text{Tr}[(\bSigma^\prime)^+ \partial_{\alpha}(\bSigma^\prime)]$ and $\bN = \bW\bn$. We can then compute:
\begin{align}\label{eq:big_nasty_FM_expression}
    \langle \Delta\hat{\lambda}^{\alpha}\Delta\hat{\lambda}^{\beta}\rangle = \frac{1}{4}([\boldsymbol{\Gamma}^{\prime}_{\alpha\rho}]^{-1})([\boldsymbol{\Gamma}^{\prime}_{\beta\sigma}]^{-1})\bigg\langle\left[\bN^T\boldsymbol{\mathcal{A}}_{\rho} \bN - \boldsymbol{\mathcal{T}}_{\rho}\right]\left[\bN^T\boldsymbol{\mathcal{A}}_{\sigma} \bN - \boldsymbol{\mathcal{T}}_{\sigma}\right]\bigg\rangle
\end{align}
Using index notation again, observe that the first term under expectation can be written as
\begin{align}
    \langle \bN^T \boldsymbol{\mathcal{A}}_{\rho}\bN\bN^T \boldsymbol{\mathcal{A}}_{\sigma}\bN \rangle &= (\boldsymbol{\mathcal{A}}_{\rho})_{ij}(\boldsymbol{\mathcal{A}}_{\sigma})_{kp}\langle \bN_{i}\bN_{j}\bN_{k}\bN_{p}\rangle
\end{align}

Applying Isserlis' theorem with $\langle \bN\bN^T\rangle = \bQ$:
\begin{align}
    (\boldsymbol{\mathcal{A}}_\rho)_{ij}(\boldsymbol{\mathcal{A}}_\sigma)_{kp}\langle \bN_i\bN_j\bN_k\bN_p\rangle 
    &= (\boldsymbol{\mathcal{A}}_\rho)_{ij}(\boldsymbol{\mathcal{A}}_\sigma)_{kp}\left[\bQ_{ij}\bQ_{kp} + \bQ_{ik}\bQ_{jp} + \bQ_{ip}\bQ_{jk}\right] \label{eq:isserlis_expansion}
\end{align}
The three terms evaluate to:
\begin{align}
    \text{Term 1:}&\quad (\boldsymbol{\mathcal{A}}_\rho)_{ij}(\boldsymbol{\mathcal{A}}_\sigma)_{kp}\bQ_{ij}\bQ_{kp} = \text{Tr}[\boldsymbol{\mathcal{A}}_\rho\bQ]\,\text{Tr}[\boldsymbol{\mathcal{A}}_\sigma\bQ]  \label{eq:term1}\\[6pt]
    \text{Term 2:}&\quad (\boldsymbol{\mathcal{A}}_\rho)_{ij}(\boldsymbol{\mathcal{A}}_\sigma)_{kp}\bQ_{ik}\bQ_{jp} = \text{Tr}[\boldsymbol{\mathcal{A}}_\rho\bQ\,\boldsymbol{\mathcal{A}}_\sigma\bQ]   \label{eq:term2}\\[6pt]
    \text{Term 3:}&\quad (\boldsymbol{\mathcal{A}}_\rho)_{ij}(\boldsymbol{\mathcal{A}}_\sigma)_{kp}\bQ_{ip}\bQ_{jk} = \text{Tr}[\boldsymbol{\mathcal{A}}_\rho\bQ\,\boldsymbol{\mathcal{A}}_\sigma\bQ]  \label{eq:term3}
\end{align}
Giving a simple expression
\begin{equation}
\langle \bN^T \boldsymbol{\mathcal{A}}_{\rho}\bN\bN^T \boldsymbol{\mathcal{A}}_{\sigma}\bN \rangle = \text{Tr}[\boldsymbol{\mathcal{A}}_\rho\bQ]\text{Tr}[\boldsymbol{\mathcal{A}}_\sigma\bQ] + 2\text{Tr}[\boldsymbol{\mathcal{A}}_\rho\bQ\,\boldsymbol{\mathcal{A}}_\sigma\bQ]
\end{equation}
The second noise terms in \eqref{eq:big_nasty_FM_expression} result in
\begin{equation}
    -\boldsymbol{\mathcal{T}}_{\mu}\langle\bN^T\boldsymbol{\mathcal{A}}_{\nu} \bN\rangle = -\boldsymbol{\mathcal{T}}_{\mu}\text{Tr}[\boldsymbol{\mathcal{A}}_{\nu}\bQ]
\end{equation}
Resulting in the final expression for the noise parameter covariance
\begin{equation}
\langle \Delta\hat{\lambda}^{\alpha}\Delta\hat{\lambda}^{\beta}\rangle = ([\boldsymbol{\Gamma}^{\prime}_{\alpha\rho}]^{-1})([\boldsymbol{\Gamma}^{\prime}_{\beta\sigma}]^{-1})\,\boldsymbol{\mathcal{I}}_{\rho\sigma}
\end{equation}
with matrix quantity
\begin{align}
    \boldsymbol{\mathcal{I}}_{\rho\sigma} &= \frac{1}{4}\left(\text{Tr}[\boldsymbol{\mathcal{A}}_\rho\bQ]\text{Tr}[\boldsymbol{\mathcal{A}}_\sigma\bQ] + 2\text{Tr}[\boldsymbol{\mathcal{A}}_\rho\bQ\,\boldsymbol{\mathcal{A}}_\sigma\bQ] - \boldsymbol{\mathcal{T}}_{\rho}\text{Tr}[\boldsymbol{\mathcal{A}}_{\sigma}\bQ] -\boldsymbol{\mathcal{T}}_{\sigma}\text{Tr}[\boldsymbol{\mathcal{A}}_{\rho}\bQ] +\boldsymbol{\mathcal{T}}_{\sigma}\boldsymbol{\mathcal{T}}_{\rho}\right) \\
    & = \frac{1}{2}\text{Tr}[\boldsymbol{\mathcal{A}}_\rho\bQ\;\boldsymbol{\mathcal{A}}_\sigma\bQ] + \frac{1}{4}\left(\text{Tr}[\boldsymbol{\mathcal{A}}_\rho\bQ] - \boldsymbol{\mathcal{T}}_\rho\right)\left(\text{Tr}[\boldsymbol{\mathcal{A}}_\sigma\bQ] - \boldsymbol{\mathcal{T}}_\sigma\right) \,,\label{eq:info_matrix}
\end{align}

When the model is correctly specified, $\bSigma' = \bQ$, the pseudo-inverse property gives $\boldsymbol{\mathcal{A}}_\rho\bQ = (\bSigma')^+\bSigma'_\rho$ and $\text{Tr}[\boldsymbol{\mathcal{A}}_\rho\bQ] = \boldsymbol{\mathcal{T}}_\rho$. The second term vanishes and $\boldsymbol{\mathcal{I}}_{\rho\sigma} = \frac{1}{2}\text{Tr}[(\bSigma')^+\bSigma'_\rho(\bSigma')^+\bSigma'_\sigma] = \Gamma'_{\rho\sigma}$, recovering the correctly-modelled result.
This gives the bias covariance
\begin{equation}
\langle\Delta\hat{\lambda}^\alpha\Delta\hat{\lambda}^\beta\rangle = (\boldsymbol{\Gamma}')^{-1}_{\alpha\rho}(\boldsymbol{\Gamma}')^{-1}_{\beta\sigma}\left\{\frac{1}{2}\text{Tr}[\boldsymbol{\mathcal{A}}_\rho\bQ\boldsymbol{\mathcal{A}}_\sigma\bQ] + \frac{1}{4}\left(\text{Tr}[\boldsymbol{\mathcal{A}}_\rho\bQ] - \boldsymbol{\mathcal{T}}_\rho\right)\left(\text{Tr}[\boldsymbol{\mathcal{A}}_\sigma\bQ] - \boldsymbol{\mathcal{T}}_\sigma\right)\right\}\,.
\end{equation}

Substituting in $\boldsymbol{\mathcal{A}}_\rho$ and $\boldsymbol{\mathcal{T}}_{\rho}$, and after some matrix algebra, we obtain the final expression

\begin{equation}
\begin{split}
\langle\Delta\hat{\lambda}^\alpha\Delta\hat{\lambda}^\beta\rangle
= (\boldsymbol{\Gamma}')^{-1}_{\alpha\rho}(\boldsymbol{\Gamma}')^{-1}_{\beta\sigma}\Bigg\{
&\frac{1}{2}\text{Tr}\!\left[(\bSigma')^+\partial_\rho(\bSigma')\,(\bSigma')^+\bQ\,(\bSigma')^+\partial_\sigma(\bSigma')\,(\bSigma')^+\bQ\right] \\
&+ \frac{1}{4}\text{Tr}\!\left[(\bSigma')^+\partial_\rho(\bSigma')\big((\bSigma')^+\bQ - \mathbb{I}\big)\right]
   \text{Tr}\!\left[(\bSigma')^+\partial_\sigma(\bSigma')\big((\bSigma')^+\bQ - \mathbb{I}\big)\right]\Bigg\}\,.
\end{split}
\end{equation}

\section{Mis-specified Maximum Likelihood: the Pseudo-true Point and the Godambe--White Covariance}
\label{app:sandwich}

In this appendix we derive the asymptotic distribution of the \emph{exact}
maximizer of the model likelihood \eqref{eq:noise_model_likelihood} under
mis-modelling, $\bSigma'(\blambda) \neq \bQ$. The results of
Sec.~\ref{sec:mismodelling_fisher_matrix} and
Appendix~\ref{app:derivation_scatter_to_width} are exact statements about the
\emph{linearized} statistic
\eqref{eq:noise_param_fluctuations_noise_mismodelling}, a quadratic form in the
data whose moments close under Isserlis' theorem. The maximizer actually
computed in an analysis (e.g.\ a numerical maximization, or the peak of a sampled posterior) is instead a nonlinear function of the data, and severe
mis-modelling may displace it beyond the quadratic neighbourhood of the truth. The appropriate description of the nonlinear
estimator is the classical asymptotic theory of mis-specified
M-estimators~\cite{huber1967behavior,white1982maximum}; here we specialise that
theory to the likelihood \eqref{eq:noise_model_likelihood} and show that every
ingredient reduces to trace functionals already introduced in this work. All model covariances
considered share the null space of $\bQ$ (the gap support is fixed and
$\blambda$-independent), so the pseudo-inverse and pseudo-determinant
identities \eqref{eq:pseudo_inv_derivative} and \eqref{eq:jacobi_inv_deriv}
may be differentiated in $\blambda$ without generating null-space terms, and
$\log\det$ is understood as restricted to the observed subspace, as in
Eq.~\eqref{eq:noise_model_likelihood}.

\subsection{The pseudo-true point}

Taking the expectation of the model log-likelihood
\eqref{eq:noise_model_likelihood} over noise realisations with the
quadratic-form identity \eqref{eq:quadratic_form_identity} gives
\begin{equation}
    \langle -\log\mathcal{L}(\blambda) \rangle
    = \frac{1}{2} K(\blambda)\,,
    \qquad
    K(\blambda) \equiv \log\det\bSigma'(\blambda)
    + \text{Tr}\big[\bSigma'(\blambda)^{+}\,\bQ\big]\,.
    \label{eq:limiting_objective}
\end{equation}
The function $K$ is, up to $\blambda$-independent constants, twice the
KL divergence from the true windowed process to the model,
\begin{equation}
    D_{\rm KL}\Big[\mathcal{N}(0, \bQ)\,\Big\|\,\mathcal{N}\big(0, \bSigma'(\blambda)\big)\Big]
    = \frac{1}{2}\Big[K(\blambda) - m - \log\det\bQ\Big]\,,
    \label{eq:kl_identification}
\end{equation}
with $m$ the dimension of the observed subspace. The log-likelihood is an
extensive sum over $N_{\text{eff}}$ weakly-correlated modes, so
$\log\mathcal{L}(\blambda)/N_{\text{eff}}$ concentrates on its mean with
fluctuations of relative order $\mathcal{O}(N_{\text{eff}}^{-1/2})$, uniformly
on compact parameter sets. By the standard consistency argument for extremum
estimators~\cite{white1982maximum}, the maximizer of $\log\mathcal{L}$
therefore converges not to the true parameters but to the maximizer of the
mean --- equivalently, by \eqref{eq:kl_identification}, to the KL-closest
member of the model family,
\begin{equation}
    \hat{\blambda} \longrightarrow \blambda^{*}
    = \arg\min_{\blambda} K(\blambda)\,.
\end{equation}
We refer to $\blambda^{*}$ as the \emph{pseudo-true} parameters: the member of
the model family closest, in the KL sense, to the true windowed
process. Differentiating $K$ with the identities
\eqref{eq:pseudo_inv_derivative} and \eqref{eq:jacobi_inv_deriv},
\begin{equation}
    \partial_{\alpha} K
    = -\,\text{Tr}\Big[(\bSigma')^{+}\bSigma'_{\alpha}
        \big((\bSigma')^{+}\bQ - \mathbb{I}\big)\Big]
    = -2\,\big\langle \partial_{\alpha}\log\mathcal{L}(\blambda) \big\rangle\,,
    \label{eq:K_gradient}
\end{equation}
so the stationarity condition defining $\blambda^{*}$ is precisely the
vanishing of the \emph{mean score},
\begin{equation}
    \big\langle \partial_{\alpha}\log\mathcal{L} \big\rangle\Big|_{\blambda^{*}}
    = \frac{1}{2}\,\text{Tr}\Big[(\bSigma')^{+}\bSigma'_{\alpha}
      \big((\bSigma')^{+}\bQ - \mathbb{I}\big)\Big]\Big|_{\blambda^{*}}
    = 0\,.
    \label{eq:pseudo_true_stationarity}
\end{equation}
This marks the pseudo-true point $\boldsymbol{\lambda}^{\star}$, the point $\boldsymbol{\hat{\lambda}} = \blambda^{\star}$ that satisfies Eq.\eqref{eq:pseudo_true_stationarity}.

In practice $\blambda^{\star}$ is obtained by numerical
minimisation of Eq.~\eqref{eq:pseudo_true}, which is cheap in the
diagonalised basis of Sec.~\ref{subsec:fullcov_trick}. Writing
$v_{j}(\blambda) = 10^{\lambda_{\rm tm}}\mu_{j} + 10^{\lambda_{\rm oms}}$ and
$q_{j} = [\boldsymbol{T}\bQ\boldsymbol{T}^{\dagger}]_{jj}$, the objective
reduces to
\begin{equation}
    \ln{\det}^{+}\bSigma'(\blambda) + \text{Tr}\big(\bSigma'(\blambda)^{+}\bQ\big)
    = \sum_{j} \left[\ln v_{j}(\blambda) + \frac{q_{j}}{v_{j}(\blambda)}\right]
    + \text{const}\,.
    \label{eq:kl_objective_diagonalised}
\end{equation}
Only the diagonal of $\boldsymbol{T}\bQ\boldsymbol{T}^{\dagger}$ is needed,
computed once. Each evaluation is then a scalar sum over the $r$ retained
modes, so the two-dimensional minimisation costs no more than a handful of
likelihood evaluations. The same quantities give $H$ and $J$ at
$\blambda^{\star}$, so the entire Godambe-White prediction follows from one
optimisation and requires no data.

\subsection{Expanding around the pseudo-true point}

The exact maximizer solves the estimating equation
$\partial_{\alpha}\log\mathcal{L}(\hat{\blambda}) = 0$. Rather than expanding
the likelihood about the truth, where the mean noise score is extensively large
under mis-modelling, we expand the noise score about the point where its
mean vanishes, $\blambda^{*}$:
\begin{equation}
    0 = \partial_{\alpha}\log\mathcal{L}(\blambda^{*})
      + \partial_{\beta}\partial_{\alpha}\log\mathcal{L}(\blambda^{*})\,
        \big(\hat{\blambda} - \blambda^{*}\big)^{\beta}
      + \mathcal{O}\big(\|\hat{\blambda} - \blambda^{*}\|^{2}\big)\,.
    \label{eq:score_expansion}
\end{equation}
We then decompose the
Hessian into its mean and a zero-mean fluctuation as in
Sec.~\ref{sec:mismodelling_fisher_matrix},
\begin{equation}
    -\partial_{\alpha}\partial_{\beta}\log\mathcal{L}(\blambda^{*})
    = H_{\alpha\beta} + \delta H_{\alpha\beta}(\bn)\,,
    \qquad
    H_{\alpha\beta}
    \equiv 
      \big\langle -\partial_{\alpha}\partial_{\beta}\log\mathcal{L} \big\rangle\Big|_{\blambda^{*}}
    = \frac{1}{2}\,\partial_{\alpha}\partial_{\beta} K\Big|_{\blambda^{*}}\,,
    \label{eq:hessian_concentration}
\end{equation}
where $\langle \delta H \rangle = 0$ with variance given by the Isserlis
calculation of Appendix~\ref{app:variance_fluctuating}, relatively
$\mathcal{O}(N_{\text{eff}}^{-1})$, so the fluctuation may be dropped at
leading order. 

We can compute the components of $H_{\alpha\beta}$ as follows. Differentiating \eqref{eq:K_gradient} once more with the
identity \eqref{eq:pseudo_inv_derivative},
\begin{align}
    H_{\alpha\beta}
    &= \frac{1}{2}\Big\{
       \text{Tr}\big[(\bSigma')^{+}\bSigma'_{\alpha}(\bSigma')^{+}\bSigma'_{\beta}(\bSigma')^{+}\bQ\big]
     + \text{Tr}\big[(\bSigma')^{+}\bSigma'_{\beta}(\bSigma')^{+}\bSigma'_{\alpha}(\bSigma')^{+}\bQ\big]
     - \text{Tr}\big[(\bSigma')^{+}\bSigma'_{\alpha\beta}(\bSigma')^{+}\bQ\big]
     \nonumber \\
    &\qquad\quad
     + \text{Tr}\big[(\bSigma')^{+}\bSigma'_{\alpha\beta}\big]
     - \text{Tr}\big[(\bSigma')^{+}\bSigma'_{\alpha}(\bSigma')^{+}\bSigma'_{\beta}\big]
       \Big\}\bigg|_{\blambda^{*}}\,,
    \label{eq:sandwich_H_explicit}
\end{align}
with $\bSigma'_{\alpha\beta} = \partial_{\alpha}\partial_{\beta}\bSigma'$.
Comparing with Sec.~\ref{sec:mismodelling_fisher_matrix}, this is nothing but
the effective (mean observed) Fisher matrix \eqref{eq:effective_fisher},
evaluated at the pseudo-true point instead of the truth:
\begin{equation}
    H_{\alpha\beta}
    = \bGamma'_{\alpha\beta} - \big\langle \bF'_{\alpha\beta} \big\rangle
      \Big|_{\blambda^{*}}
    = \bGamma'_{\alpha\beta}
    + \frac{1}{2}\,\text{Tr}\big[\boldsymbol{\mathcal{C}}'_{\alpha\beta}
      (\bQ - \bSigma')\big]\Big|_{\blambda^{*}}\,,
    \label{eq:sandwich_H_effective}
\end{equation}
with $\boldsymbol{\mathcal{C}}'_{\alpha\beta}$ defined in
Eq.~\eqref{eq:calC_definition}.

Note that $H$ is automatically positive semi-definite: it is
(half) the curvature of $K$ at its own minimum. Second, the score at
$\blambda^{*}$ is a \emph{zero-mean} quadratic form in the Gaussian data
[Eq.~\eqref{eq:pseudo_true_stationarity}] --- a weighted sum over many
weakly-correlated modes --- and therefore obeys a central limit theorem with
covariance
$J_{\alpha\beta} = \mathrm{Cov}\big[\partial_{\alpha}\log\mathcal{L},\,
\partial_{\beta}\log\mathcal{L}\big]_{\blambda^{*}}$. This is the entire
reason for expanding about $\blambda^{*}$ rather than the truth: it is the
unique point at which the leading term of \eqref{eq:score_expansion} is
centred, so that the remaining fluctuations are genuinely
$\mathcal{O}(N_{\text{eff}}^{-1/2})$ and the truncation is controlled.
Inverting \eqref{eq:score_expansion},
\begin{equation}
    \big(\hat{\blambda} - \blambda^{*}\big)^{\alpha}
    \simeq (H^{-1})^{\alpha\beta}\, \partial_{\beta}\log\mathcal{L}(\blambda^{*})
    \qquad\Longrightarrow\qquad
    \mathrm{Cov}\big[\hat{\blambda}\big] \simeq H^{-1} J\, H^{-1}\,,
    \label{eq:sandwich_derived}
\end{equation}
i.e.\ the estimator is asymptotically Gaussian, centred on $\blambda^{*}$,
with the Godambe--White ``sandwich'' covariance.
Under correct specification the information identity $J = H = \bGamma$
holds~\cite{godambe1960optimum} and the sandwich collapses to the familiar
$\bGamma^{-1}$; mis-modelling breaks precisely this identity, which is why $H$ (the mean model curvature) and $J$ (the actual score
fluctuations under the true process) must be tracked separately.

To compute the components of $J$, we focus on the noise score $\partial_{\rho}\log\mathcal{L}$. We
retain the notation
$\boldsymbol{\mathcal{A}}_{\alpha} = (\bSigma')^+ \bSigma'_{\alpha} (\bSigma')^+$,
$\boldsymbol{\mathcal{T}}_{\alpha} = \text{Tr}[(\bSigma')^+ \bSigma'_{\alpha}]$
and $\bN = \bW\bn$, $\langle \bN \bN^T \rangle = \bQ$, of
Appendix~\ref{app:derivation_scatter_to_width}. The noise score is the quadratic form
$\partial_{\rho}\log\mathcal{L} = \frac{1}{2}(\bN^{T}\boldsymbol{\mathcal{A}}_{\rho}\bN
- \boldsymbol{\mathcal{T}}_{\rho})$, so its central covariance follows from
the connected Isserlis contractions \eqref{eq:term2} and \eqref{eq:term3}:
\begin{equation}
    J_{\rho\sigma}
    = \frac{1}{2}\,\text{Tr}\big[\boldsymbol{\mathcal{A}}_{\rho}\,\bQ\,
      \boldsymbol{\mathcal{A}}_{\sigma}\,\bQ\big]\Big|_{\blambda^{*}}\,.
    \label{eq:sandwich_J_general}
\end{equation}
This is exactly the first term of the linearized scatter
\eqref{eq:info_matrix}, evaluated at $\blambda^{*}$ rather than at the truth;
the second term of \eqref{eq:info_matrix} --- the outer product of the mean
scores --- is absent here by the stationarity condition
\eqref{eq:pseudo_true_stationarity}. For the diagonal-family models
$\boldsymbol{\mathcal{A}}_{\rho} = \mathrm{diag}\big[\boldsymbol{A}^{*}_{\rho}\big]$
with $A^{*}_{\rho, j} = \big[(\bSigma')^{+}\bSigma'_{\rho}(\bSigma')^{+}\big]_{jj}
\big|_{\blambda^{*}}$, and \eqref{eq:sandwich_J_general} may be evaluated in
the one-sided complex frequency-domain representation: each positive-frequency
bin carries a proper (circularly-symmetric) complex residual with
$\mathrm{Cov}\big(|r_{j}|^{2}, |r_{k}|^{2}\big) = |\tilde{Q}_{jk}|^{2}$, and
the $\pm f$ pairing absorbs the factor $\frac{1}{2}$, giving
\begin{equation}
    J_{\rho\sigma}
    = \sum_{j,k} A^{*}_{\rho, j}\, \big|\tilde{Q}_{jk}\big|^{2}\,
      A^{*}_{\sigma, k}\,,
    \label{eq:sandwich_J_diag}
\end{equation}
with the sums over the positive-frequency bins of each independent data
channel. (The one-sided reduction drops the improper $\pm f$-coupling terms
$|\tilde{Q}_{j\,\overline{-k}}|^{2}$ generated by window leakage across DC and
Nyquist; for typical bands and gap durations these are negligible, and
\eqref{eq:sandwich_J_general} remains available whenever they are not.) We
emphasise the structure of \eqref{eq:sandwich_J_diag}: the \emph{model} enters
only through the diagonal weights $\boldsymbol{A}^{*}$, but the data fluctuate
with the \emph{true} covariance, and the off-diagonal $|\tilde{Q}_{jk}|^{2}$, the window-leakage correlations the diagonal model omits, contribute to
$J$ regardless. A diagonal model therefore cannot satisfy the information
identity on windowed data: the mis-modelling survives in the meat even when
the diagonal variances are matched exactly.

When the model family contains the truth, the KL
divergence \eqref{eq:kl_identification} attains its global minimum of zero at
the point where $\bSigma'(\blambda^{*}) = \bQ$: the pseudo-true parameters
coincide with the true ones and the estimator is asymptotically unbiased.
There the pseudo-inverse identities reduce both
\eqref{eq:sandwich_J_general} and \eqref{eq:sandwich_H_explicit} to
$\frac{1}{2}\text{Tr}(\bQ^{+}\bQ_{\alpha}\bQ^{+}\bQ_{\beta}) =
\bGamma_{\alpha\beta}$, and \eqref{eq:sandwich_derived} collapses to
$\bGamma^{-1}$, recovering the correctly-specified result. Conversely, the
linearized theory of Sec.~\ref{subsec:metrics_scatter_to_width} is recovered
as the small-displacement limit of the sandwich: replacing
$(\blambda^{*},\, H,\, J)$ by (one Newton step from the truth,
$\bGamma'$, the first term of \eqref{eq:info_matrix} at the truth) reproduces
Eqs.~\eqref{eq:biased_noise_params_mismodelling} and
\eqref{eq:noise_mismodelling_params_covariance}. The domain of validity of the
sandwich itself is set by the truncation of \eqref{eq:score_expansion}: the
model likelihood must be approximately quadratic in $\blambda$ over the
scatter scale $H^{-1} J H^{-1}$ around $\blambda^{*}$. Beyond that regime,
third-derivative (Edgeworth-type) corrections of higher-order M-estimation
theory would be required, which we do not pursue.

\section{The Determinant of $\bSigma$ Using the Schur Complement}
\label{app:schur_det}

This appendix derives Eq.~\eqref{eq:complement_identity}, the identity used
in Sec.~\ref{subsec:td_determinant} to evaluate $\log\det\bSigma_{OO}$ at
the cost of the gapped samples rather than the observed ones. We work
throughout with the block partition of Eq.~\eqref{eq:block_partition}, in
which the $m = |O|$ observed samples come first and the $g = |G| = N - m$
gapped samples second.

Provided $\bSigma_{OO}$ is invertible, which it is by
Sec.~\ref{subsec:td_exact}, this block matrix admits the exact
factorisation~\cite{horn2012matrix,zhang2005schur}
\begin{equation}
    \bSigma =
    \begin{pmatrix} \mathbb{I}_{m} & \boldsymbol{0} \\
        \bSigma_{GO}\bSigma_{OO}^{-1} & \mathbb{I}_{g}\end{pmatrix}
    \begin{pmatrix} \bSigma_{OO} & \boldsymbol{0} \\
        \boldsymbol{0} & \boldsymbol{S} \end{pmatrix}
    \begin{pmatrix} \mathbb{I}_{m} & \bSigma_{OO}^{-1}\bSigma_{OG} \\
        \boldsymbol{0} & \mathbb{I}_{g}\end{pmatrix}\,,
    \qquad
    \boldsymbol{S} \equiv \bSigma_{GG} -
    \bSigma_{GO}\bSigma_{OO}^{-1}\bSigma_{OG}\,,
    \label{eq:block_ldu}
\end{equation}
where $\boldsymbol{S}$ is the Schur complement of $\bSigma_{OO}$ in
$\bSigma$. The outer factors are unit triangular and so have unit
determinant. Taking determinants of \eqref{eq:block_ldu} therefore leaves
\begin{equation}
    \det\bSigma = \det\bSigma_{OO}\,\det\boldsymbol{S}\,.
    \label{eq:schur_det}
\end{equation}
Inverting \eqref{eq:block_ldu} factor by factor, and using that the inverse
of a unit triangular matrix is obtained by negating its off-diagonal block,
\begin{equation}
    \bSigma^{-1} =
    \underbrace{\begin{pmatrix} \mathbb{I}_{m} & -\bSigma_{OO}^{-1}\bSigma_{OG} \\
        \boldsymbol{0} & \mathbb{I}_{g}\end{pmatrix}}_{\boldsymbol{U}^{-1}}
    \underbrace{\begin{pmatrix} \bSigma_{OO}^{-1} & \boldsymbol{0} \\
        \boldsymbol{0} & \boldsymbol{S}^{-1} \end{pmatrix}}_{\boldsymbol{D}^{-1}}
    \underbrace{\begin{pmatrix} \mathbb{I}_{m} & \boldsymbol{0} \\
        -\bSigma_{GO}\bSigma_{OO}^{-1} & \mathbb{I}_{g}\end{pmatrix}}_{\boldsymbol{L}^{-1}}\,.
    \label{eq:block_inverse}
\end{equation}
To extract the $GG$ block of \eqref{eq:block_inverse}, let $\bR_{G} \in
\{0,1\}^{g \times N}$ select the gapped indices, in the same sense as the
$\bR$ of Sec.~\ref{subsec:td_missing}, so that
$\big[\bSigma^{-1}\big]_{GG} = \bR_{G}\bSigma^{-1}\bR_{G}^{T}$. The last $g$
rows of $\boldsymbol{U}^{-1}$ are $(\boldsymbol{0}\;\;\mathbb{I}_{g})$ and
the last $g$ columns of $\boldsymbol{L}^{-1}$ are
$(\boldsymbol{0}\;\;\mathbb{I}_{g})^{T}$, which is to say
\begin{equation}
    \bR_{G}\boldsymbol{U}^{-1} = \bR_{G}\,, \qquad
    \boldsymbol{L}^{-1}\bR_{G}^{T} = \bR_{G}^{T}\,.
    \label{eq:triangular_drop}
\end{equation}
The triangular factors therefore drop out of this block entirely, and since
$\boldsymbol{D}^{-1}$ is block diagonal with $\boldsymbol{S}^{-1}$ in its
$GG$ slot,
\begin{equation}
    \big[\bSigma^{-1}\big]_{GG}
    = \bR_{G}\boldsymbol{U}^{-1}\boldsymbol{D}^{-1}\boldsymbol{L}^{-1}\bR_{G}^{T}
    = \bR_{G}\boldsymbol{D}^{-1}\bR_{G}^{T}
    = \boldsymbol{S}^{-1}\,,
    \label{eq:gg_block_is_schur_inverse}
\end{equation}
so that $\det\boldsymbol{S} = 1/\det\big[\bSigma^{-1}\big]_{GG}$.
Substituting this into \eqref{eq:schur_det} gives
$\det\bSigma = \det\bSigma_{OO}\big/\det\big[\bSigma^{-1}\big]_{GG}$, and
taking logarithms returns Eq.~\eqref{eq:complement_identity} of the main
text. 
\section{Full-Covariance Posteriors Across the Three Scenarios}\label{app:abc_corner}
This appendix collects, for reference, the posteriors of the \emph{correct}
(full windowed covariance) analysis in the three gap scenarios, overlaid on
the same axes (Fig.~\ref{fig:corner_full_ABC}). All three are exactly
calibrated ($\Upsilon = \Xi = 1$), so every difference between them is
information lost to the gap pattern itself, not mis-modelling; the figure
is referenced at the start of Sec.~\ref{sec:results}, where the approximate
models are measured against these reference posteriors.

\begin{figure*}[t!]
\centering
\includegraphics[width=0.92\textwidth]{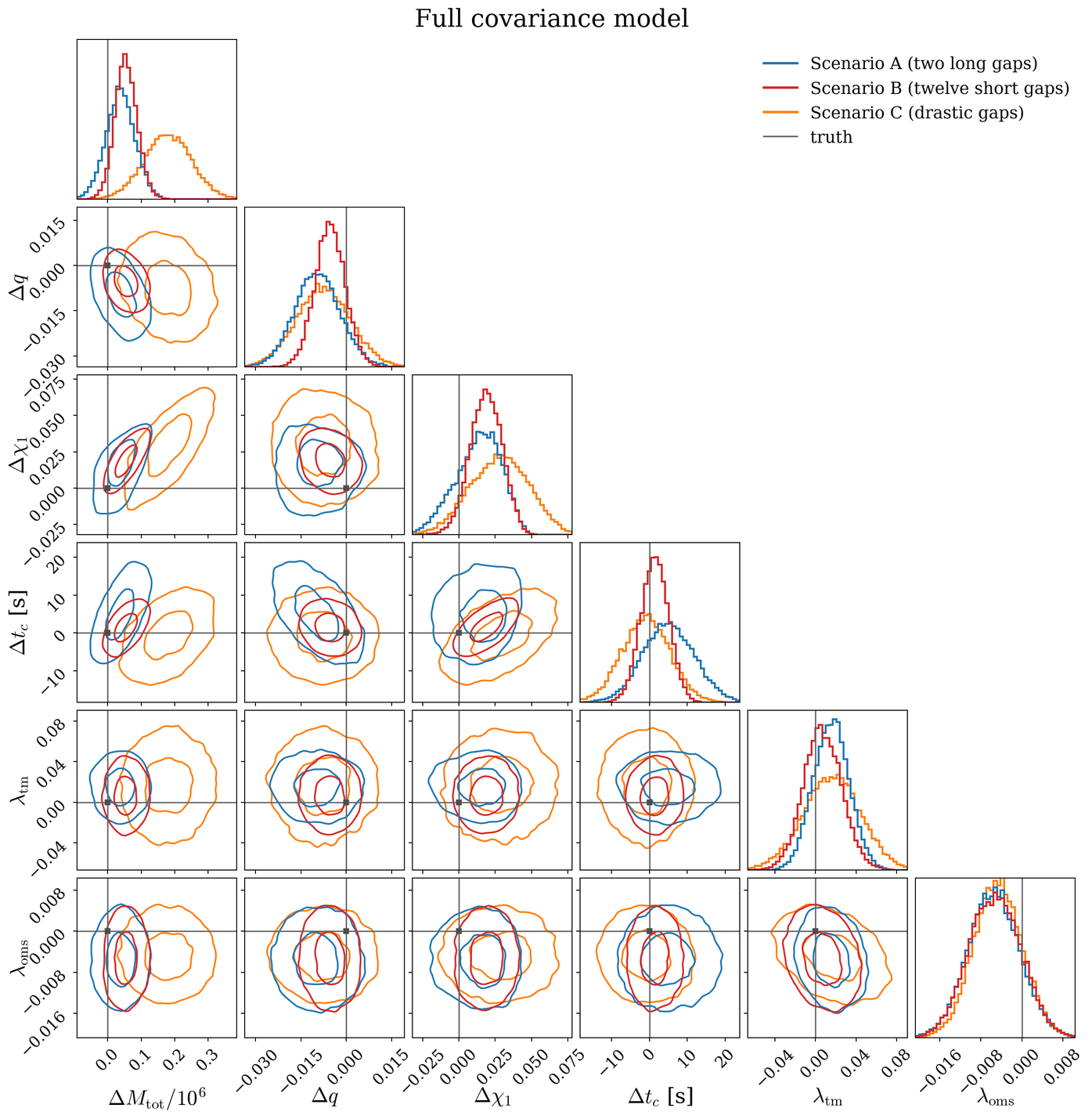}
\caption{What the \emph{correct} analysis attains in each scenario: the
full-covariance posteriors of A, B and C overlaid (key parameters, shifts
from the truth). All three are exactly calibrated ($\Upsilon = \Xi = 1$), so
every difference is information lost to the gap pattern, not mis-modelling.
The signal widths follow the SNR ordering (B, $743$; A, $585$; C, $319$),
with one instructive exception: $\Delta t_c$ is wider in A than in B despite
the similar SNRs, because the clipped merger carries the timing information.
In the noise sector $\lambda_{\rm oms}$ is indistinguishable across the
scenarios --- the OMS amplitude is measured by the loud high-frequency bins,
which every gap pattern retains --- while $\lambda_{\rm tm}$ widens in C,
where the aliased floor buries the quiet low-frequency bins that carry the
TM information. The visible offsets are realization scatter and are
correlated across scenarios, which share a single noise realization. Produced by \srclink{plot_ABC.py}.}
\label{fig:corner_full_ABC}
\end{figure*}

\end{widetext}

\bibliographystyle{apsrev4-2}
\bibliography{manual_refs,sample}

\end{document}